%% file: neurips_2025.tex
\documentclass[table]{article}

\usepackage[preprint]{neurips_2025}

\usepackage[utf8]{inputenc} 
\usepackage[T1]{fontenc}    
\usepackage[hyphens]{url}
\usepackage{hyperref}       
\usepackage{url}            
\usepackage{booktabs}       
\usepackage{amsfonts}       
\usepackage{nicefrac}       
\usepackage{microtype}      
\usepackage{amsmath,amssymb,amsfonts,graphicx}
\usepackage{bm}
\usepackage{mathtools}
\usepackage{pgffor}
\usepackage{geometry}
\usepackage{tabularx}
\usepackage[table]{xcolor}
\usepackage{booktabs}
\usepackage{tabularx}
\usepackage{makecell}
\usepackage{float}
\usepackage{placeins}
\usepackage{subcaption} 
\usepackage{makecell}

\title{
SSIMBaD: Sigma Scaling with SSIM-Guided Balanced Diffusion for AnimeFace Colorization
}

\author{
  Junpyo Seo \\
  Department of Computer Science \\
  Seoul National University \\
  \texttt{jpseo99@snu.ac.kr}
  \And
  Hanbin Koo \\
  Department of Computer Science \\
  Seoul National University \\
  \texttt{nagnebin@snu.ac.kr}
  \And
  Jieun Yook \\
  Department of Computer Science \\
  Seoul National University \\
  \texttt{yookje@snu.ac.kr}
  \And
  Byung-Ro Moon \\
  Department of Computer Science \\
  Seoul National University \\
  \texttt{moon@snu.ac.kr} \\
}

\begin{document}

\maketitle

\input{sections/abstract}
\input{sections/introduction}
\input{sections/relatedwork}
\input{sections/background}

\input{sections/proposition}

\input{sections/experiments}
\input{sections/conclusion}
\clearpage
\bibliography{ref}
\input{sections/appendix}

\end{document}

%% file: sections/abstract.tex
\begin{abstract}

We propose a novel diffusion-based framework for automatic colorization of Anime-style facial sketches, which preserves the structural fidelity of the input sketch while effectively transferring stylistic attributes from a reference image. Our approach builds upon recent continuous-time diffusion models, but departs from traditional methods that rely on predefined noise schedules, which often fail to maintain perceptual consistency across the generative trajectory. To address this, we introduce \textbf{SSIMBaD} (\textbf{S}igma \textbf{S}caling with SS\textbf{IM}-Guided \textbf{Ba}lanced \textbf{D}iffusion), a sigma-space transformation that ensures linear alignment of perceptual degradation, as measured by structural similarity. This perceptual scaling enforces uniform visual difficulty across timesteps, enabling more balanced and faithful reconstructions. Experiments on a large-scale Anime face dataset show that our method significantly outperforms state-of-the-art (SOTA) models in terms of both pixel-level accuracy and perceptual quality, while generalizing robustly to diverse styles and structural variations.
Code and implementation details are available at \footnote{\url{https://github.com/Giventicket/SSIMBaD-Sigma-Scaling-with-SSIM-Guided-Balanced-Diffusion-for-AnimeFace-Colorization}}.


\end{abstract}

%% file: sections/introduction.tex
\section{Introduction}

The rapid growth of content industries such as webtoons, animation, and virtual avatars has intensified the demand for automatic generation of high-quality Anime-style images. Among the various sub-tasks, colorizing sketch images remains a labor-intensive step in the content creation pipeline, as line art lacks shading and color information, requiring significant manual effort from artists. Automating this process not only enhances production efficiency but also ensures visual consistency across frames and styles \cite{Ci2018AnimeColorization, Furusawa2017Comicolorization}.

Early colorization models have been predominantly based on Generative Adversarial Networks (GANs). For instance,~\cite{Ci2018AnimeColorization, Sangkloy2017Scribbler, Zhang2018TwoStage} leveraged conditional GANs guided by sparse color scribbles as user-privided inputs.
However, these methods rely heavily on user-provided color hints and are sensitive to scribble placement and spatial correspondence. To alleviate this, Lee et al.~\cite{Lee2020ReferenceColorization} proposed reference-based colorization using a Spatially Corresponding Feature Transfer (SCFT) module that extracts semantic correspondences between the sketch and reference images. Yet, their approach struggles under large domain gaps or structural mismatches, a challenge that persists across reference-guided generation settings \cite{Li2022GradientConflict}.

Recently, diffusion models have emerged as a powerful class of generative models capable of producing high-fidelity images while avoiding common GAN pitfalls such as mode collapse and training instability \cite{Dhariwal2021DiffusionGANs, Ho2020DDPM, SohlDickstein2015Thermo, Song2019NCSN}. In particular, \cite{Cao2024AnimeDiffusion} was the first to apply Denoising Diffusion Probabilistic Models \cite{Ho2020DDPM} to anime face colorization. By leveraging pixel-level supervision and multi-scale structural similarity, they achieved noticeable improvements in Peak Signal-to-Noise Ratio (PSNR), Multi-Scale Structural Similarity Index (MS-SSIM)~\cite{Wang2004SSIM}, and Fréchet Inception Distance (FID)~\cite{heusel2017gans} compared to GAN-based baselines. However,  \cite{Cao2024AnimeDiffusion}'s discrete cosine-based forward noise schedule was shown to rapidly degrade SSIM in early timesteps while flattening in later steps, yielding uneven difficulty across the trajectory. This non-uniform perceptual degradation complicates reverse trajectory learning, making it harder to recover fine-grained details such as color textures \cite{Song2020ImprovedScore}.

Elucidated diffusion models (EDM) \cite{Karras2022EDM} introduced a continuous-time noise formulation in $\sigma$-space, allowing finer-grained control over corruption levels and enabling improved sample quality across noise scales. While EDM has demonstrated SOTA performance in image synthesis tasks, its use of logarithmic $\sigma$ sampling results in non-uniform perceptual changes in colorization tasks, where perceptual consistency over the trajectory is crucial. 

To address this issue, we propose a novel noise schedule, SSIM-aligned sigma-space scaling, which ensures that SSIM degradation progresses uniformly over time. Specifically, we introduce a novel monotonic squash-based transformation $\phi^*(\sigma)$ that maps the $\sigma$-space to perceptual difficulty space. This transformation yields a noise schedule that enforces equidistant degradation in SSIM space, which we integrate into the EDM framework to construct a continuous sigma-space diffusion model tailored for anime face colorization.

Our proposed schedule is utilized throughout both training and trajectory refining phases. During training, it encourages the model to learn from perceptually uniform noise levels, avoiding overfitting to either extreme noise or near-clean regimes. During inference, the same schedule ensures consistent reconstruction fidelity across all sampling steps. Unlike prior methods that optimize reverse fidelity purely empirically, we explicitly align the forward and reverse diffusion dynamics by leveraging SSIM as a perceptual anchor \cite{Wang2004SSIM, Zhang2018LPIPS}.

\vspace{1mm}
\noindent Our main contributions are summarized as follows:
\begin{itemize}
    \item \textbf{A Novel Unified Framework for Perceptually Balanced Diffusion} : We
propose \textbf{SSIMBaD} (\textbf{S}igma \textbf{S}caling with SS\textbf{IM}-Guided \textbf{Ba}lanced \textbf{D}iffusion, a pioneering framework that balances structural and stylistic
fidelity in anime face colorization. Unlike prior approaches that suffer
from inconsistencies in perceptual quality, SSIMBaD integrates
perceptual schedule alignment, training-time consistency, and trajectory
refinement to achieve stable and high-quality generation.


    \item \textbf{A Perceptual Sigma-Space Transformation for Enhanced Stability and
Consistency} : We propose a novel sigma-space transformation, $\phi^*(\sigma)$,
as a core innovation within SSIMBaD. By linearly aligning SSIM
degradation across diffusion timesteps, this perceptual rescaling
mechanism significantly improves step-wise consistency during the
generation process, ensuring consistent perceptual generation, and
overcoming the limitations of conventional noise schedules that often
bias towards low or high-frequency details.


    \item \textbf{State-of-the-Art Performance Validated by Comprehensive
Experimentation} : Extensive experiments on the Danbooru AnimeFace
dataset~\cite{Cao2024AnimeDiffusion} demonstrate that SSIMBaD substantially advances the state of
the art on PSNR, MS-SSIM, and FID metrics. Rigorous ablation studies
confirm that each component of SSIMBaD—the EDM backbone, sigma-
space scaling, and trajectory refinement— contributes to establishing a
new state of the art in anime colorization under both same-reference
and cross-reference settings.
\end{itemize}

%% file: sections/relatedwork.tex
\section{Related Works}

\input{sections/related_work/related_work1}
\input{sections/related_work/related_work2}
\input{sections/related_work/related_work3}
\input{sections/related_work/related_work4}

%% file: sections/related_work/related_work1.tex
\paragraph{GAN-Based Sketch Colorization}
Early colorization models primarily relied on GANs, guided by user-provided inputs such as sparse color scribbles~\cite{Ci2018AnimeColorization, Sangkloy2017Scribbler, Zhang2018TwoStage}. While effective, these approaches are highly sensitive to scribble placement and often fail to generalize. To mitigate this, Lee et al.~\cite{Lee2020ReferenceColorization} proposed a reference-based method using SCFT module, which extracts semantic alignments between sketches and reference images. However, SCFT remains vulnerable to domain gaps and structural mismatches~\cite{Li2022GradientConflict}. Other works explored semi-automatic pipelines~\cite{Furusawa2017Comicolorization} and two-stage GANs for flat-filling and shading~\cite{Zhang2018TwoStage}, or incorporated text tags for semantic guidance~\cite{Kim2019Tag2Pix}, but challenges in consistency and stability persist. 

%% file: sections/related_work/related_work2.tex
\paragraph{Generative Diffusion Models}
Diffusion models have emerged as powerful generative frameworks that address key limitations of GANs, including mode collapse and training instability~\cite{Ho2020DDPM, Song2021DDIM, Nichol2021ImprovedDiffusion, Dhariwal2021DiffusionGANs, Kingma2021VDM, Salimans2022ProgressiveDistillation}. By learning to reverse a gradual noising process, they enable stable training and high-quality image synthesis. Nichol and Dhariwal~\cite{Nichol2021ImprovedDiffusion} demonstrated that well-tuned diffusion models can outperform GANs across diverse benchmarks.

Subsequent advancements have improved their flexibility and performance. Song et al.~\cite{Song2019NCSN, Song2020ImprovedScore, Song2021SDE} introduced a score-based formulation using stochastic differential equations (SDEs), enabling continuous-time generation and principled control over sampling dynamics. In parallel, several works have proposed deterministic sampling methods based on ordinary differential equations (ODEs), such as PNDM~\cite{Liu2022PNDM} and DPM-Solver~\cite{Lu2022DPMSolver}, which accelerate inference while maintaining generation quality. Karras et al.~\cite{Karras2022EDM} extended this with EDM, which operate in continuous $\sigma$-space and decouple noise level selection from timestep scheduling. EDM achieves state-of-the-art results on high-resolution datasets such as FFHQ~\cite{Karras2019StyleGAN} and ImageNet~\cite{Deng2009ImageNet}.

%% file: sections/related_work/related_work3.tex
\paragraph{Reference-Guided Diffusion Colorization}

Diffusion models have shown strong potential for image colorization by conditioning the denoising process on inputs such as sketches or reference images. Techniques like classifier guidance~\cite{Dhariwal2021DiffusionGANs}, cross-attention, and adaptive normalization enable fine-grained control. User-guided methods such as SDEdit~\cite{Meng2022SDEdit} and DiffusArt~\cite{carrillo2023diffusart} leverage partial noise or scribbles for controllable generation, but often require carefully crafted inputs. ILVR~\cite{choi2021ilvr} and ControlNet~\cite{zhang2023controlnet} improve precision via reference alignment and auxiliary signals, yet depend on heavy Stable Diffusion backbones. In contrast, our approach maintains controllability within a lightweight architecture, making it more suitable for efficient deployment.

%% file: sections/related_work/related_work4.tex
\paragraph{AnimeDiffusion}
Cao et al.\cite{Cao2024AnimeDiffusion} pioneered the use of denoising diffusion probabilistic models (DDPMs)\cite{Ho2020DDPM} for anime face colorization. Leveraging pixel-wise supervision and multi-scale structural similarity (MS-SSIM)\cite{wang2003multiscale}, their method significantly improved PSNR, MS-SSIM, and FID compared to GAN-based baselines. However, like many diffusion models, the choice of noise schedule can introduce varying levels of perceptual distortion at different timesteps. If not carefully designed, this could potentially lead to uneven learning difficulty across the generative trajectory, which might affect the model's ability to reconstruct fine-grained texture and color details with uniform quality.

%% file: sections/background.tex
\section{Background: Elucidating the Design Space of Diffusion-Based Generative Models}
\label{sec:edm}

\input{sections/background/background1.tex}

%% file: sections/background/background1.tex
The EDM framework~\cite{Karras2022EDM} generalizes DDPM by introducing a continuous-time formulation of the forward noising process based on a scale variable $\sigma \in [\sigma_{\min}, \sigma_{\max}]$, which replaces the discrete timestep index $t$. Under this formulation, a clean image $x_0$ is perturbed into a noisy observation $x$ using a continuous noise level:

\begin{equation}
\label{eq:corruption}
x = x_0 + \sigma \cdot \epsilon, \quad \epsilon \sim \mathcal{N}(0, \mathbf{I}).
\end{equation}
This allows the model to learn over a continuous spectrum of corruption strengths, offering greater flexibility than DDPM’s fixed timestep schedule.

To stabilize training and ensure scale-invariant learning, the noisy input $x$ is preconditioned using the noise level $\sigma$ and a fixed constant $\sigma_{\text{data}}$ (typically 0.5). The network $F_\theta$ takes $x$ and $\sigma$ as input and produces a denoised estimate. The final prediction $D_\theta(x; \sigma)$ is computed using noise-dependent skip connections, as defined by:

\begin{equation}
D_\theta(x; \sigma) = c_{\text{skip}}(\sigma) \cdot x + c_{\text{out}}(\sigma) \cdot F_\theta(c_{\text{in}}(\sigma) \cdot x, \sigma),
\end{equation}
where $c_{\text{skip}}$, $c_{\text{in}}$, and $c_{\text{out}}$ are predefined scaling coefficients derived from $\sigma$.

At inference time, EDM defines the generative process as a reverse-time probability flow ODE, derived from the SDE framework in score-based diffusion models~\cite{Song2021SDE}:
\begin{equation}
\frac{d\mathbf{x}}{dt} = -\frac{1}{\sigma} \left(D_\theta(x, \sigma) - x\right).
\end{equation}
This ODE is numerically integrated using Euler or higher-order methods such as Heun or Runge-Kutta.

To discretize this continuous formulation, EDM introduces a $\rho$-parameterized noise schedule:
\begin{equation}
\label{edm_sigmas}
\sigma_i = \left( \sigma_{\max}^{1/\rho} + \frac{i}{N - 1}(\sigma_{\min}^{1/\rho} - \sigma_{\max}^{1/\rho}) \right)^{\rho}, \quad i = 0, \dots, N-1.
\end{equation}
By adjusting $\rho$, sampling steps can be concentrated in low- or high-noise regions. Most constants and scheduling heuristics in this formulation are directly adopted from the original EDM framework~\cite{Karras2022EDM}.

%% file: sections/proposition.tex
\section{SSIMBad : Sigma Scaling with SSIM-Guided Balanced Diffusion for AnimeFace Colorization}
\label{section4}
We propose \textbf{SSIMBaD}, which incorporates a perceptually grounded noise schedule into the EDM~\cite{Karras2022EDM}. Unlike prior log-based schemes, SSIMBaD aligns forward and reverse trajectories using a transformation that ensures perceptually uniform SSIM degradation.

The model conditions on $I_{\text{cond}} \in \mathbb{R}^{H \times W \times 4}$, formed by concatenating a TPS-warped reference image $I_{\text{ref}}$ (with rotation) and an XDoG-style sketch $I_{\text{sketch}}$. The clean target $I_{\text{gt}} \in \mathbb{R}^{H \times W \times 3}$ is corrupted with Gaussian noise to produce $I_{\text{noise}}$, which is denoised over time conditioned on $I_{\text{cond}}$. We now describe the key components of SSIMBaD, with full implementation details and EDM adaptation provided in Appendix~\ref{appendix:proposition_details}.

\input{sections/proposition/proposition2}
\input{sections/proposition/proposition1}

%% file: sections/proposition/proposition2.tex
\subsection{SSIM-aligned Sigma-Space Scaling}
\label{sec:sigmascale}

The perceptual quality of diffusion models is highly sensitive to how noise is distributed across the denoising trajectory. In EDM, inference uses a $\rho$-parameterized schedule \eqref{edm_sigmas} to sample noise levels in a nonlinear manner, typically concentrating steps near low-noise regions. In contrast, training samples $\ln \sigma$ from a log-normal distribution $\mathcal{N}(P_{\text{mean}}, P_{\text{std}}^2)$, implicitly assuming a different transformation. This discrepancy implies that the transformation applied during training, $\phi_{\text{train}}(\sigma) = \log(\sigma)$, differs from that used in inference, $\phi_{\text{inference}}(\sigma) = \sigma^{1/\rho}$—resulting in a perceptual misalignment between forward and reverse trajectories.

To resolve this, we propose \textbf{SSIM-aligned sigma-space scaling}—a perceptually motivated strategy that defines a shared transformation $\phi: \mathbb{R}_+ \rightarrow \mathbb{R}$ used consistently across both training and inference. This transformation maps the noise scale $\sigma$ to a perceptual difficulty axis, ensuring visually uniform degradation throughout the diffusion process. Based on this transformation, we construct the noise schedule by interpolating linearly in the $\phi$-space:
\begin{equation}
\sigma_i = \phi^{-1} \left( \phi(\sigma_{\min}) + \frac{i}{N - 1} \left( \phi(\sigma_{\max}) - \phi(\sigma_{\min}) \right) \right), \quad i = 0, 1, \dots, N - 1.
\label{eq:scaled_sigma_schedule}
\end{equation}

To identify the optimal $\phi^*$, we consider a diverse candidate set $\Phi$ of analytic and squash-like transformations:
\[
\Phi = \left\{
\begin{aligned}
&\sigma, \quad \log(\sigma), \quad \log(1+\sigma), \quad \sigma^2, \quad \frac{1}{\sigma}, \quad \frac{1}{\sigma^2}, \quad \operatorname{arcsinh}(\sigma), \quad \tanh(\sigma), \\
&\mathrm{sigmoid}(\sigma), \quad \frac{\sigma}{\sigma + c}, \quad \frac{\sigma^p}{\sigma^p + 1}, \quad \log(\sigma^2 + 1), \quad \arctan(\sigma)
\end{aligned}
\right\}
\]
where $c > 0$ and $p > 0$ are tunable constants. Each $\phi$ is evaluated by how linearly its induced noise schedule aligns with perceptual degradation, measured by SSIM. Specifically, we compute the coefficient of determination ($R^2$) between $\sigma_i^\phi$ and SSIM degradation under additive noise:
\begin{equation}
\phi^* = \underset{\phi \in \Phi}{\arg\max} \ 
\mathbb{E}_{I_{\text{gt}}, \bm{n}} \left[
R^2\left( 
\left\{ \left( \sigma_i^\phi,\ \text{SSIM}\left( I_{\text{gt}} + \sigma_i^\phi \cdot \bm{n},\ I_{\text{gt}} \right) \right) \right\}_{i=0}^{N-1}
\right)
\right]
\end{equation}
where $(I_{\text{gt}}, \bm{n})$ are drawn from the data distribution and Gaussian noise, respectively.

Our empirical search reveals that $\phi^*(\sigma) = \frac{\sigma}{\sigma + 0.3}$ yields the highest $R^2$ and near-linear SSIM degradation. We adopt this transformation consistently in both training and inference, unifying the sampling dynamics across the diffusion process.

In addition, we replace the conventional $\log(\sigma)$ noise embedding with $c_{\text{noise}} = \phi^*(\sigma)$ to align temporal conditioning with the perceptual trajectory. This alignment stabilizes training, improves reconstruction fidelity, and enhances generalization across diverse reference domains (see Section~\ref{sec:experiments_scaling}).

%% file: sections/proposition/proposition1.tex
\subsection{Framework of SSIMBaD}

\paragraph{Denoising Network}
The denoising model $D_\theta$ follows a preconditioned residual design adapted from EDM~\cite{Karras2022EDM}, where the noisy input is scaled and fused with a learned residual correction. Distinctively, we replace the conventional $\log(\sigma)$ noise embedding with a perceptually grounded squash function $c_{\text{noise}}(\sigma) = \phi^*(\sigma) = \frac{\sigma}{\sigma + 0.3}$, ensuring better alignment with visual difficulty across the noise trajectory.

Formally, the denoiser is defined as:
\[
D_\theta(I_{\text{noise}}, I_{\text{cond}}; \sigma) = 
c_{\text{skip}}(\sigma) \cdot I_{\text{noise}} +
c_{\text{out}}(\sigma) \cdot F_\theta\left( c_{\text{in}}(\sigma) \cdot I_{\text{noise}},\ I_{\text{cond}};\ \phi^*(\sigma) \right),
\]
where the coefficients $(c_{\text{skip}}, c_{\text{out}}, c_{\text{in}})$ are derived from $\sigma$ using the EDM preconditioning formulation (Appendix~\ref{appendix:proposition_details}).

\paragraph{Training}
To expose the model to a perceptually balanced distribution of noise scales, we sample $\sigma$ such that $\phi^*(\sigma)$ is uniformly distributed over $[\phi^*(\sigma_{\min}), \phi^*(\sigma_{\max})]$. The noise embedding $c_{\text{noise}}$ is set to $\phi^*(\sigma)$, replacing traditional log-variance encodings. Given noisy input $x = I_{\text{gt}} + \bm{n}$ with $\bm{n} \sim \mathcal{N}(0, \sigma^2 \mathbf{I})$, the pretraining loss is:
\begin{equation}
\mathcal{L}_{\text{train}} = 
\mathbb{E}_{\phi^*(\sigma) \sim \mathcal{U}[\phi^*(\sigma_{\min}), \phi^*(\sigma_{\max})]} \, 
\mathbb{E}_{I_{\text{gt}} \sim p_{\text{data}}} \, 
\mathbb{E}_{\bm{n} \sim \mathcal{N}(0, \sigma^2 \mathbf{I})} 
\left\| D_\theta(I_{\text{gt}} + \bm{n}, I_{\text{cond}}; \sigma) - I_{\text{gt}} \right\|^2.
\end{equation}

\paragraph{Trajectory Refinement}
To further enhance perceptual fidelity, we apply trajectory refinement. The reverse diffusion process is initialized from a pure Gaussian noise sample $I^{(N-1)} \sim \mathcal{N}(0, \mathbf{I})$, and integrated backward using a perceptually scaled sigma schedule $\{\sigma_i\}_{i=0}^{N-1}$ derived from $\phi^*(\sigma)$. For each denoising step $i = N-1$ down to $0$ ($\sigma_{-1} = 0$), we perform \textbf{Euler} updates as:

\begin{equation}
I^{(i-1)} = I^{(i)} - \frac{\Delta t_i}{\sigma_i} \left( D_\theta(I^{(i)}, I_{\text{cond}}; \sigma_i) - I^{(i)} \right), \quad \Delta t_i = \sigma_i - \sigma_{i-1}.
\end{equation}

We optimize the entire reverse trajectory via:
\begin{equation}
\label{eq:trajectory_refinement}
\mathcal{L}_{\text{trajectory refinement}} = 
\mathbb{E}_{I_{\text{gt}} \sim p_{\text{data}}} \,
\mathbb{E}_{\bm{n} \sim \mathcal{N}(0, \mathbf{I})}
\left\| \textbf{Euler}(\bm{n}, I_{\text{cond}}, \{\sigma_i\}) - I_{\text{gt}} \right\|^2.
\end{equation}

\paragraph{Inference}
During inference, we reuse the same $\phi^*(\sigma)$ transformation and construct a deterministic schedule:
\begin{equation}
\sigma_i = \left( \phi^* \right)^{-1} \left[ 
    \phi^*(\sigma_{\min}) 
    + \frac{i}{N - 1} \cdot 
    \left( \phi^*(\sigma_{\max}) - \phi^*(\sigma_{\min}) \right)
\right], \quad i = 0, \dots, N - 1.
\end{equation}
We then apply the same Heun's method (i.e., improved Euler integration, a second-order Runge-Kutta method) as in trajectory refinement to produce the final image from pure noise.

%% file: sections/experiments.tex
\section{Experiments}
\label{section5}

\input{sections/experiments/experiment1}
\input{sections/experiments/experiment2}

\subsection{Experimental Results}
\input{sections/experiments/experiment3_1}
\input{sections/experiments/experiment3_2}

\input{sections/experiments/experiment3_3}

\input{sections/experiments/experiment4}

%% file: sections/experiments/experiment1.tex
\subsection{Dataset Description}

We evaluate our method on a benchmark dataset introduced by~\cite{Cao2024AnimeDiffusion}, specifically curated for reference-guided anime face colorization. The dataset comprises $31{,}696$ sketch--color training pairs and $579$ test samples, all resized to a resolution of $256 \times 256$ pixels. Each training instance consists of a ground-truth color image $I_{\text{gt}}$ and its corresponding sketch $I_{\text{sketch}}$, generated via an edge detection operator such as XDoG~\cite{winnemoller2012xdog}. The sketch images serve as the structural input, while the reference images provide appearance cues such as color and style.

We evaluate model robustness under two test settings. In the \textbf{same-reference} scenario, the reference image is a perturbed version of the ground-truth, sharing the same structural input as $I_{\text{sketch}}$. In the \textbf{cross-reference} scenario, the reference is randomly sampled from other test images, introducing variations in both color and facial attributes. This dual setup enables evaluation of reconstruction fidelity under ideal conditions and generalization under domain shift.

%% file: sections/experiments/experiment2.tex
\subsection{Evaluation Metrics}

We evaluate colorization performance using three standard quantitative metrics: PSNR, MS-SSIM, and FID. PSNR assesses pixel-level accuracy via mean squared error, though it aligns poorly with human perception for semantic or stylistic tasks.
MS-SSIM improves upon SSIM by incorporating multi-scale luminance, contrast, and structure comparisons, making it suitable for colorization with structural constraints.
FID measures the Fréchet distance between generated and real image features, capturing both realism and semantic fidelity.
These metrics collectively assess fidelity, structural consistency, and perceptual realism, and are reported under both same-reference and cross-reference settings. For full implementation details, please refer to Appendix~\ref{implementation_details}. 

%% file: sections/experiments/experiment3_1.tex
\subsubsection{Empirical Evaluation of SSIM-Aligned Sigma-Space Scaling Functions}
\label{sec:experiments_scaling}

To ensure perceptual consistency across the generative trajectory, we construct the noise schedule by uniformly sampling in a transformed $\phi(\sigma)$ space and applying its inverse. We empirically select $\phi(\sigma) = \frac{\sigma}{\sigma + 0.3}$ based on its near-linear SSIM degradation behavior. Full analysis details are provided in Appendix~\ref{app:phi_schedule}.

To construct a perceptually uniform noise schedule, we empirically analyze the relationship between SSIM degradation and transformed noise levels $\phi(\sigma)$ for various candidate functions. For each transformation $\phi$, a clean image $I_{\text{clean}}$ is corrupted at $N = 50$ different noise levels by adding scaled Gaussian noise as defined in \eqref{eq:corruption}.

\begin{table}[H]
\centering
\scriptsize
\setlength{\tabcolsep}{5pt}
\renewcommand{\arraystretch}{1.4}
\caption{Transformation functions $\phi(\sigma)$ sorted by $R^2$ linearity with SSIM degradation. Bounded squash functions yield the highest perceptual alignment.}
\label{tab:R_square_comparison}
\vspace{0.5em}

\resizebox{\textwidth}{!}{%
\begin{tabular}{lcccccccc}
\toprule
$\phi(\sigma)$ &
$\sigma^2$ &
$\frac{1}{\sigma^2}$ &
$\sigma$ &
$\frac{1}{\sigma}$ &
$\log(\sigma^2 + 1)$ &
$\log1p(\sigma)$ &
$\operatorname{arcsinh}(\sigma)$ &
$\frac{\sigma^p}{\sigma^p + 1}$ \\
$R^2$ &
0.0616 &
0.0624 &
0.0768 &
0.1183 &
0.2225 &
0.3754 &
0.4001 &
0.7332 \\
\midrule
$\phi(\sigma)$ &
$\operatorname{sigmoid}(\sigma)$ &
$\frac{\sigma}{\sigma + 0.9}$ &
$\tanh(\sigma)$ &
$\frac{\sigma}{\sigma + 0.7}$ &
$\log(\sigma)$ &
$\frac{\sigma}{\sigma + 0.1}$ &
$\frac{\sigma}{\sigma + 0.5}$ &
$\frac{\sigma}{\sigma + 0.3}$ \\
$R^2$ &
0.6837 &
0.8196 &
0.8650 &
0.8710 &
0.8972 &
0.9275 &
0.9277 &
\cellcolor{red!10}\textbf{0.9793} \\
\bottomrule
\end{tabular}
}
\end{table}

%% file: sections/experiments/experiment3_2.tex
\subsubsection{Evaluation under Same and Cross Reference Scenarios}

Table~\ref{tab:all_metrics}, Figure~\ref{fig:same_comparison}, and Figure~\ref{fig:cross_comparison} demonstrate that, due to SSIM-aligned sigma-space scaling, SSIMBaD with trajectory refinement outperforms existing SOTA methods under both same-reference and cross-reference scenarios.

Notably, the finetuned model from~\cite{Cao2024AnimeDiffusion} shows overall inferior performance compared to our trajectory-refined SSIMBad, achieving a lower PSNR of 13.32 and a lower MS-SSIM of 0.7001. Particularly striking is the difference in FID scores, with their model scoring 135.12, markedly worse than our model’s 34.98. Although the finetuning stage used in \cite{Cao2024AnimeDiffusion} improves PSNR and MS-SSIM compared to their pretrained model, it introduces a trade-off that adversely affects FID values. In contrast, our trajectory refinement consistently enhances all three metrics—PSNR, MS-SSIM, and FID—compared to our model without refinement. The finetuning method proposed by \cite{Cao2024AnimeDiffusion} performs reconstruction in both the forward and reverse processes. In contrast, our trajectory refinement conducts reconstruction only in the reverse process, using \eqref{eq:trajectory_refinement} as the objective function. Nevertheless, thanks to the diffusion schedule induced by SSIM-aligned sigma-space scaling, our method strongly enforces perceptual linearity along the generative trajectory. As a result, SSIMBaD with trajectory refinement achieves superior performance compared to the finetuned AnimeDiffusion.

Under the same-reference scenario illustrated in Figure~\ref{fig:same_comparison}-(h), trajectory refined SSIMBad consistently generates results that are visually more faithful and stylistically coherent compared to GAN-based methods and the variants proposed by~\cite{Cao2024AnimeDiffusion}, excelling in preserving facial structures, consistently applying reference colors, and effectively avoiding artifacts such as blurring or mode collapse. In the more challenging cross-reference scenario shown in Figure~\ref{fig:cross_comparison}-(h), while other methods tend to either excessively emphasize or ignore the reference image style, our model demonstrates robust generalization capabilities by successfully preserving both the structural integrity and color consistency, even when presented with unfamiliar reference images.

\begin{table}[H]
\centering
\scriptsize
\setlength{\tabcolsep}{5pt}
\renewcommand{\arraystretch}{1.3}
\caption{Quantitative comparison under both same-reference and cross-reference settings.}
\resizebox{\textwidth}{!}{%
\begin{tabular}{lcc|cc|cc|cc}
\toprule
\textbf{Method} & \textbf{Training} & 
& \multicolumn{2}{c|}{\textbf{PSNR (↑)}} 
& \multicolumn{2}{c|}{\textbf{MS-SSIM (↑)}} 
& \multicolumn{2}{c}{\textbf{FID (↓)}} \\
& & & \textbf{Same} & \textbf{Cross} 
  & \textbf{Same} & \textbf{Cross} 
  & \textbf{Same} & \textbf{Cross} \\
\midrule
SCFT~\cite{Lee2020ReferenceColorization} & 300 epochs & & 17.17 & 15.47 & 0.7833 & 0.7627 & 43.98 & 45.18 \\
AnimeDiffusion~\cite{Cao2024AnimeDiffusion} (pretrained)	 & 300 epochs & & 11.39 & 11.39 & 0.6748 & 0.6721 & 46.96 & 46.72 \\
AnimeDiffusion~\cite{Cao2024AnimeDiffusion} (finetuned)  & 300 + 10 epochs & & 13.32 & 12.52 & 0.7001 & 0.5683 & 135.12 & 139.13 \\
SSIMBaD (w/o trajectory refinement) & 300 epochs & & 15.15 & 13.04 & 0.7115 & 0.6736 & 53.33 & 55.18 \\
\rowcolor{red!10}
\textbf{SSIMBaD (w/ trajectory refinement)} & \textbf{300 + 10 epochs} & 
& \textbf{18.92} & \textbf{15.84} 
& \textbf{0.8512} & \textbf{0.8207} 
& \textbf{34.98} & \textbf{37.10} \\
\bottomrule
\end{tabular}
}
\label{tab:all_metrics}
\end{table}

\subsubsection{Comparison of Diffusion Schedules in DDPM, EDM, and EDM with SSIM-Aligned Sigma-Space Scaling}

Figure~\ref{fig:corruption_schedule} illustrates the behavior of the forward diffusion process for a single training image under different noise schedules. Specifically, it plots how SSIM values change across timesteps ($N=25$) and visualizes a series of 50 corrupted images corresponding to each timestep, allowing intuitive assessment of the degree of corruption. These findings emphasize the crucial role of scheduling in aligning diffusion dynamics with perceptual difficulty.



The DDPM baseline employs a cosine-based schedule, designed to increase noise linearly across discrete timesteps. As seen in the graph in Figure~\ref{fig:corruption_schedule}-(a), DDPM introduces minimal noise during early steps but abruptly escalates noise levels in later stages, resulting in uneven SSIM degradation(noise levels) across timesteps. This leads to difficulty in reconstruction during the reverse process.

EDM improves upon DDPM by interpolating noise levels in $\sigma$-space via a $\rho$-parameterized schedule, yielding a smoother degradation curve (Figure~\ref{fig:corruption_schedule}-(b)). However, SSIM changes are concentrated in the mid-$\sigma$ range, with saturation at both ends. As a result, only a portion of the schedule contributes effectively to training, reducing overall efficiency and biasing learning toward the central region.

As shown in Figure~\ref{fig:corruption_schedule}-(c), the proposed $\phi^*(\sigma)$ schedule, which employs SSIM-aligned sigma-space scaling, is designed so that SSIM degradation becomes linear with respect to the transformation of $\sigma$. The images corresponding to each timestep demonstrate that, at no stage, is there an excessive SSIM degradation; rather, smooth and balanced noise is introduced at every step. This uniformity ensures that all diffusion stages become equally important, thereby improving reconstruction reconstruction fidelity across all frequencies. Furthermore, it enables more stable training and interpretable sampling behavior.

\input{sections/discussion/same_reference_table}
\input{sections/discussion/cross_reference_table}

%% file: sections/discussion/same_reference_table.tex
\begin{figure*}[t]
\centering
\renewcommand{\arraystretch}{0.5}
\setlength{\tabcolsep}{2pt}

\begin{tabular}{cccccccc}
\textbf{(a)} & \textbf{(b)} & \textbf{(c)} & \textbf{(d)} & \textbf{(e)} & \textbf{(f)} & \textbf{(g)} & \textbf{(h)} \\

\raisebox{-.5\height}{\includegraphics[width=0.11\linewidth]{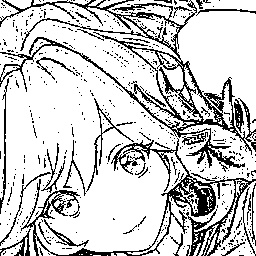}}
 & \raisebox{-.5\height}{\includegraphics[width=0.11\linewidth]{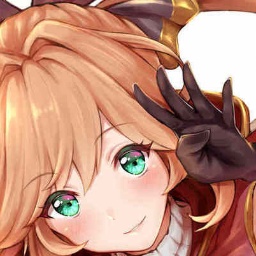}}
 & \raisebox{-.5\height}{\includegraphics[width=0.11\linewidth]{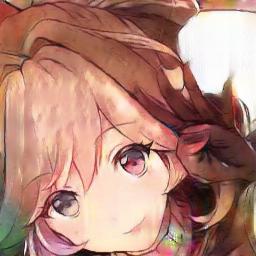}}
 & \raisebox{-.5\height}{\includegraphics[width=0.11\linewidth]{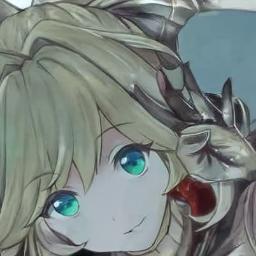}}
 & \raisebox{-.5\height}{\includegraphics[width=0.11\linewidth]{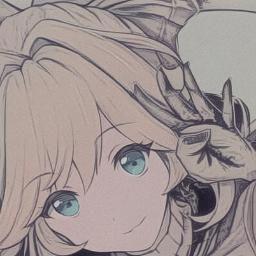}}
 & \raisebox{-.5\height}{\includegraphics[width=0.11\linewidth]{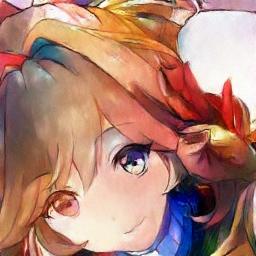}}
 & \raisebox{-.5\height}{\includegraphics[width=0.11\linewidth]{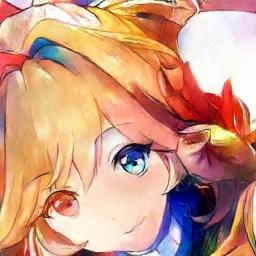}}
 & \raisebox{-.5\height}{\includegraphics[width=0.11\linewidth]{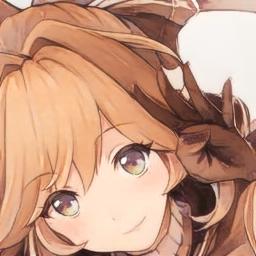}}
 \\
\raisebox{-.5\height}{\includegraphics[width=0.11\linewidth]{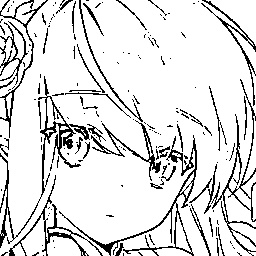}}
 & \raisebox{-.5\height}{\includegraphics[width=0.11\linewidth]{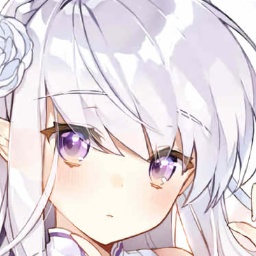}}
 & \raisebox{-.5\height}{\includegraphics[width=0.11\linewidth]{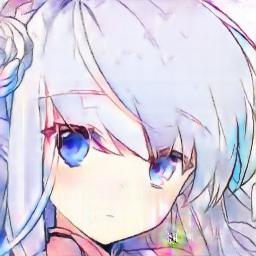}}
 & \raisebox{-.5\height}{\includegraphics[width=0.11\linewidth]{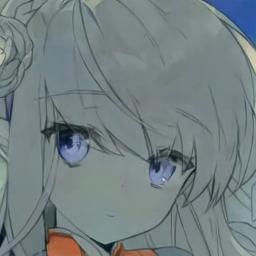}}
 & \raisebox{-.5\height}{\includegraphics[width=0.11\linewidth]{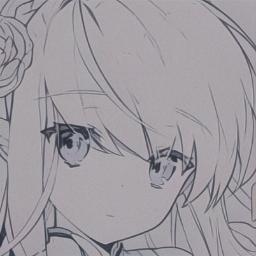}}
 & \raisebox{-.5\height}{\includegraphics[width=0.11\linewidth]{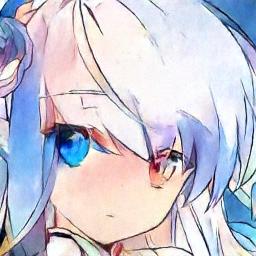}}
 & \raisebox{-.5\height}{\includegraphics[width=0.11\linewidth]{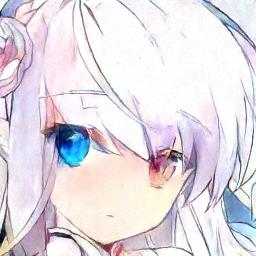}}
 & \raisebox{-.5\height}{\includegraphics[width=0.11\linewidth]{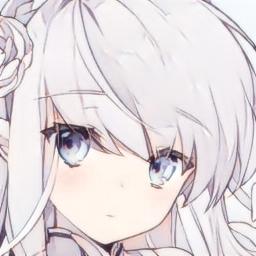}}
 \\
\raisebox{-.5\height}{\includegraphics[width=0.11\linewidth]{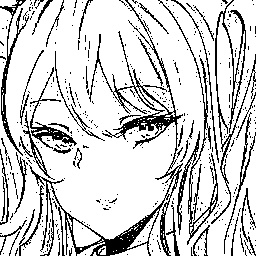}}
 & \raisebox{-.5\height}{\includegraphics[width=0.11\linewidth]{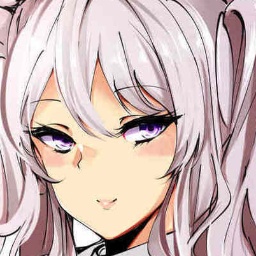}}
 & \raisebox{-.5\height}{\includegraphics[width=0.11\linewidth]{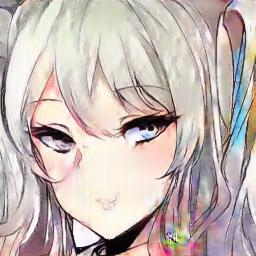}}
 & \raisebox{-.5\height}{\includegraphics[width=0.11\linewidth]{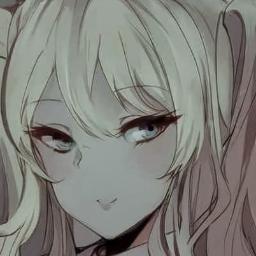}}
 & \raisebox{-.5\height}{\includegraphics[width=0.11\linewidth]{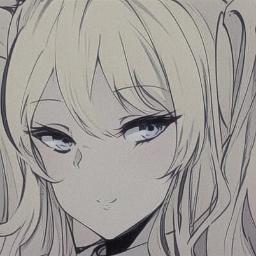}}
 & \raisebox{-.5\height}{\includegraphics[width=0.11\linewidth]{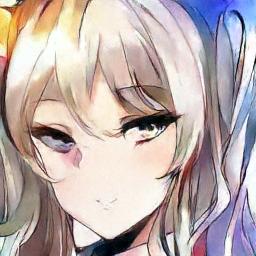}}
 & \raisebox{-.5\height}{\includegraphics[width=0.11\linewidth]{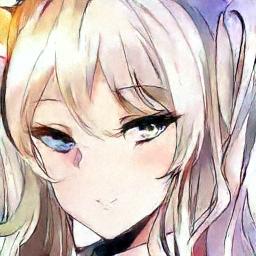}}
 & \raisebox{-.5\height}{\includegraphics[width=0.11\linewidth]{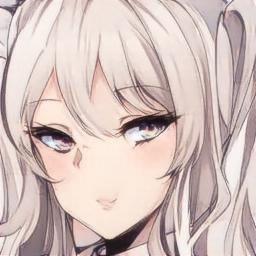}}
 \\
\raisebox{-.5\height}{\includegraphics[width=0.11\linewidth]{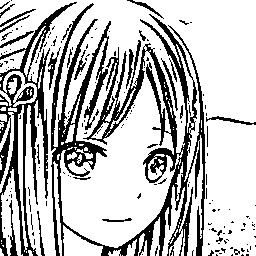}}
 & \raisebox{-.5\height}{\includegraphics[width=0.11\linewidth]{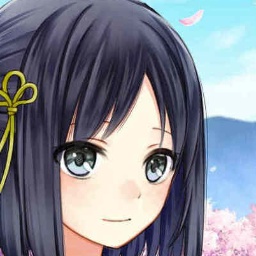}}
 & \raisebox{-.5\height}{\includegraphics[width=0.11\linewidth]{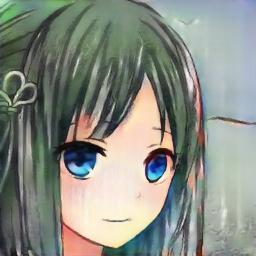}}
 & \raisebox{-.5\height}{\includegraphics[width=0.11\linewidth]{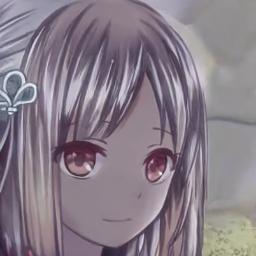}}
 & \raisebox{-.5\height}{\includegraphics[width=0.11\linewidth]{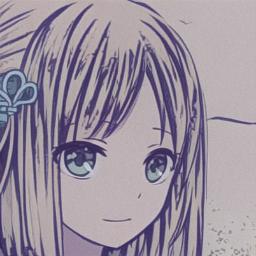}}
 & \raisebox{-.5\height}{\includegraphics[width=0.11\linewidth]{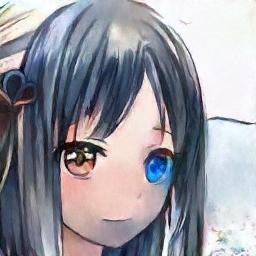}}
 & \raisebox{-.5\height}{\includegraphics[width=0.11\linewidth]{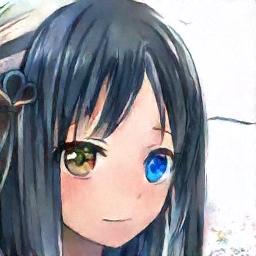}}
 & \raisebox{-.5\height}{\includegraphics[width=0.11\linewidth]{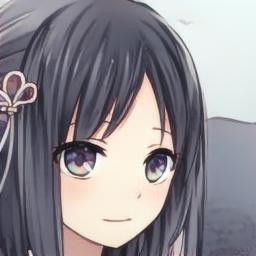}}
 \\
\end{tabular}
\caption{
\textbf{Qualitative comparison of colorization results under the same-reference scenario.}
From left to right: (a) Sketch input. (b) Reference image. (c) SCFT~\cite{Lee2020ReferenceColorization}. (d) AnimeDiffusion~\cite{Cao2024AnimeDiffusion} (pretrained). (e) AnimeDiffusion~\cite{Cao2024AnimeDiffusion} (finetuned). 
(f) AnimeDiffusion (EDM backbone, default $\sigma$-schedule). (g) Our model (w/o trajectory refinement). (h) Our model (w/ trajectory refinement).
}

\label{fig:same_comparison}
\end{figure*}

%% file: sections/discussion/cross_reference_table.tex
\begin{figure*}[t]
\centering
\renewcommand{\arraystretch}{0.5}
\setlength{\tabcolsep}{2pt}

\begin{tabular}{cccccccc}
\textbf{(a)} & \textbf{(b)} & \textbf{(c)} & \textbf{(d)} & \textbf{(e)} & \textbf{(f)} & \textbf{(g)} & \textbf{(h)} \\

\raisebox{-.5\height}{\includegraphics[width=0.11\linewidth]{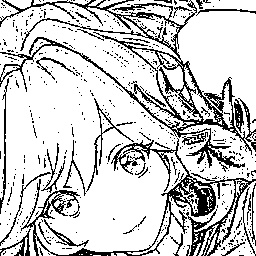}}
 & \raisebox{-.5\height}{\includegraphics[width=0.11\linewidth]{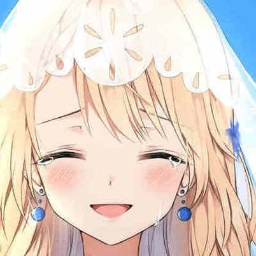}}
 & \raisebox{-.5\height}{\includegraphics[width=0.11\linewidth]{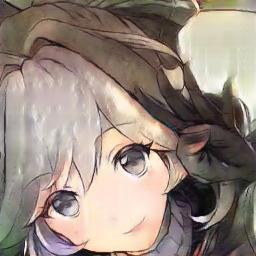}}
 & \raisebox{-.5\height}{\includegraphics[width=0.11\linewidth]{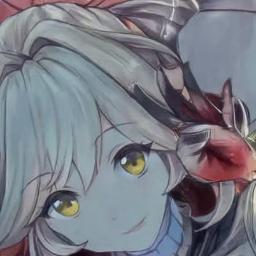}}
 & \raisebox{-.5\height}{\includegraphics[width=0.11\linewidth]{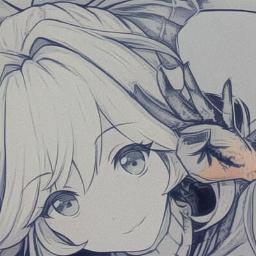}}
 & \raisebox{-.5\height}{\includegraphics[width=0.11\linewidth]{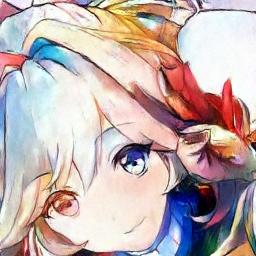}}
 & \raisebox{-.5\height}{\includegraphics[width=0.11\linewidth]{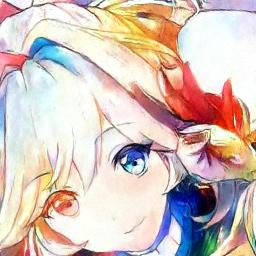}}
 & \raisebox{-.5\height}{\includegraphics[width=0.11\linewidth]{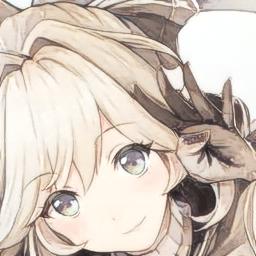}}
 \\
\raisebox{-.5\height}{\includegraphics[width=0.11\linewidth]{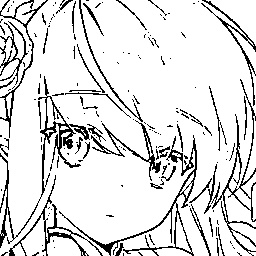}}
 & \raisebox{-.5\height}{\includegraphics[width=0.11\linewidth]{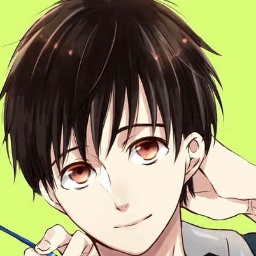}}
 & \raisebox{-.5\height}{\includegraphics[width=0.11\linewidth]{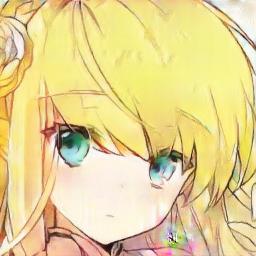}}
 & \raisebox{-.5\height}{\includegraphics[width=0.11\linewidth]{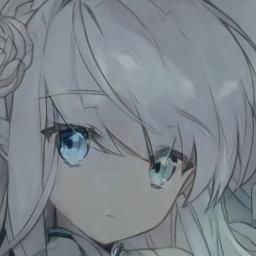}}
 & \raisebox{-.5\height}{\includegraphics[width=0.11\linewidth]{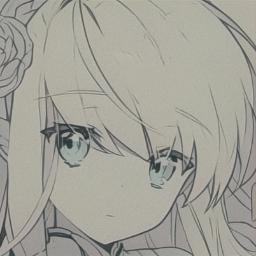}}
 & \raisebox{-.5\height}{\includegraphics[width=0.11\linewidth]{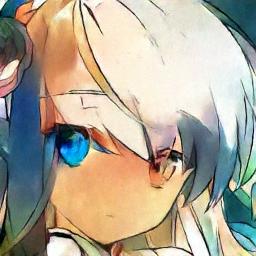}}
 & \raisebox{-.5\height}{\includegraphics[width=0.11\linewidth]{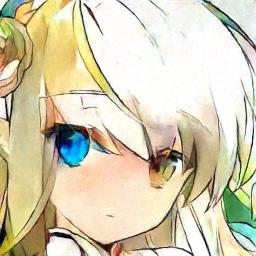}}
 & \raisebox{-.5\height}{\includegraphics[width=0.11\linewidth]{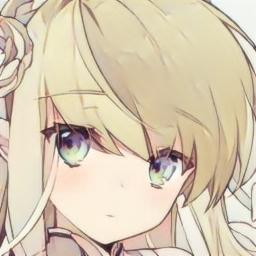}}
 \\
\raisebox{-.5\height}{\includegraphics[width=0.11\linewidth]{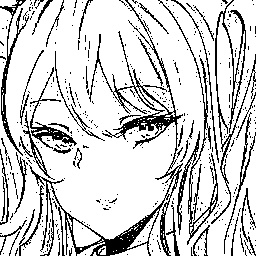}}
 & \raisebox{-.5\height}{\includegraphics[width=0.11\linewidth]{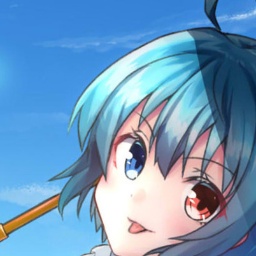}}
 & \raisebox{-.5\height}{\includegraphics[width=0.11\linewidth]{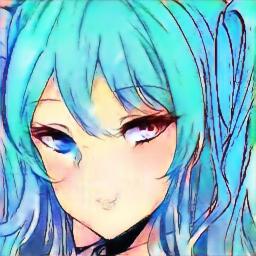}}
 & \raisebox{-.5\height}{\includegraphics[width=0.11\linewidth]{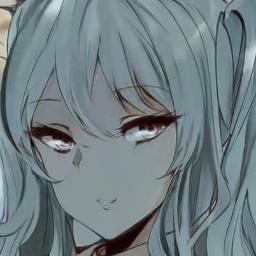}}
 & \raisebox{-.5\height}{\includegraphics[width=0.11\linewidth]{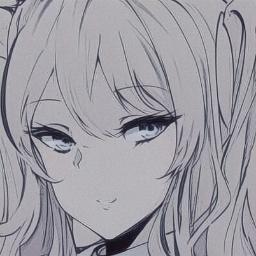}}
 & \raisebox{-.5\height}{\includegraphics[width=0.11\linewidth]{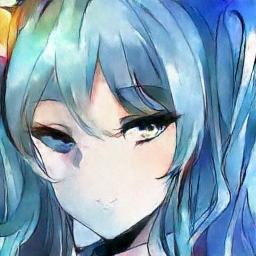}}
 & \raisebox{-.5\height}{\includegraphics[width=0.11\linewidth]{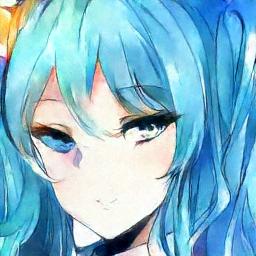}}
 & \raisebox{-.5\height}{\includegraphics[width=0.11\linewidth]{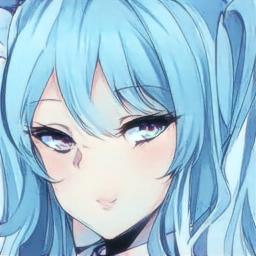}}
 \\
\raisebox{-.5\height}{\includegraphics[width=0.11\linewidth]{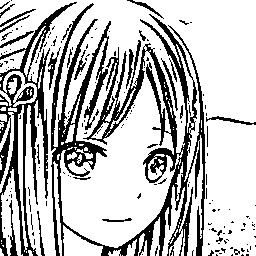}}
 & \raisebox{-.5\height}{\includegraphics[width=0.11\linewidth]{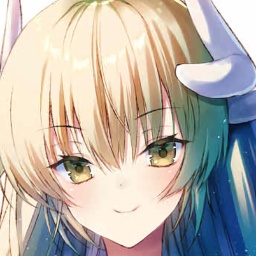}}
 & \raisebox{-.5\height}{\includegraphics[width=0.11\linewidth]{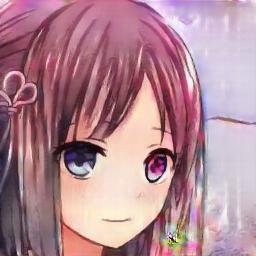}}
 & \raisebox{-.5\height}{\includegraphics[width=0.11\linewidth]{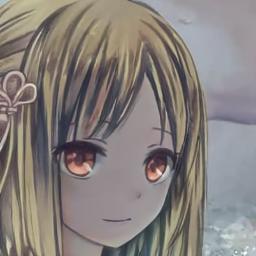}}
 & \raisebox{-.5\height}{\includegraphics[width=0.11\linewidth]{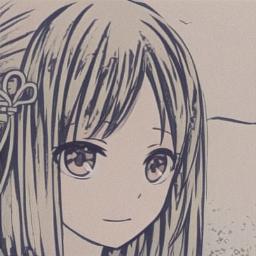}}
 & \raisebox{-.5\height}{\includegraphics[width=0.11\linewidth]{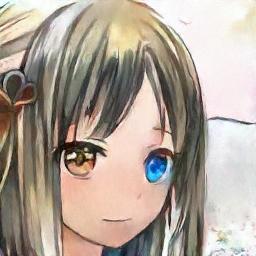}}
 & \raisebox{-.5\height}{\includegraphics[width=0.11\linewidth]{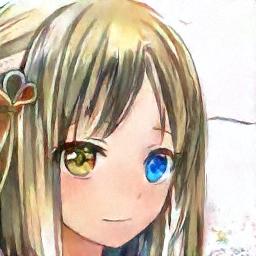}}
 & \raisebox{-.5\height}{\includegraphics[width=0.11\linewidth]{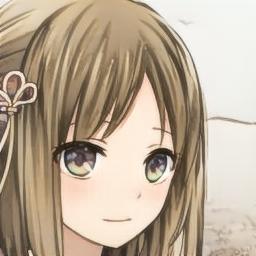}}
 \\
\end{tabular}
\caption{
\textbf{Qualitative comparison of colorization results under the cross-reference scenario.}
(a) Sketch input. (b) Reference image. (c) SCFT~\cite{Lee2020ReferenceColorization}. (d) AnimeDiffusion~\cite{Cao2024AnimeDiffusion} (pretrained). (e) AnimeDiffusion~\cite{Cao2024AnimeDiffusion} (finetuned). 
(f) AnimeDiffusion (EDM backbone, default $\sigma$-schedule).
(g) Our model (w/o trajectory refinement). (h) Our model (w/ trajectory refinement).
}

\label{fig:cross_comparison}
\end{figure*}

%% file: sections/experiments/experiment3_3.tex
\subsection{Ablation Study}


Table~\ref{tab:ablation_metrics} demonstrates how the major components of SSIMBaD, the EDM architecture, SSIM-aligned sigma-space scaling, and trajectory refinement-are cumulatively integrated into the baseline~\cite{Cao2024AnimeDiffusion}(which does not include the finetuning process), contributing to progressive performance improvements. In particular, SSIM-aligned sigma-space scaling substantially enhances perceptual quality, while the final trajectory refinement stage further increases fidelity and realism. 

When EDM is introduced, a slight initial decrease in MS-SSIM and FID is observed, likely due to the need for more training steps resulting from its continuous-time formulation. However, as additional modules were introduced, the balanced SSIM degradation across all diffusion timesteps effectively compensated for this initial decline.



The diffusion schedule induced by SSIM-aligned sigma-space scaling enables performance improvements with a simplified trajectory refinement, as compared to the finetuning approach proposed by ~\cite{Cao2024AnimeDiffusion}, by aligning SSIM degradation. As highlighted in Table~\ref{tab:ablation_metrics}, the proposed refinement stage not only stabilizes the sampling process but also brings consistent improvements across PSNR, MS-SSIM, and FID. These results underscore the importance of perceptual scheduling in structure-aware generation tasks.

\begin{table}[H]
\caption{Cumulative ablation study under both same- and cross-reference settings. Each added component incrementally improves model performance across all metrics and settings.}
\centering
\scriptsize
\renewcommand{\arraystretch}{1.3}
\setlength{\tabcolsep}{4pt}
\resizebox{\textwidth}{!}{%
\begin{tabular}{cccc|cc|cc|cc}
\toprule
\textbf{Base} & \textbf{+ EDM} & \makecell{\textbf{SSIM-aligned} \\ \textbf{sigma-space} \\ \textbf{scaling}} & \makecell{\textbf{+ Trajectory} \\ \textbf{Refinement}} &
\multicolumn{2}{c|}{\textbf{PSNR (↑)}} &
\multicolumn{2}{c|}{\textbf{MS-SSIM (↑)}} &
\multicolumn{2}{c}{\textbf{FID (↓)}} \\
& & & &
\textbf{Same} & \textbf{Cross} &
\textbf{Same} & \textbf{Cross} &
\textbf{Same} & \textbf{Cross} \\
\midrule
\checkmark & -- & -- & -- & 11.39 & 11.39 & 0.6748 & 0.6721 & 46.96 & 46.72 \\
\checkmark & \checkmark & -- & -- & 13.30 & 12.11 & 0.6426 & 0.6219 & 52.18 & 53.60 \\
\checkmark & \checkmark & \checkmark & -- & 15.15 & 13.04 & 0.7115 & 0.6736 & 53.33 & 55.18 \\
\rowcolor{red!10}
\checkmark & \checkmark & \checkmark & \checkmark &
\textbf{18.92} & \textbf{15.84} &
\textbf{0.8512} & \textbf{0.8207} &
\textbf{34.98} & \textbf{37.10} \\
\bottomrule
\end{tabular}
}
\label{tab:ablation_metrics}
\end{table}


\section{Conclusion}
This study proposes a novel framework, \textbf{SSIMBaD}, to address the perceptual inconsistency that has been overlooked in conventional diffusion-based models for anime-style face colorization. The core of SSIMBaD lies in the \textbf{SSIM-aligned sigma-space scaling}, which scales the noise parameter $\sigma$ to align with SSIM degradation. This approach ensures that structural corruption and restoration occur uniformly at each stage of the diffusion process, thereby enabling perceptually consistent training and inference. The SSIM-aligned sigma-space scaling is fully integrated into the EDM framework, replacing heuristic noise schedules and effectively leveraging continuous-time noise representation and preconditioning strategies. As a result, the proposed method achieves consistent colorization outcomes with respect to the reference images.

Experimental results on the Danbooru Anime Face dataset demonstrate that SSIMBaD outperforms benchmarks such as SCFT~\cite{Lee2020ReferenceColorization} and AnimeDiffusion~\cite{Cao2024AnimeDiffusion} under both same-reference and cross-reference conditions. Notably, the superior generalization ability of SSIMBaD is evident in the cross-reference condition. However, baselines, including ours, still have limitations in restoring fine details such as eye color. 

Furthermore, SSIMBaD enhances restoration fidelity in both low- and high-frequency regions, provides more interpretable sampling behavior, and unifies training, trajectory refinement, and inference under a single perceptual trajectory. This integration thereby reduces the risks of mismatch and overfitting. Beyond anime colorization, the proposed SSIM-algined sigma-space scaling offers potential for extension to various image generation tasks where spatial fidelity and perceptual balance are crucial, such as sketch-based synthesis, medical imaging, and controllable content generation.

\input{sections/discussion/various_corruption_figure}
\FloatBarrier

%% file: sections/discussion/various_corruption_figure.tex
\begin{figure}[t]
\centering

\begin{minipage}[b]{0.32\linewidth}
  \centering
  \includegraphics[width=\linewidth]{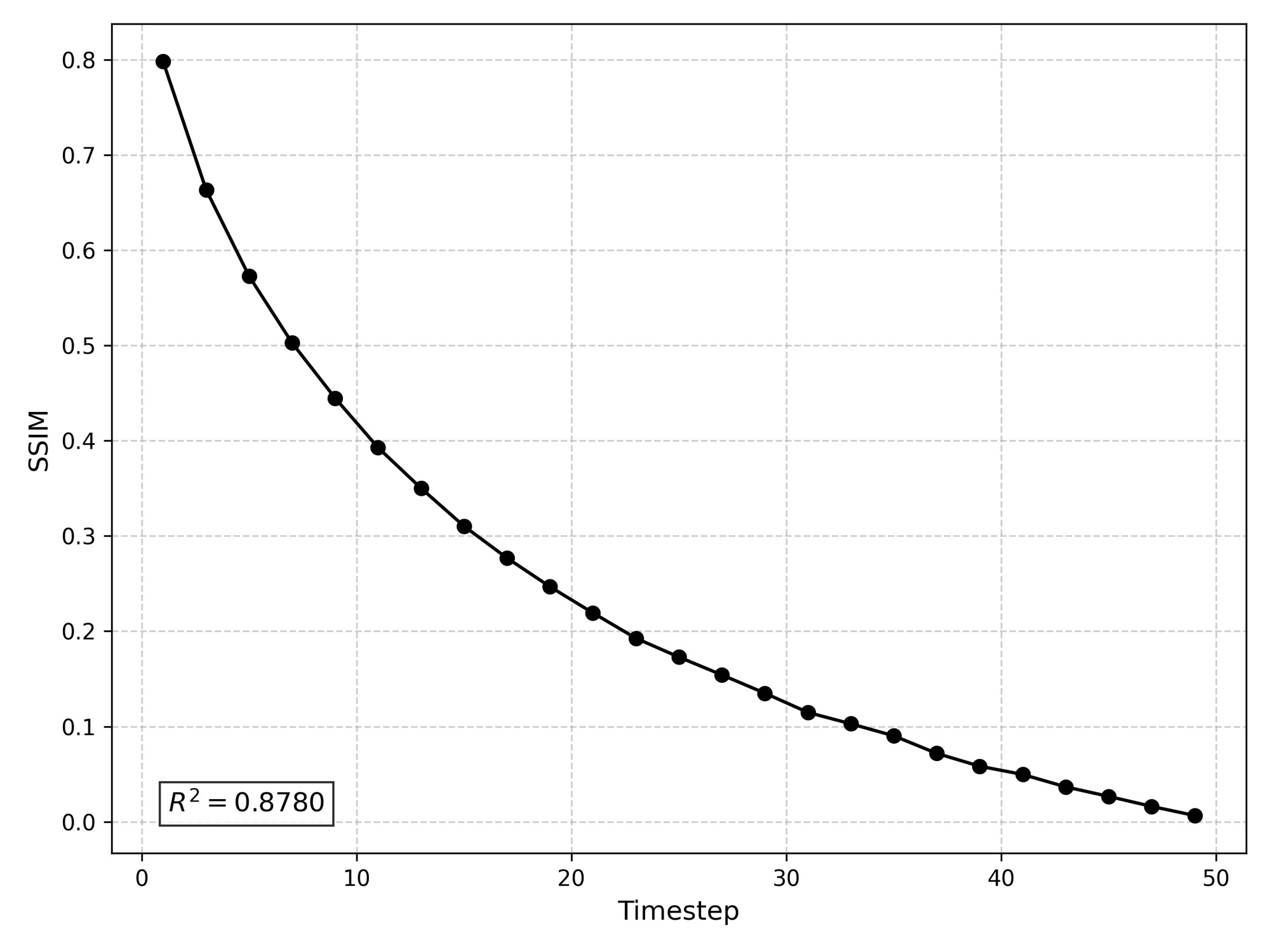}
  \vspace{0.3em}
  \textbf{(a)} SSIM Curve (DDPM)
\end{minipage}
\hfill
\begin{minipage}[b]{0.32\linewidth}
  \centering
  \includegraphics[width=\linewidth]{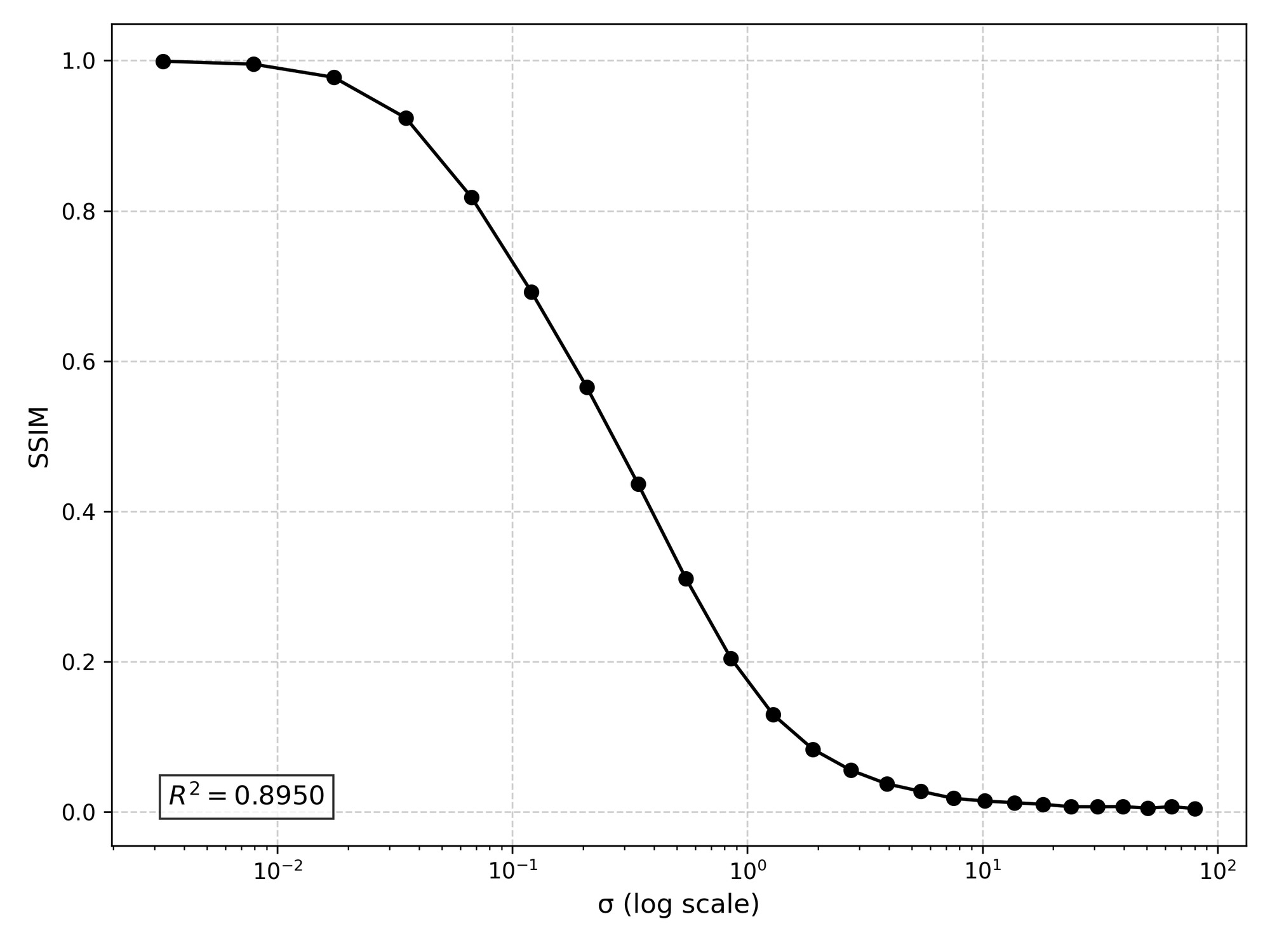}
  \vspace{0.3em}
  \textbf{(b)} SSIM Curve (EDM)
\end{minipage}
\hfill
\begin{minipage}[b]{0.32\linewidth}
  \centering
  \includegraphics[width=\linewidth]{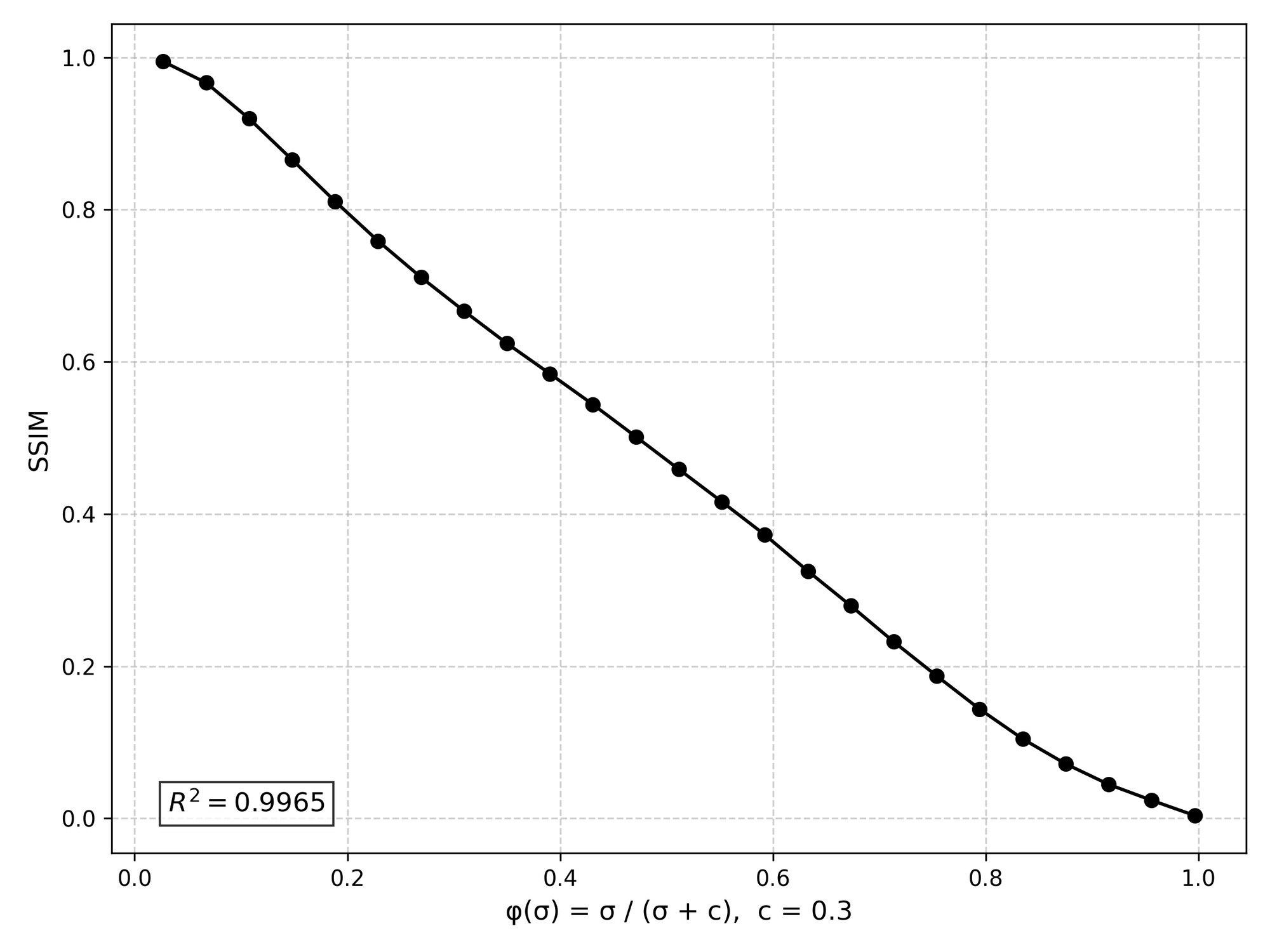}
  \vspace{0.3em}
  \textbf{(c)} SSIM Curve (SSIMBaD)
\end{minipage}

\vspace{1.5em}

\begin{minipage}[b]{0.32\linewidth}
  \centering
  \includegraphics[width=\linewidth]{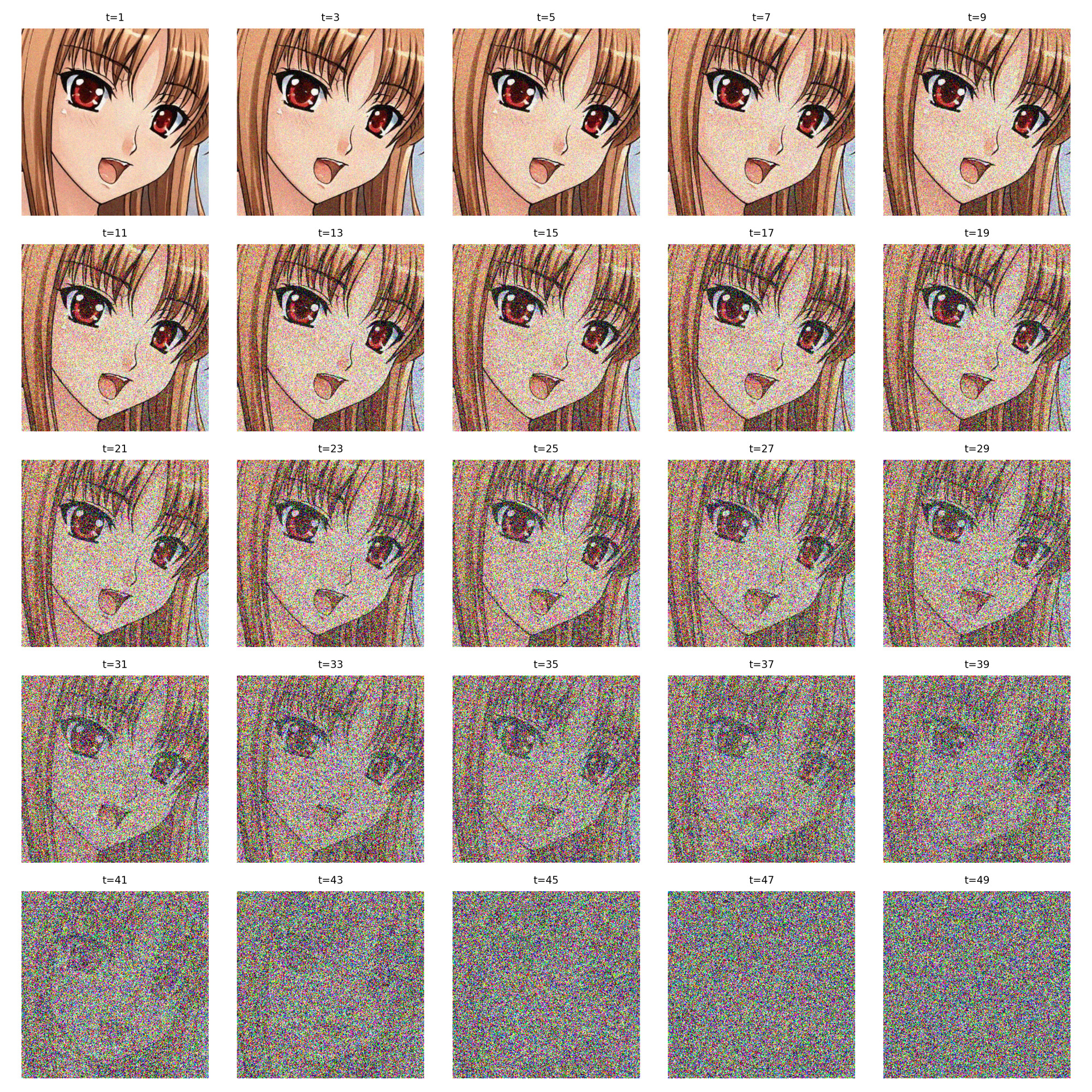}
  \vspace{0.3em}
  \textbf{(d)} Noisy Grid (DDPM)
\end{minipage}
\hfill
\begin{minipage}[b]{0.32\linewidth}
  \centering
  \includegraphics[width=\linewidth]{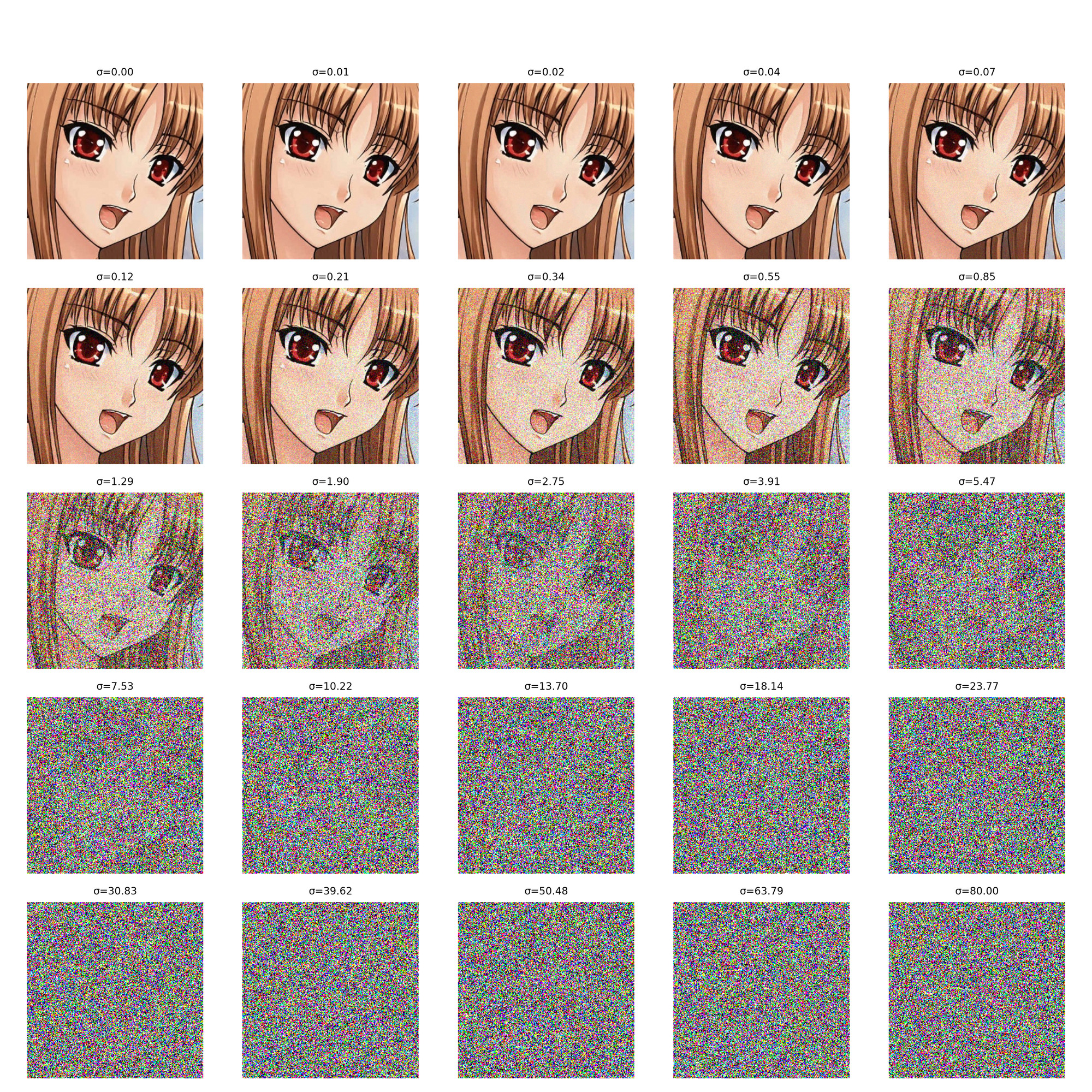}
  \vspace{0.3em}
  \textbf{(e)} Noisy Grid (EDM)
\end{minipage}
\hfill
\begin{minipage}[b]{0.32\linewidth}
  \centering
  \includegraphics[width=\linewidth]{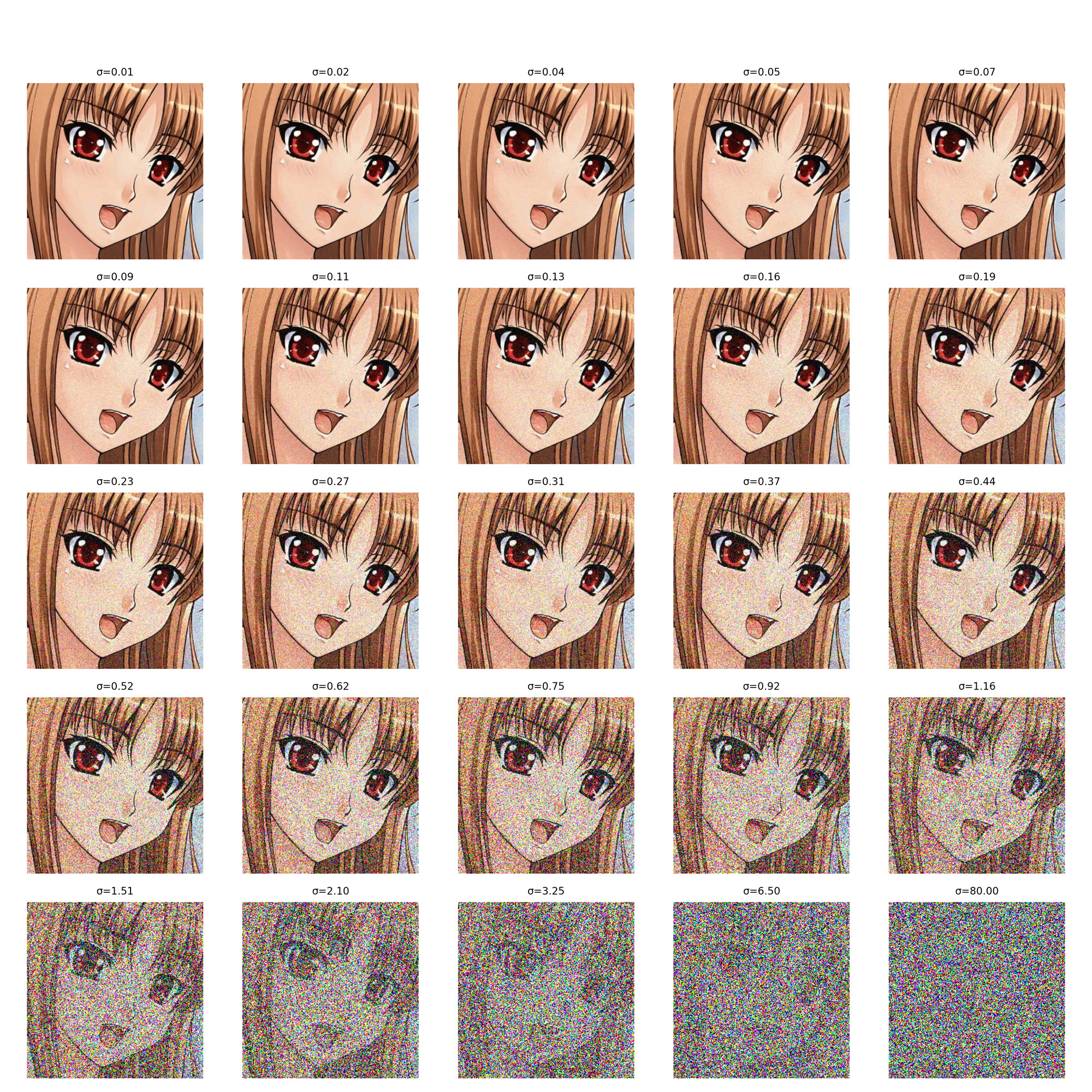}
  \vspace{0.3em}
  \textbf{(f)} Noisy Grid (SSIMBaD)
\end{minipage}

\vspace{1.5em}

\caption{
\textbf{Comparison of forward diffusion schedules.}
Top: SSIM curves for DDPM \textbf{(a)}, EDM \textbf{(b)}, and our schedule $\phi^{\ast}(\sigma)$ \textbf{(c)}.
Bottom: $5 \times 5$ corrupted grids \textbf{(d)}–\textbf{(f)} show each schedule’s visual effect.
Our method yields perceptually uniform degradation across timesteps.
}
\label{fig:corruption_schedule}
\end{figure}

%% file: sections/appendix.tex
\newpage

\appendix
\section*{Appendix}

\section{Details of the Proposed Framework}
\label{appendix:proposition_details}
\input{sections/appendix/appendix_proposition1}
\input{sections/appendix/appendix_proposition2}

\section{Details on SSIM-Aligned Sigma-Space Scaling}
\label{app:phi_schedule}

\input{sections/appendix/ssim_forward_corruption}
\input{sections/appendix/content_sigma_scaling}

\section{Extended Qualitative Comparisons}

To complement our main results, we present qualitative comparisons in both same-reference and cross-reference scenarios (Figures~\ref{fig:more_same_comparison} and~\ref{fig:more_cross_comparison}). In the same-reference scenario, our model produces visually faithful results that align well with both structure and style. In the cross-reference scenario, it generalizes robustly to unseen references, avoiding oversaturation and preserving content. These results highlight the benefit of SSIM-aligned sigma-space scaling and trajectory refinement in achieving perceptually consistent generation. 

\newpage
\subsection{Same-Reference Scenario}
\FloatBarrier
\label{app:same_ref}
\input{sections/appendix/more_same_reference_table}

\newpage
\subsection{Cross-Reference Scenario}
\FloatBarrier
\label{app:cross_ref}
\input{sections/appendix/more_cross_reference_table}


\newpage
\section{Why Did We Add Rotation to TPS?}
\begin{table}[H]
\centering
\scriptsize
\setlength{\tabcolsep}{5pt}
\renewcommand{\arraystretch}{1.3}
\caption{Quantitative results without TPS rotation under both same-reference and cross-reference settings. Finetuning improves visual fidelity in both conditions.}
\resizebox{\textwidth}{!}{%
\begin{tabular}{l|cc|cc|cc}
\toprule
\textbf{Method} 
& \multicolumn{2}{c|}{\textbf{PSNR ↑}} 
& \multicolumn{2}{c|}{\textbf{MS-SSIM ↑}} 
& \multicolumn{2}{c}{\textbf{FID ↓}} \\
& \textbf{Same} & \textbf{Cross} 
& \textbf{Same} & \textbf{Cross} 
& \textbf{Same} & \textbf{Cross} \\
\midrule
SSIMBaD (w/o trajectory refinement) & 20.55 & 11.34 & 0.8446 & 0.5996 & 56.18 & 65.69 \\
\rowcolor{red!10}
\textbf{SSIMBaD (w/ trajectory refinement)} & \textbf{23.10} & \textbf{14.00} & \textbf{0.9190} & \textbf{0.7714} & \textbf{24.35} & \textbf{40.73} \\
\bottomrule
\end{tabular}
}
\label{tab:no_rotation_metrics}
\end{table}

Despite visually plausible results in Figure~\ref{fig:no_rotation}, especially after trajectory refinement, Table~\ref{tab:no_rotation_metrics} reveals a significant performance gap between same- and cross-reference scenarios. For instance, PSNR drops from 23.10~dB to 14.00~dB, and MS-SSIM from 0.9190 to 0.7714, highlighting limited referential generalization. To address this, we introduce a lightweight affine rotation into the TPS pipeline, improving alignment between the sketch and reference. As shown in Table~\ref{tab:all_metrics}, incorporating TPS rotation reduces the PSNR and MS-SSIM gaps from 9.1~dB and 0.1476 to 3.08~dB and 0.0305, respectively. FID also improves, and our method surpasses all baselines under cross-reference scenario while retaining strong performance in the same-reference scenario.


\section{Does SSIM Behave as Intended During Generation?}
\label{app:ssim_finetune}

\begin{figure}[h]
\centering
\begin{minipage}[t]{0.48\linewidth}
    \centering
    \includegraphics[width=\linewidth]{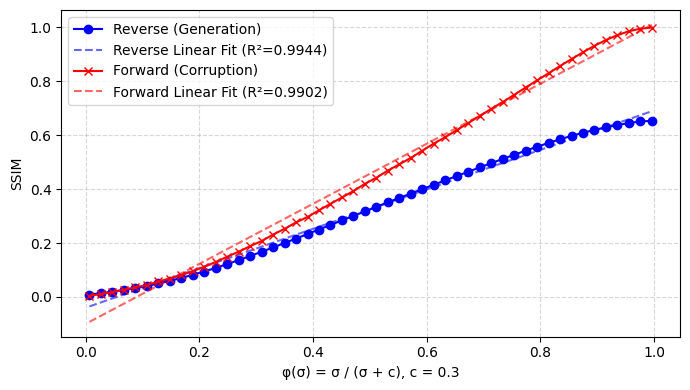}
    \subcaption{Ours (w/o trajectory refinement)}
    \label{fig:ssim_vs_phi_default}
\end{minipage}
\hfill
\begin{minipage}[t]{0.48\linewidth}
    \centering
    \includegraphics[width=\linewidth]{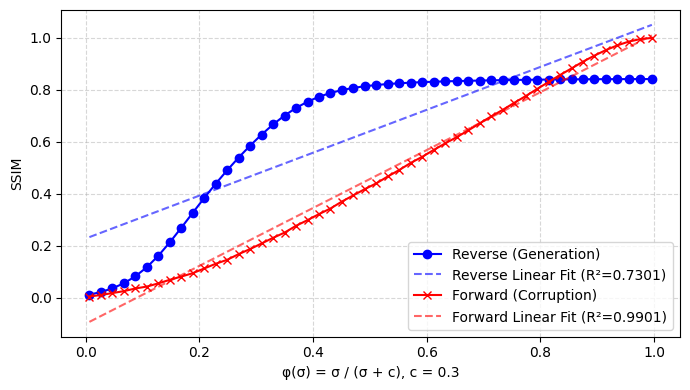}
    \subcaption{Ours (w/ trajectory refinement)}
    \label{fig:ssim_vs_phi_finetune}
\end{minipage}
\vspace{0.6em}
\caption{
SSIM vs $\phi(\sigma)$ curves for the same input image under forward (corruption, red) and reverse (generation, blue) processes.
Finetuning improves perceptual linearity in certain regions, but quickly saturates due to existing generation dynamics.
The model nonetheless maintains an overall perceptually stable trajectory, suggesting potential for further improvement through trajectory-aware objectives.
}
\label{fig:ssim_curve_comparison}
\end{figure}

To visually examine how closely the model's generation aligns with the intended noise schedule, we plot SSIM against $\phi(\sigma)^*$ for both the forward (corruption) and reverse (generation) processes, using the same input image and schedule.

Figure~\ref{fig:ssim_curve_comparison} compares this alignment before and after trajectory refinement. In both cases, the forward trajectory (red) shows near-perfect linear SSIM degradation, serving as a perceptual baseline. Notably, the reverse trajectory (blue) already exhibits a fair degree of linearity even before trajectory refinement, suggesting that the model implicitly learns to follow the $\phi(\sigma)^*$ path.

Importantly, trajectory refinement does not disrupt this linearity, preserving perceptual consistency while improving sample quality. These results highlight the robustness of our noise schedule and suggest that further improvements may be possible by designing more principled refinement objectives, which we leave for future work.


\newpage
\input{sections/appendix/no_rotation_table}

\newpage
\input{sections/appendix/implementation_details}

%% file: sections/appendix/appendix_proposition1.tex
\subsection{Conditional Input Construction}

Let $I_{\text{gt}} \in \mathbb{R}^{H \times W \times 3}$ denote the RGB ground-truth anime image, where $H$ and $W$ are the spatial resolution of the image. To form a pair of conditioning signals that guide both structure and style reconstruction, we derive two distinct inputs from $I_{\text{gt}}$: a structural sketch and a perturbed reference.

\paragraph{Sketch Extraction} 
The sketch $I_{\text{sketch}}$ is extracted via the extended Difference-of-Gaussians (XDoG) operator~\cite{winnemoller2012xdog}, which enhances edge-like regions through nonlinear contrast enhancement. Formally:
\begin{equation}
I_{\text{sketch}} = \text{XDoG}(I_{\text{gt}}) \in \mathbb{R}^{H \times W \times 1}.
\end{equation}
This 1-channel sketch preserves high-frequency structure such as contours and character outlines, serving as a strong spatial constraint during generation.

\paragraph{Reference Transformation}
To simulate reference-guided generation under diverse style domains, we construct a distorted version of $I_{\text{gt}}$ using a sequence of geometric transformations. First, a Thin Plate Spline (TPS) deformation is applied to introduce local warping, followed by random global rotations to inject non-aligned style cues:
\begin{equation}
I_{\text{ref}} = \text{Rotate}(\text{TPS}(I_{\text{gt}})) \in \mathbb{R}^{H \times W \times 3}.
\end{equation}
This 3-channel reference encodes the target color palette and texture, potentially with mild spatial misalignments.

\paragraph{Channel-Wise Conditioning}
The final conditional input is formed by concatenating the sketch and reference along the channel dimension:
\begin{equation}
I_{\text{cond}} = [I_{\text{ref}} \,\|\, I_{\text{sketch}}] \in \mathbb{R}^{H \times W \times 4},
\end{equation}
where $\|$ denotes channel-wise concatenation. This composite input retains both semantic layout and color style information, enabling the network to model structural consistency and stylization jointly. Note that $I_{\text{cond}}$ is held fixed throughout each diffusion trajectory to serve as a conditioning context for the denoiser.

%% file: sections/appendix/appendix_proposition2.tex
\subsection{Incorporating EDM}

We reformulate~\cite{Cao2024AnimeDiffusion} within the continuous-time framework of EDM~\cite{Karras2022EDM}, preserving its U-Net-based conditional denoiser $F_\theta$ while adopting a noise-level parameterization based on a continuous scale $\sigma$ rather than a discrete timestep $t$. This transition from discrete to continuous noise coordinates enables finer-grained modeling of the forward and reverse processes, as well as improved control over perceptual degradation across the diffusion trajectory.

Under the EDM formulation, the forward process perturbs a ground-truth image $I_{\text{gt}}$ into a noisy observation $I_{\text{noise}}$ by adding Gaussian noise of standard deviation $\sigma$:
\begin{equation}
p_{\sigma}(I_{\text{noise}} \mid I_{\text{cond}}) = \int_{\mathbb{R}^{H \times W \times 3}} \mathcal{N}\left(I_{\text{noise}};\ I_{\text{gt}},\ \sigma^2 \mathbf{I} \right)\, p_{\text{data}}(I_{\text{gt}} \mid I_{\text{cond}})\, dI_{\text{gt}},
\end{equation}
where $I_{\text{cond}}$ is a fixed conditioning tensor (e.g., reference and sketch) and $\mathbf{I}$ denotes the identity matrix. This parameterization allows the model to operate over a continuous spectrum of noise intensities, removing the timestep discretization bottleneck of DDPM~\cite{Ho2020DDPM}.

\paragraph{Noise-Aware Preconditioning}
To stabilize training and normalize feature magnitudes across varying $\sigma$, EDM applies a noise-aware preconditioning scheme~\cite{Karras2022EDM}. The denoiser $D_\theta$ is constructed as a residual mapping composed of pre-scaled input/output paths:
\begin{equation}
D_\theta(I_{\text{noise}}, I_{\text{cond}}; \sigma) = c_{\text{skip}}(\sigma) I_{\text{noise}} + c_{\text{out}}(\sigma) \cdot F_\theta(c_{\text{in}}(\sigma) I_{\text{noise}}, I_{\text{cond}};\ c_{\text{noise}}(\sigma)),
\end{equation}
where $c_{\text{in}}$, $c_{\text{out}}$, and $c_{\text{skip}}$ are scale-dependent coefficients defined as:
\[
c_{\text{skip}} = \frac{\sigma_{\text{data}}^2}{\sigma^2 + \sigma_{\text{data}}^2}, \quad
c_{\text{out}} = \frac{\sigma}{\sqrt{\sigma^2 + \sigma_{\text{data}}^2}}, \quad
c_{\text{in}} = \frac{1}{\sqrt{\sigma^2 + \sigma_{\text{data}}^2}}, \quad
c_{\text{noise}} = \frac{1}{4} \ln \sigma.
\]
This formulation ensures that input features have consistent scale, preventing signal collapse at low noise or amplification at high noise levels. In practice, we use $\sigma_{\text{data}} = 0.5$.

\paragraph{Training Objective}
Unlike DDPM which samples timesteps $t \in \{1, ..., T\}$, EDM samples $\ln \sigma$ from a normal distribution $\mathcal{N}(P_{\text{mean}}, P_{\text{std}}^2)$. The training loss is defined over random $\sigma$ as:
\begin{equation}
\mathcal{L} = \mathbb{E}_{\ln \sigma \sim \mathcal{N}(P_{\text{mean}}, P_{\text{std}}^2)} \, \mathbb{E}_{I_{\text{gt}} \sim p_{\text{data}}} \, \mathbb{E}_{\bm{n} \sim \mathcal{N}(0, \sigma^2 \mathbf{I})}
\left\| D_\theta(I_{\text{gt}} + \bm{n}, I_{\text{cond}}; \sigma) - I_{\text{gt}} \right\|^2.
\end{equation}

\paragraph{Sampling via Reverse-Time ODE}
At inference time, EDM uses a score-based formulation to define a reverse-time ordinary differential equation (ODE) that approximates the likelihood gradient with the denoiser output:
\begin{equation}
\nabla_{I_{\text{noise}}} \log p(I_{\text{noise}} \mid I_{\text{cond}}; \sigma) \approx \frac{D_\theta(I_{\text{noise}}, I_{\text{cond}}; \sigma) - I_{\text{noise}}}{\sigma^2},
\end{equation}
leading to the continuous reverse-time dynamics:
\begin{equation}
\frac{dI_{\text{noise}}}{dt} = -\frac{1}{\sigma} \left( D_\theta(I_{\text{noise}}, I_{\text{cond}}; \sigma) - I_{\text{noise}} \right).
\end{equation}

\paragraph{Sigma Schedule and Discretization}
To discretize this process, we apply the Euler method using a $\rho$-parameterized sigma schedule:
\begin{equation}
\sigma_i = \left[ \sigma_{\max}^{1/\rho} + \frac{i}{N - 1} \left( \sigma_{\min}^{1/\rho} - \sigma_{\max}^{1/\rho} \right) \right]^{\rho}, \quad i = 0, 1, \dots, N - 1.
\end{equation}
We initialize the trajectory from pure noise $I^{(N-1)} \sim \mathcal{N}(0, \mathbf{I})$ and integrate the ODE in reverse over the precomputed $\{\sigma_i\}$ sequence. The denoising step at each index $i$ is performed as:
\begin{equation}
I^{(i-1)} = I^{(i)} - \frac{\Delta t_i}{\sigma_i} \left( D_\theta(I^{(i)}, I_{\text{cond}}; \sigma_i) - I^{(i)} \right), \quad \Delta t_i = \sigma_i - \sigma_{i-1}.
\end{equation}

This continuous-time formulation enables~\cite{Cao2024AnimeDiffusion} to benefit from the architectural and sampling improvements of EDM, while retaining its original conditioning and loss structure. In Section~\ref{sec:sigmascale}, we further extend this pipeline by introducing a perceptual scaling of $\sigma$ to ensure uniform SSIM degradation across steps.

%% file: sections/appendix/ssim_forward_corruption.tex
\begin{figure*}[t]
\centering

\begin{subfigure}[b]{0.32\textwidth}
    \includegraphics[width=\linewidth]{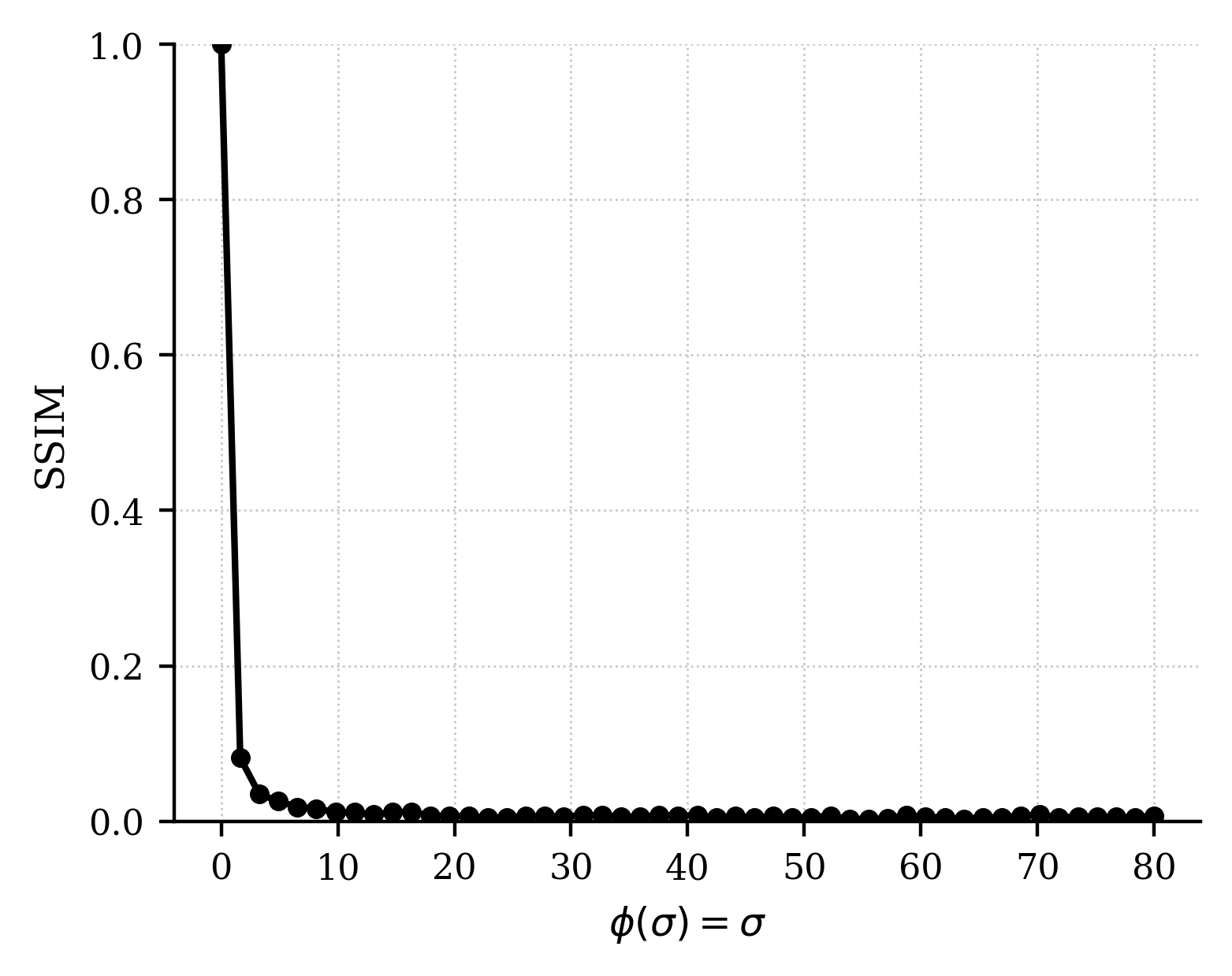}
\end{subfigure}
\begin{subfigure}[b]{0.32\textwidth}
    \includegraphics[width=\linewidth]{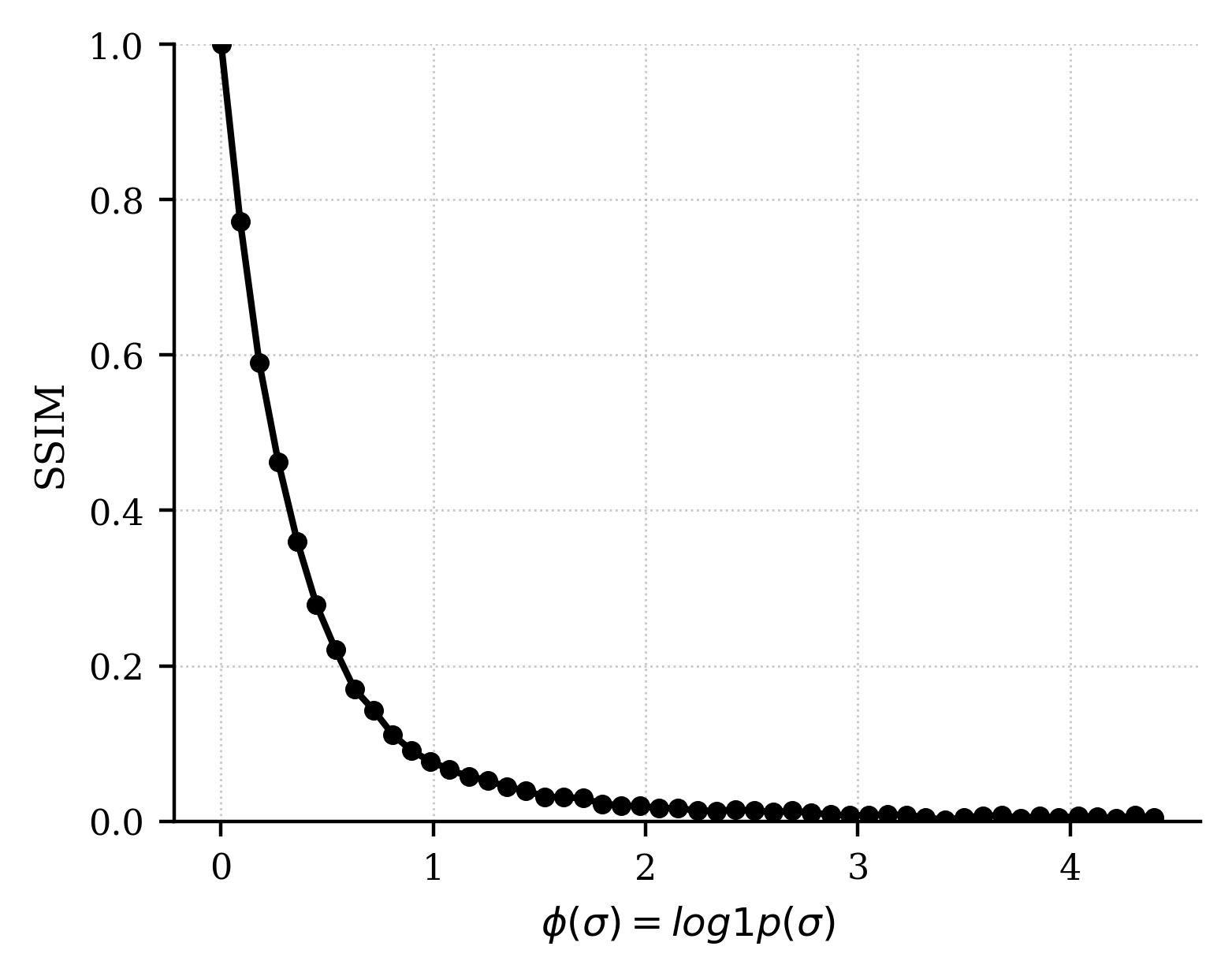}
\end{subfigure}
\begin{subfigure}[b]{0.32\textwidth}
    \includegraphics[width=\linewidth]{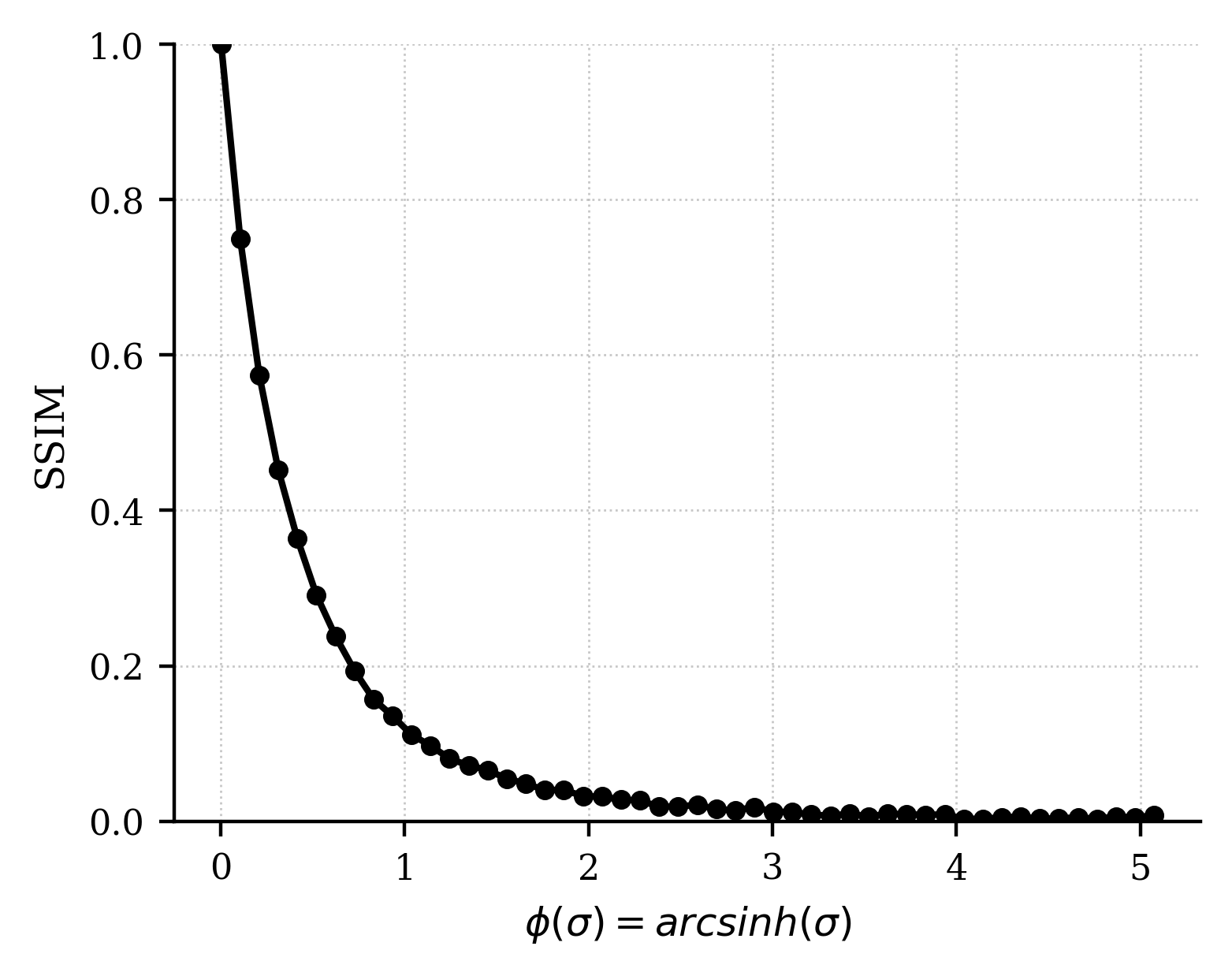}
\end{subfigure}
\vspace{2mm}

\begin{subfigure}[b]{0.32\textwidth}
    \includegraphics[width=\linewidth]{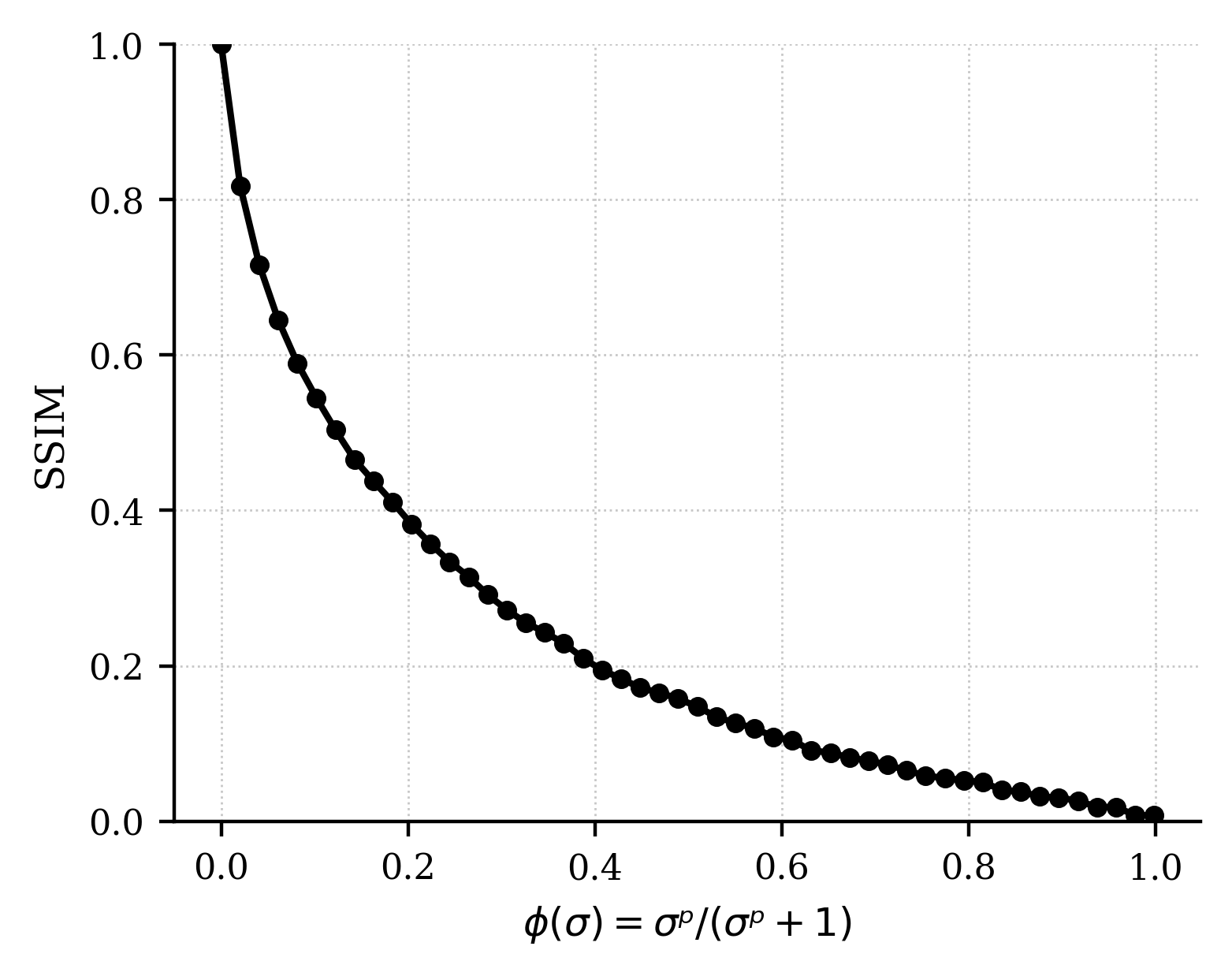}
\end{subfigure}
\begin{subfigure}[b]{0.32\textwidth}
    \includegraphics[width=\linewidth]{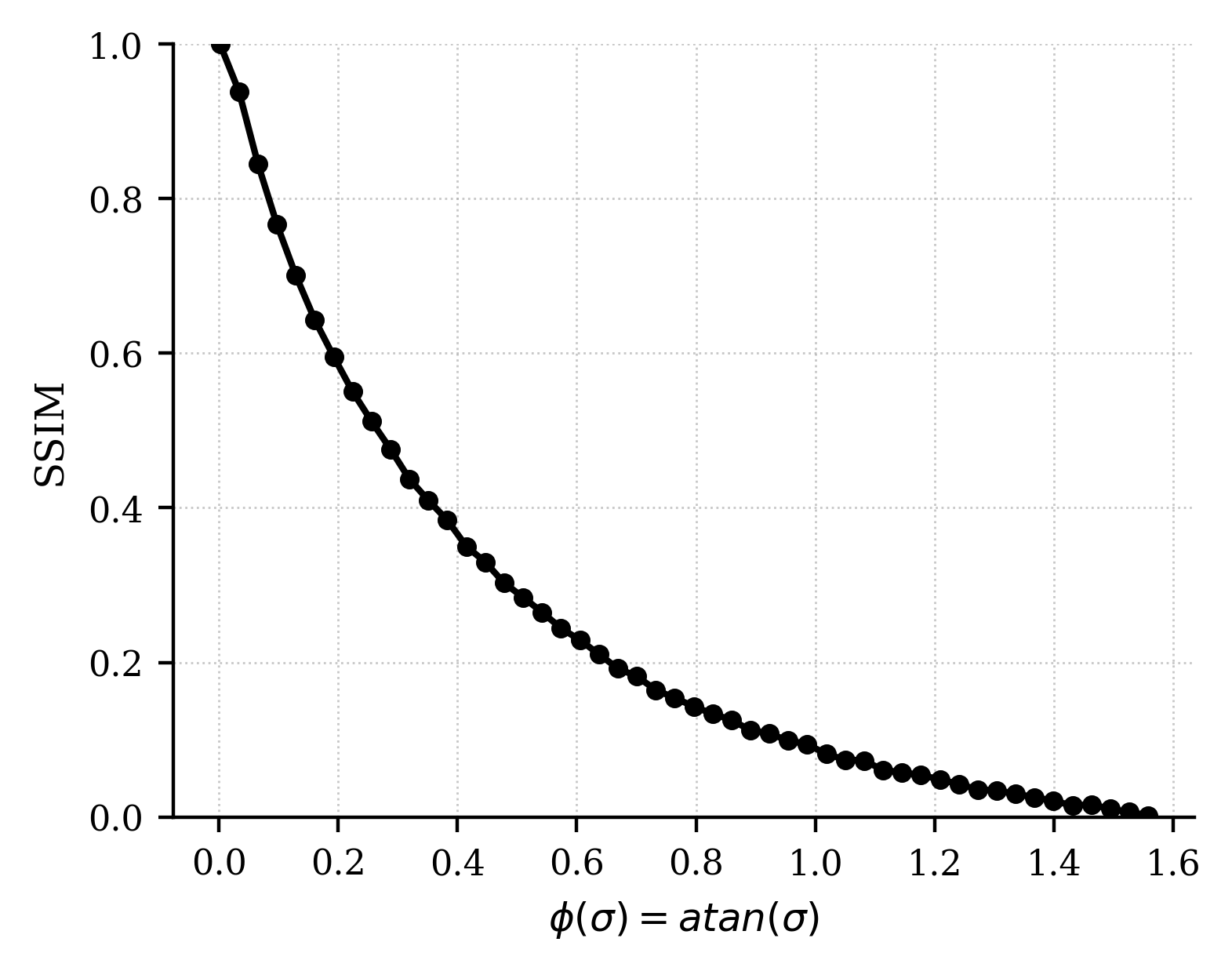}
\end{subfigure}
\begin{subfigure}[b]{0.32\textwidth}
    \includegraphics[width=\linewidth]{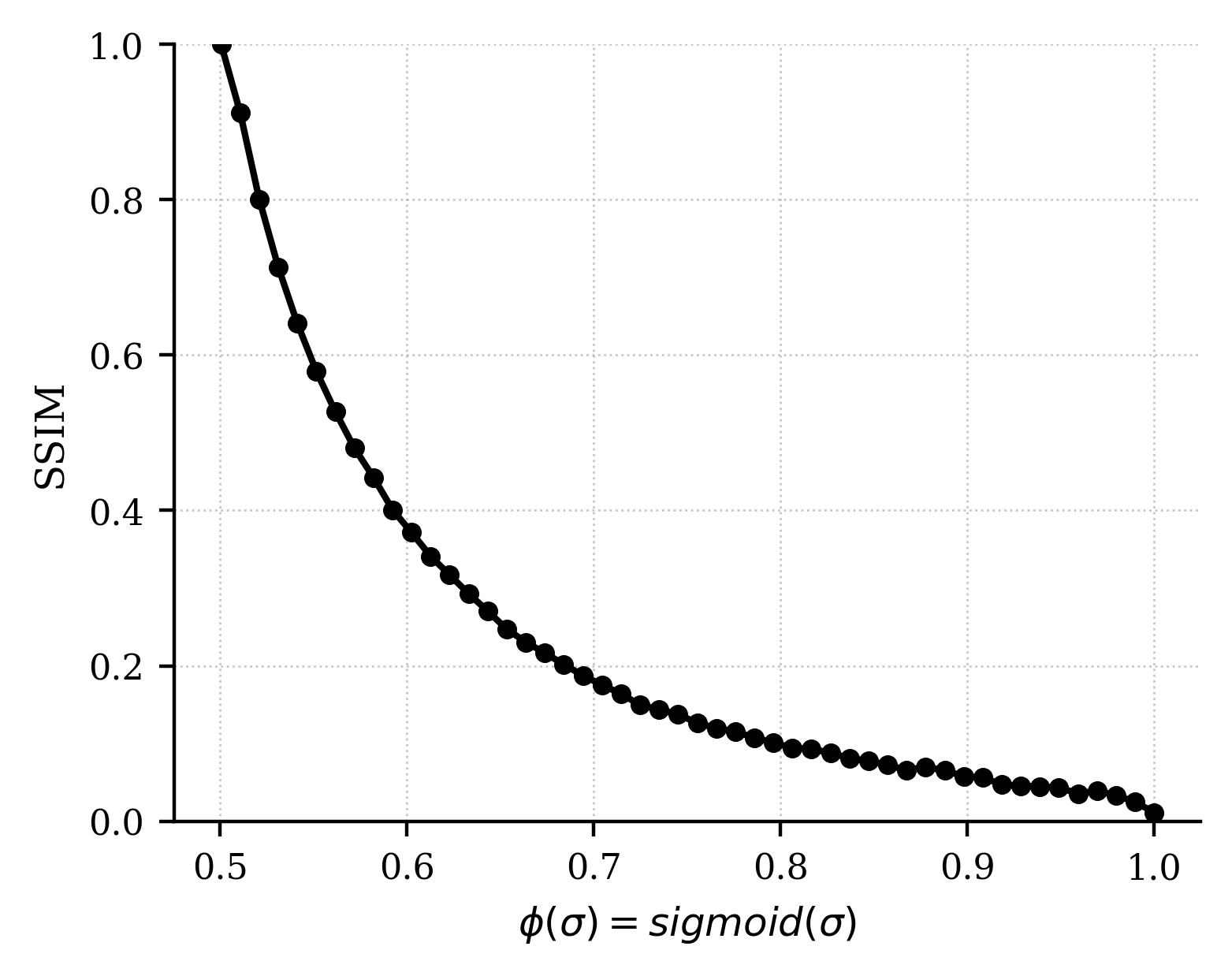}
\end{subfigure}
\vspace{2mm}

\begin{subfigure}[b]{0.32\textwidth}
    \includegraphics[width=\linewidth]{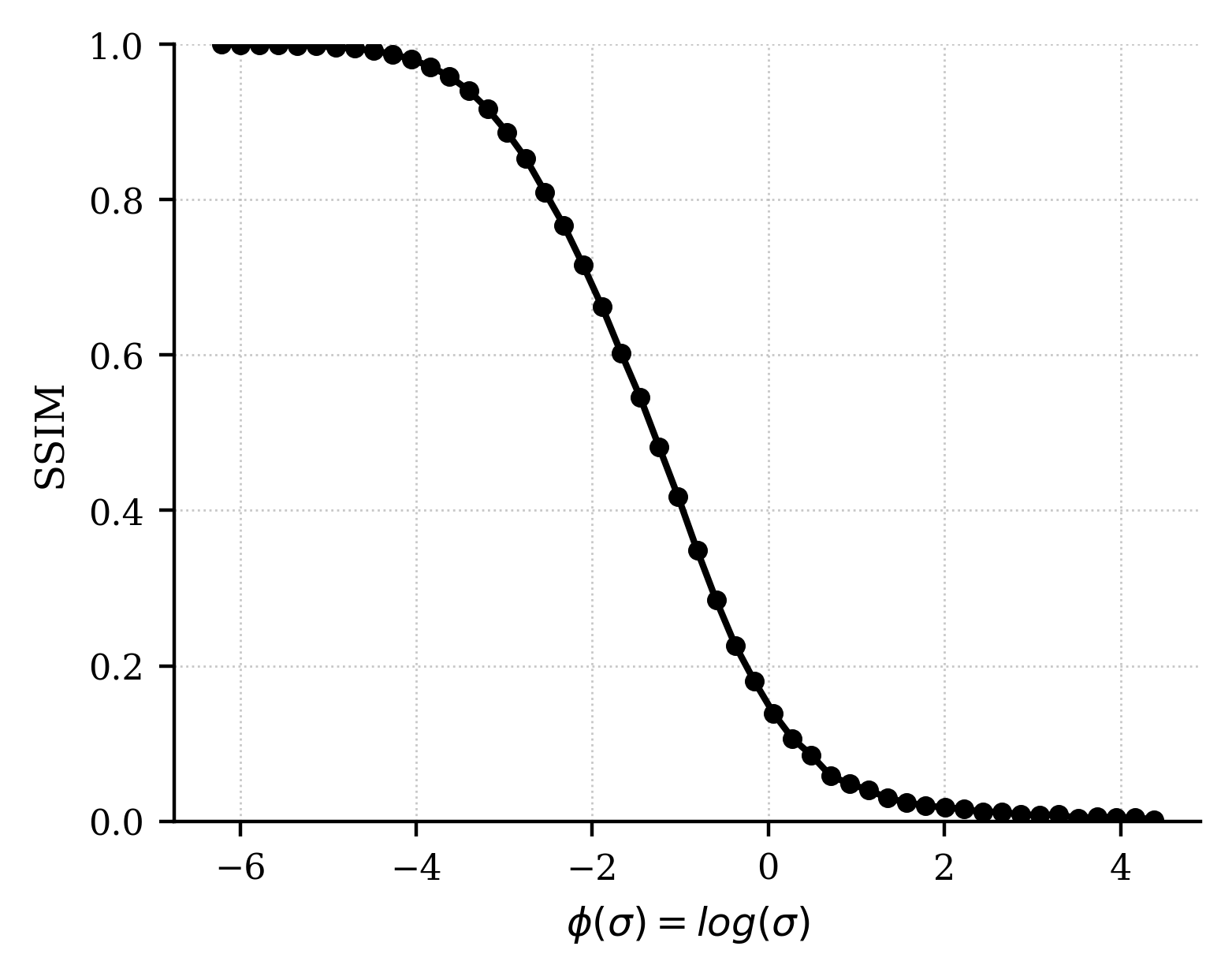}
\end{subfigure}
\begin{subfigure}[b]{0.32\textwidth}
    \includegraphics[width=\linewidth]{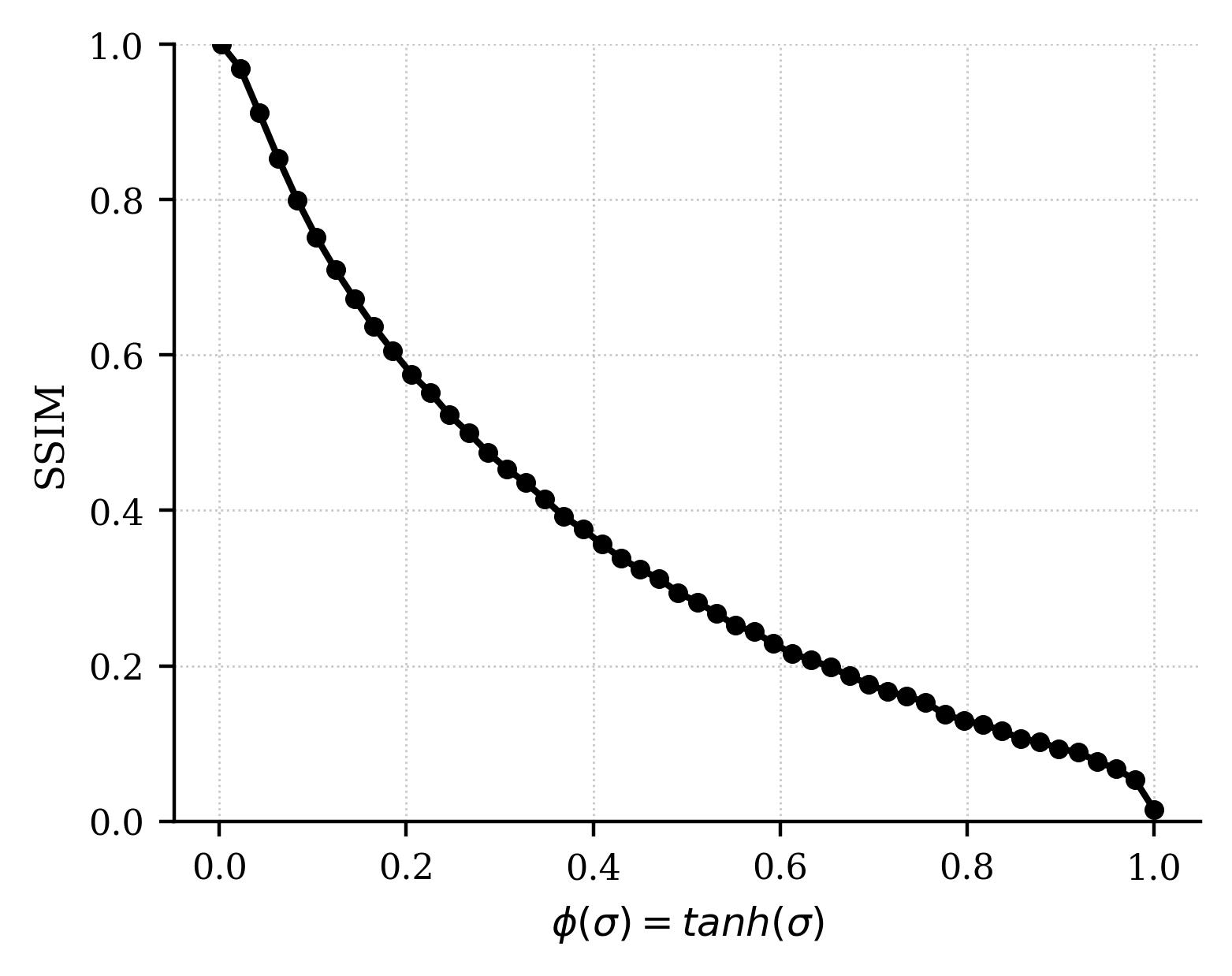}
\end{subfigure}
\begin{subfigure}[b]{0.32\textwidth}
    \includegraphics[width=\linewidth]{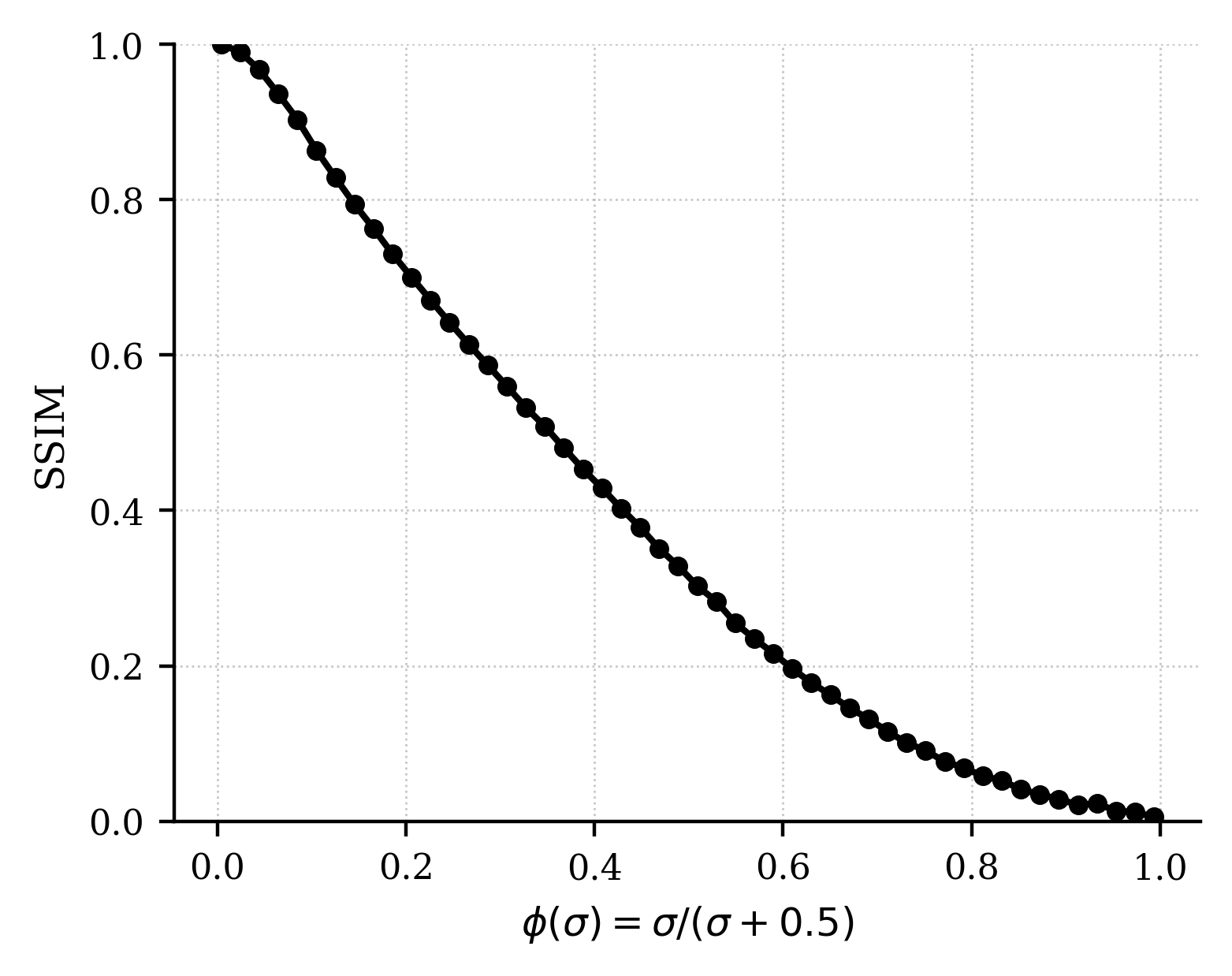}
\end{subfigure}
 
\vspace{2mm}

\caption{
\textbf{SSIM degradation across various transformations $\phi(\sigma)$.}
Each curve shows the SSIM between the clean image and its noisy counterpart as the noise level $\sigma$ increases, under a specific transformation $\phi$. The transformations are ordered by increasing linearity ($R^2$). Among them, bounded squash functions of the form $\phi(\sigma) = \frac{\sigma}{\sigma + c}$ exhibit the most linear degradation trends. In particular, $\phi(\sigma) = \frac{\sigma}{\sigma + 0.3}$ achieves near-perfect linearity, making it well-suited for constructing perceptually uniform sigma schedules. For clarity, we visualize a representative subset of the evaluated transformations.
}

\label{fig:ssim_forward_corruption}
\end{figure*}

%% file: sections/appendix/content_sigma_scaling.tex
To design a perceptually uniform noise schedule, we empirically analyze the relationship between SSIM degradation and transformed noise levels $\phi(\sigma)$ across various candidate functions. For each transformation $\phi$, a clean image $I_{\text{clean}}$ is corrupted at $N=50$ distinct noise levels by adding scaled Gaussian noise as described in \eqref{eq:corruption}. We then compute the SSIM between each noisy image and its clean counterpart to obtain a degradation curve. To quantify the perceptual consistency of each transformation, we plot SSIM values against $\phi(\sigma)$ and measure the linearity of the resulting curve using the coefficient of determination ($R^2$). This procedure is applied to 1\% of randomly sampled training images, each undergoing 50 corruption steps, yielding a comprehensive perceptual degradation profile across a wide range of noise intensities.

As illustrated in Figure~\ref{fig:ssim_forward_corruption}, plotting SSIM against $\phi(\sigma)$ reveals that certain transformations induce nearly linear degradation. In particular, bounded squash functions of the form
\[
\phi(\sigma) = \frac{\sigma}{\sigma + c}
\]
produce the most perceptually uniform trends. Among these, $\phi(\sigma) = \frac{\sigma}{\sigma + 0.3}$ achieves near-perfect linearity with an $R^2$ value of 0.9949. Based on this result, we adopt this transformation as our default scaling function in sigma-space. Table~\ref{tab:R_square_comparison} summarizes the $R^2$ values for representative candidate functions.

Finally, we construct our noise schedule by uniformly sampling steps in the transformed $\phi$-space and applying the inverse of the selected transformation to compute the corresponding $\sigma$ values, as defined in \eqref{eq:scaled_sigma_schedule}. This perceptually aligned schedule ensures that each diffusion step contributes uniformly to structural degradation, which is critical for achieving balanced and stable restoration during generation.

%% file: sections/appendix/more_same_reference_table.tex
\begin{figure}[H]
\centering
\renewcommand{\arraystretch}{0.5}
\setlength{\tabcolsep}{2pt}

\begin{tabular}{cccccccc}
\textbf{(a)} & \textbf{(b)} & \textbf{(c)} & \textbf{(d)} & \textbf{(e)} & \textbf{(f)} & \textbf{(g)} & \textbf{(h)} \\

\raisebox{-.5\height}{\includegraphics[width=0.11\linewidth]{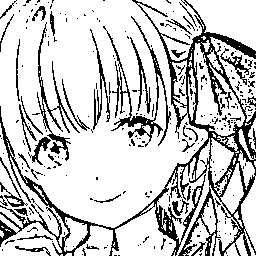}} &
\raisebox{-.5\height}{\includegraphics[width=0.11\linewidth]{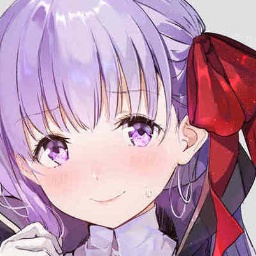}} &
\raisebox{-.5\height}{\includegraphics[width=0.11\linewidth]{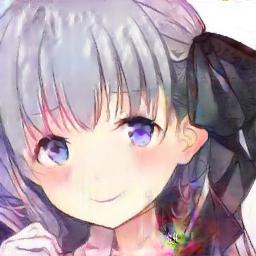}} &
\raisebox{-.5\height}{\includegraphics[width=0.11\linewidth]{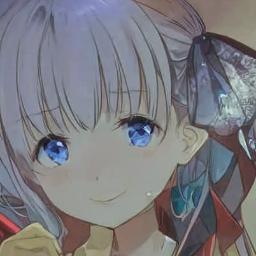}} &
\raisebox{-.5\height}{\includegraphics[width=0.11\linewidth]{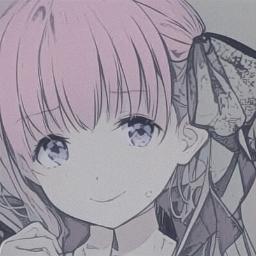}} &
\raisebox{-.5\height}{\includegraphics[width=0.11\linewidth]{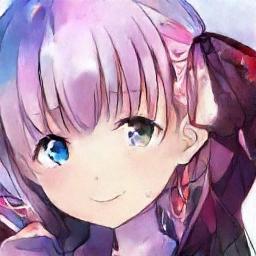}} &
\raisebox{-.5\height}{\includegraphics[width=0.11\linewidth]{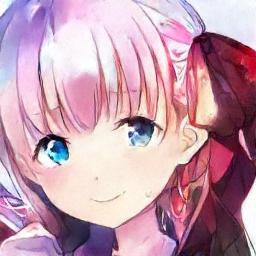}} &
\raisebox{-.5\height}{\includegraphics[width=0.11\linewidth]{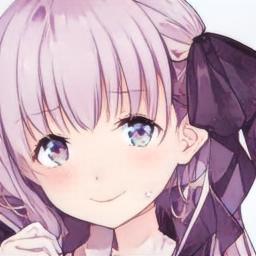}} \\

\raisebox{-.5\height}{\includegraphics[width=0.11\linewidth]{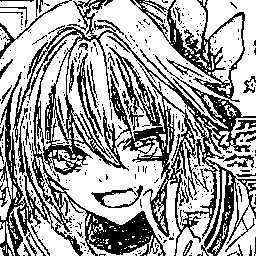}} &
\raisebox{-.5\height}{\includegraphics[width=0.11\linewidth]{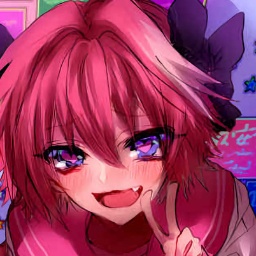}} &
\raisebox{-.5\height}{\includegraphics[width=0.11\linewidth]{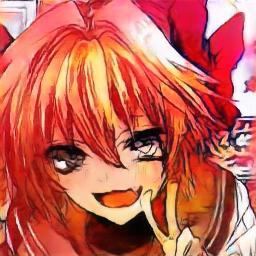}} &
\raisebox{-.5\height}{\includegraphics[width=0.11\linewidth]{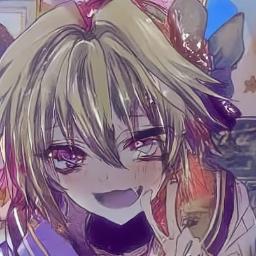}} &
\raisebox{-.5\height}{\includegraphics[width=0.11\linewidth]{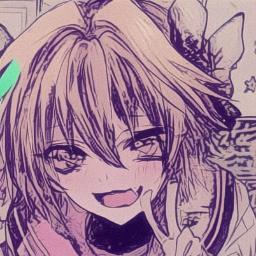}} &
\raisebox{-.5\height}{\includegraphics[width=0.11\linewidth]{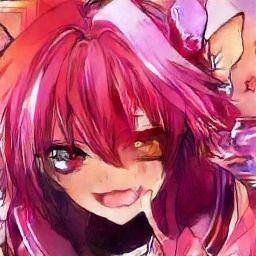}} &
\raisebox{-.5\height}{\includegraphics[width=0.11\linewidth]{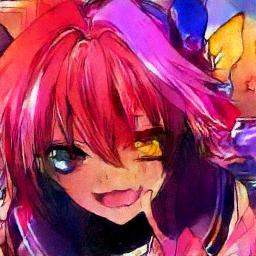}} &
\raisebox{-.5\height}{\includegraphics[width=0.11\linewidth]{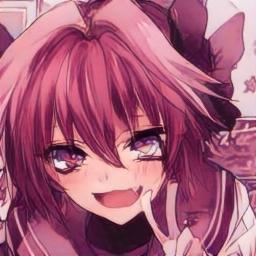}} \\

\raisebox{-.5\height}{\includegraphics[width=0.11\linewidth]{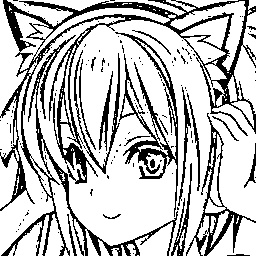}} &
\raisebox{-.5\height}{\includegraphics[width=0.11\linewidth]{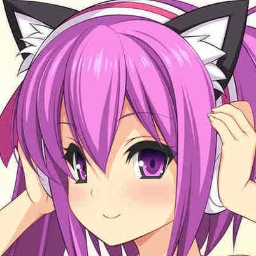}} &
\raisebox{-.5\height}{\includegraphics[width=0.11\linewidth]{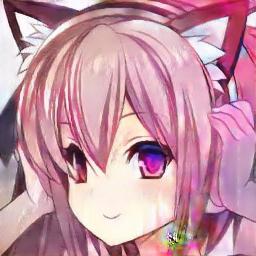}} &
\raisebox{-.5\height}{\includegraphics[width=0.11\linewidth]{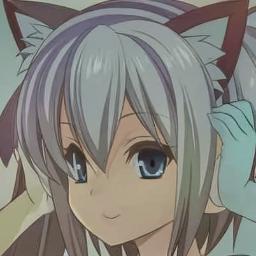}} &
\raisebox{-.5\height}{\includegraphics[width=0.11\linewidth]{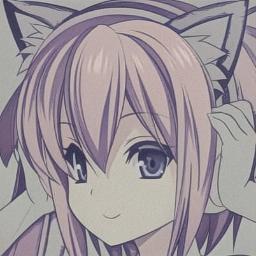}} &
\raisebox{-.5\height}{\includegraphics[width=0.11\linewidth]{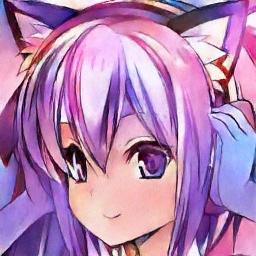}} &
\raisebox{-.5\height}{\includegraphics[width=0.11\linewidth]{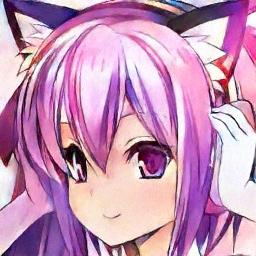}} &
\raisebox{-.5\height}{\includegraphics[width=0.11\linewidth]{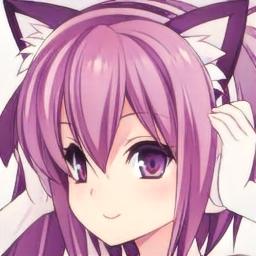}} \\

\raisebox{-.5\height}{\includegraphics[width=0.11\linewidth]{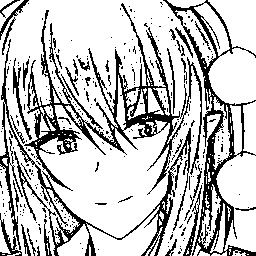}} &
\raisebox{-.5\height}{\includegraphics[width=0.11\linewidth]{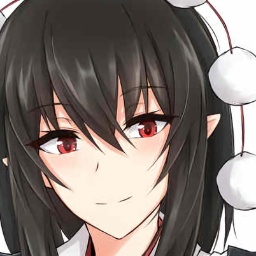}} &
\raisebox{-.5\height}{\includegraphics[width=0.11\linewidth]{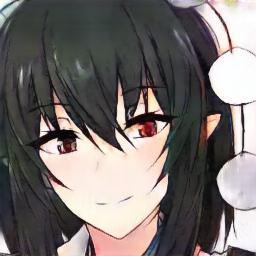}} &
\raisebox{-.5\height}{\includegraphics[width=0.11\linewidth]{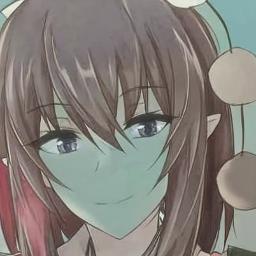}} &
\raisebox{-.5\height}{\includegraphics[width=0.11\linewidth]{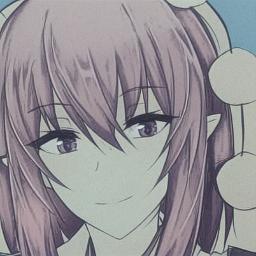}} &
\raisebox{-.5\height}{\includegraphics[width=0.11\linewidth]{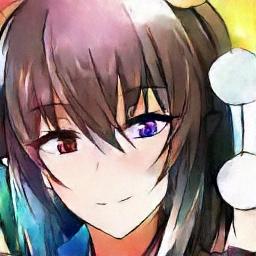}} &
\raisebox{-.5\height}{\includegraphics[width=0.11\linewidth]{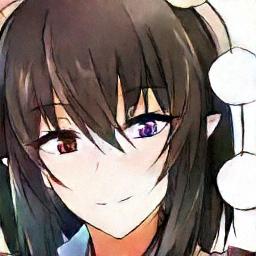}} &
\raisebox{-.5\height}{\includegraphics[width=0.11\linewidth]{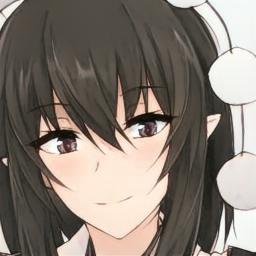}} \\

\raisebox{-.5\height}{\includegraphics[width=0.11\linewidth]{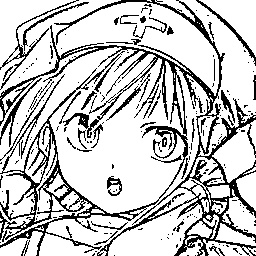}} &
\raisebox{-.5\height}{\includegraphics[width=0.11\linewidth]{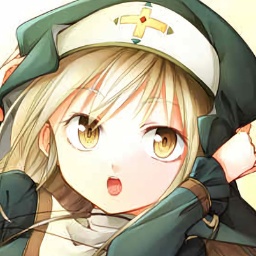}} &
\raisebox{-.5\height}{\includegraphics[width=0.11\linewidth]{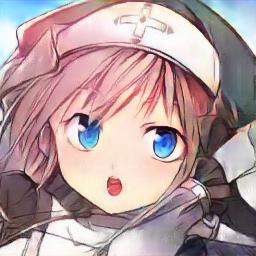}} &
\raisebox{-.5\height}{\includegraphics[width=0.11\linewidth]{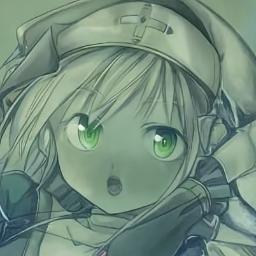}} &
\raisebox{-.5\height}{\includegraphics[width=0.11\linewidth]{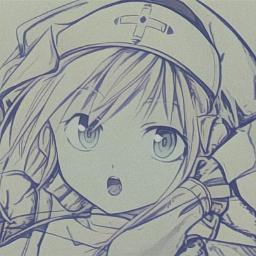}} &
\raisebox{-.5\height}{\includegraphics[width=0.11\linewidth]{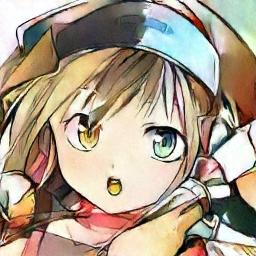}} &
\raisebox{-.5\height}{\includegraphics[width=0.11\linewidth]{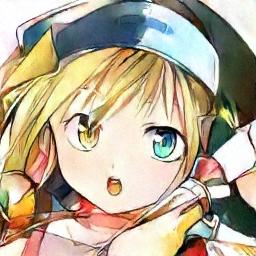}} &
\raisebox{-.5\height}{\includegraphics[width=0.11\linewidth]{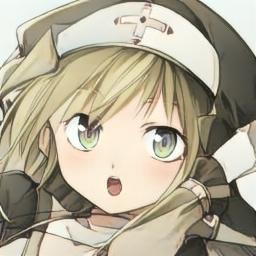}} \\

\raisebox{-.5\height}{\includegraphics[width=0.11\linewidth]{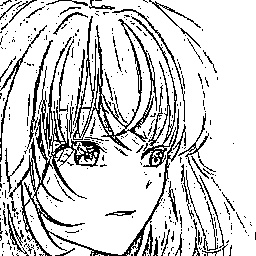}} &
\raisebox{-.5\height}{\includegraphics[width=0.11\linewidth]{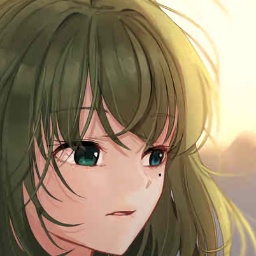}} &
\raisebox{-.5\height}{\includegraphics[width=0.11\linewidth]{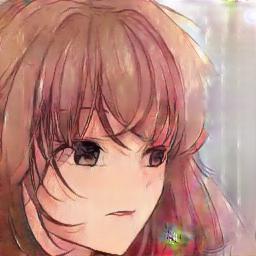}} &
\raisebox{-.5\height}{\includegraphics[width=0.11\linewidth]{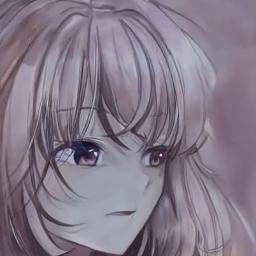}} &
\raisebox{-.5\height}{\includegraphics[width=0.11\linewidth]{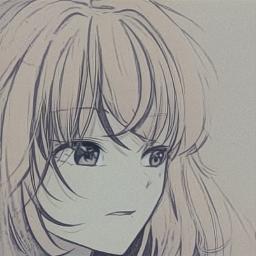}} &
\raisebox{-.5\height}{\includegraphics[width=0.11\linewidth]{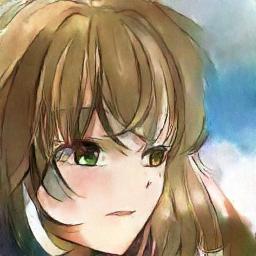}} &
\raisebox{-.5\height}{\includegraphics[width=0.11\linewidth]{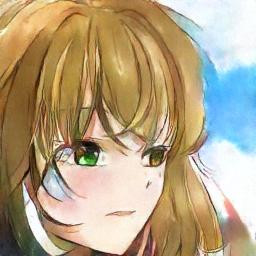}} &
\raisebox{-.5\height}{\includegraphics[width=0.11\linewidth]{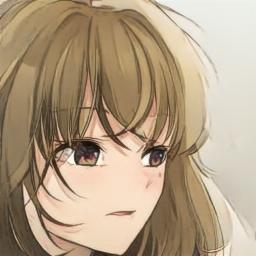}} \\

\raisebox{-.5\height}{\includegraphics[width=0.11\linewidth]{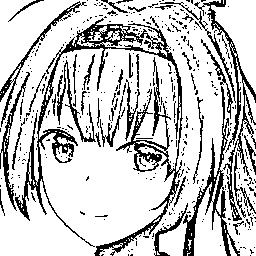}} &
\raisebox{-.5\height}{\includegraphics[width=0.11\linewidth]{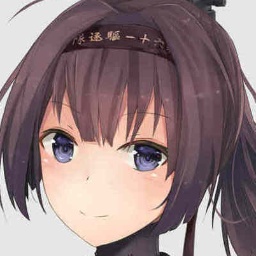}} &
\raisebox{-.5\height}{\includegraphics[width=0.11\linewidth]{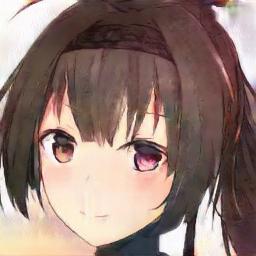}} &
\raisebox{-.5\height}{\includegraphics[width=0.11\linewidth]{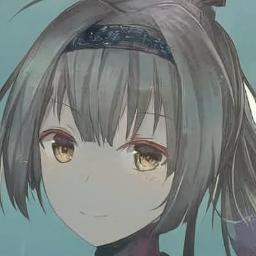}} &
\raisebox{-.5\height}{\includegraphics[width=0.11\linewidth]{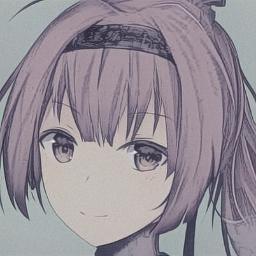}} &
\raisebox{-.5\height}{\includegraphics[width=0.11\linewidth]{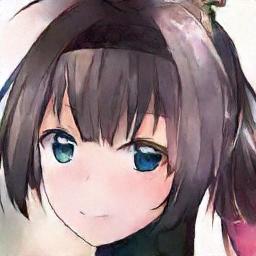}} &
\raisebox{-.5\height}{\includegraphics[width=0.11\linewidth]{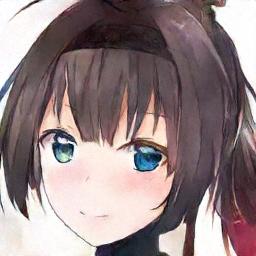}} &
\raisebox{-.5\height}{\includegraphics[width=0.11\linewidth]{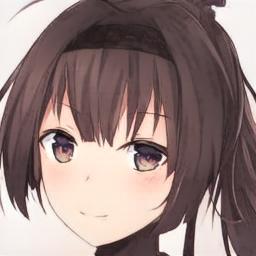}} \\

\raisebox{-.5\height}{\includegraphics[width=0.11\linewidth]{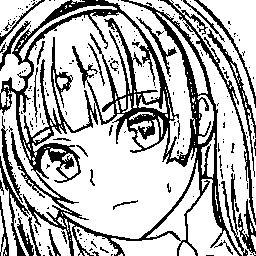}} &
\raisebox{-.5\height}{\includegraphics[width=0.11\linewidth]{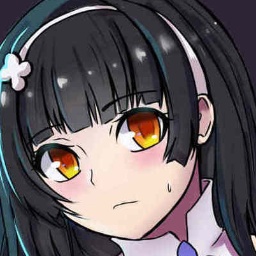}} &
\raisebox{-.5\height}{\includegraphics[width=0.11\linewidth]{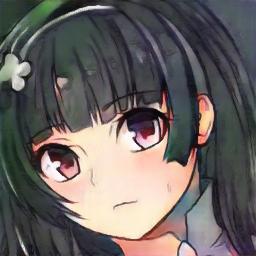}} &
\raisebox{-.5\height}{\includegraphics[width=0.11\linewidth]{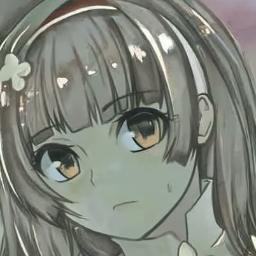}} &
\raisebox{-.5\height}{\includegraphics[width=0.11\linewidth]{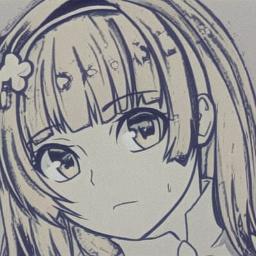}} &
\raisebox{-.5\height}{\includegraphics[width=0.11\linewidth]{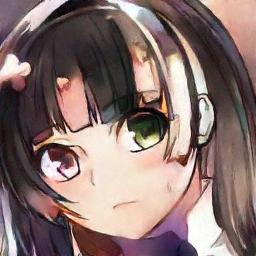}} &
\raisebox{-.5\height}{\includegraphics[width=0.11\linewidth]{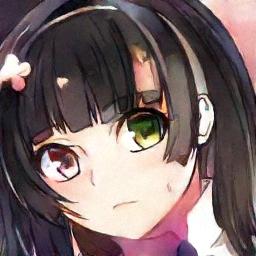}} &
\raisebox{-.5\height}{\includegraphics[width=0.11\linewidth]{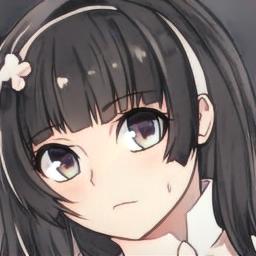}} \\

\raisebox{-.5\height}{\includegraphics[width=0.11\linewidth]{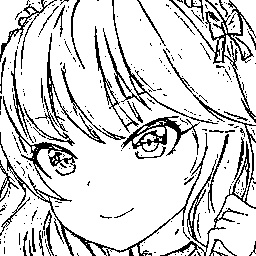}} &
\raisebox{-.5\height}{\includegraphics[width=0.11\linewidth]{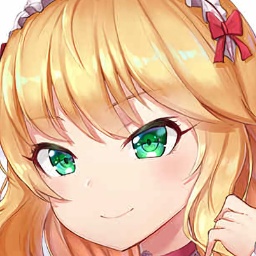}} &
\raisebox{-.5\height}{\includegraphics[width=0.11\linewidth]{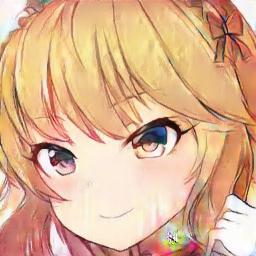}} &
\raisebox{-.5\height}{\includegraphics[width=0.11\linewidth]{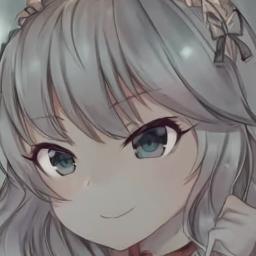}} &
\raisebox{-.5\height}{\includegraphics[width=0.11\linewidth]{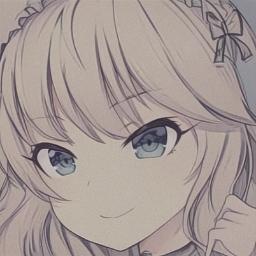}} &
\raisebox{-.5\height}{\includegraphics[width=0.11\linewidth]{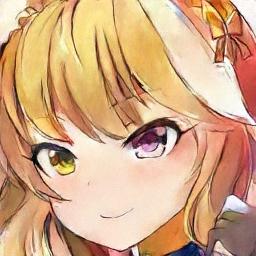}} &
\raisebox{-.5\height}{\includegraphics[width=0.11\linewidth]{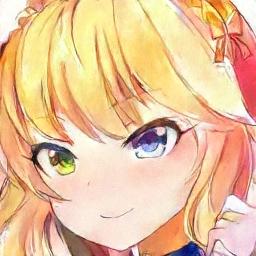}} &
\raisebox{-.5\height}{\includegraphics[width=0.11\linewidth]{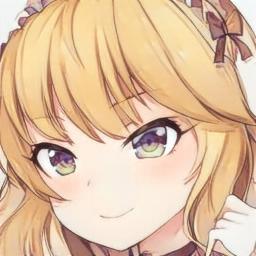}} \\

\raisebox{-.5\height}{\includegraphics[width=0.11\linewidth]{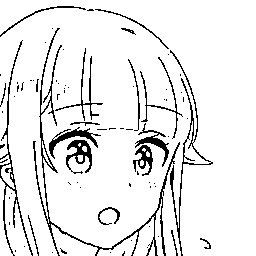}} &
\raisebox{-.5\height}{\includegraphics[width=0.11\linewidth]{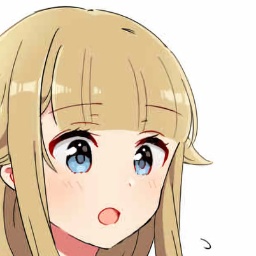}} &
\raisebox{-.5\height}{\includegraphics[width=0.11\linewidth]{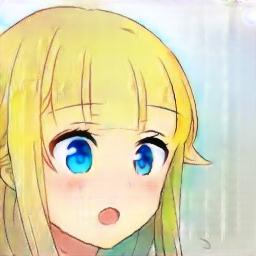}} &
\raisebox{-.5\height}{\includegraphics[width=0.11\linewidth]{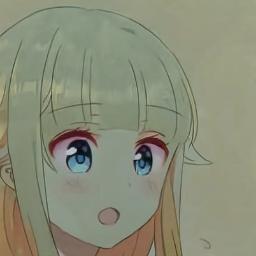}} &
\raisebox{-.5\height}{\includegraphics[width=0.11\linewidth]{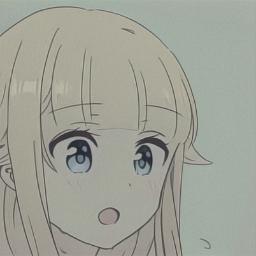}} &
\raisebox{-.5\height}{\includegraphics[width=0.11\linewidth]{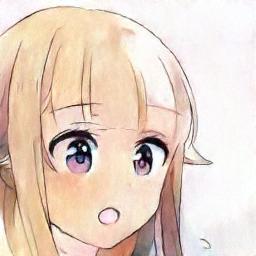}} &
\raisebox{-.5\height}{\includegraphics[width=0.11\linewidth]{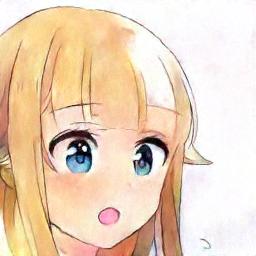}} &
\raisebox{-.5\height}{\includegraphics[width=0.11\linewidth]{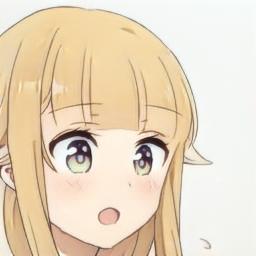}} \\

\raisebox{-.5\height}{\includegraphics[width=0.11\linewidth]{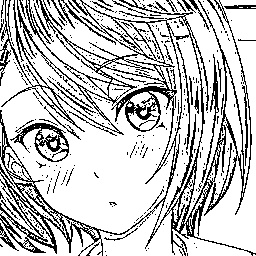}} &
\raisebox{-.5\height}{\includegraphics[width=0.11\linewidth]{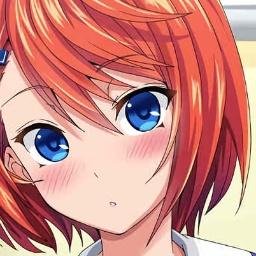}} &
\raisebox{-.5\height}{\includegraphics[width=0.11\linewidth]{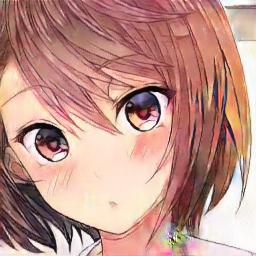}} &
\raisebox{-.5\height}{\includegraphics[width=0.11\linewidth]{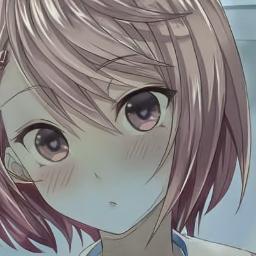}} &
\raisebox{-.5\height}{\includegraphics[width=0.11\linewidth]{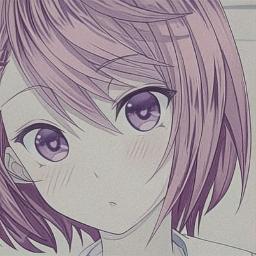}} &
\raisebox{-.5\height}{\includegraphics[width=0.11\linewidth]{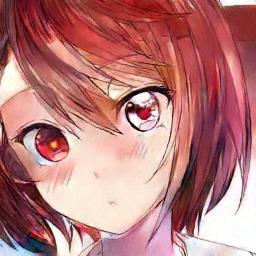}} &
\raisebox{-.5\height}{\includegraphics[width=0.11\linewidth]{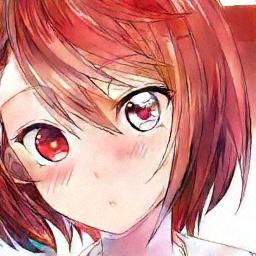}} &
\raisebox{-.5\height}{\includegraphics[width=0.11\linewidth]{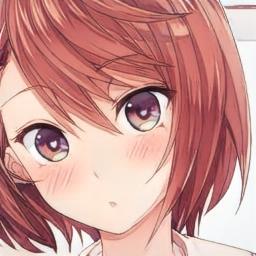}} \\

\end{tabular}

\caption{
\textbf{Qualitative comparison under the same-reference scenario.}
(a) Sketch input. (b) Reference image. (c) SCFT~\cite{Lee2020ReferenceColorization}. (d) AnimeDiffusion~\cite{Cao2024AnimeDiffusion} (pretrained). (e) AnimeDiffusion~\cite{Cao2024AnimeDiffusion} (finetuned). 
(f) AnimeDiffusion (EDM backbone, default $\sigma$-schedule). 
(g) Our model (w/o trajectory refinement). (h) Our model (w/ trajectory refinement).
}

\label{fig:more_same_comparison}
\end{figure}

%% file: sections/appendix/more_cross_reference_table.tex
\begin{figure}[H]
\centering
\renewcommand{\arraystretch}{0.5}
\setlength{\tabcolsep}{2pt}

\begin{tabular}{cccccccc}
\textbf{(a)} & \textbf{(b)} & \textbf{(c)} & \textbf{(d)} & \textbf{(e)} & \textbf{(f)} & \textbf{(g)} & \textbf{(h)} \\

\raisebox{-.5\height}{\includegraphics[width=0.11\linewidth]{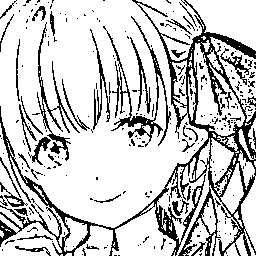}} &
\raisebox{-.5\height}{\includegraphics[width=0.11\linewidth]{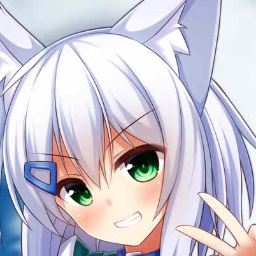}} &
\raisebox{-.5\height}{\includegraphics[width=0.11\linewidth]{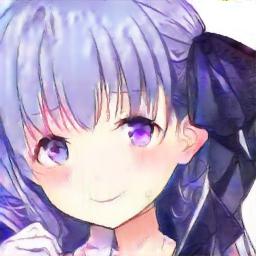}} &
\raisebox{-.5\height}{\includegraphics[width=0.11\linewidth]{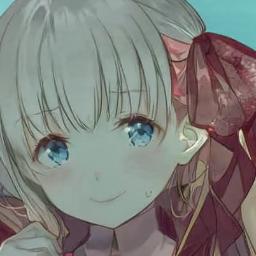}} &
\raisebox{-.5\height}{\includegraphics[width=0.11\linewidth]{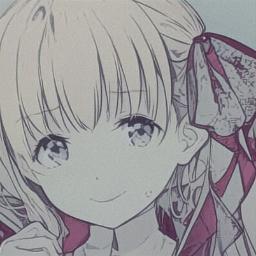}} &
\raisebox{-.5\height}{\includegraphics[width=0.11\linewidth]{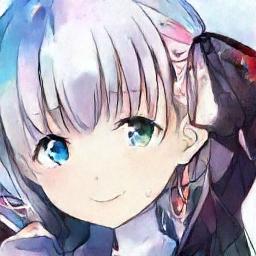}} &
\raisebox{-.5\height}{\includegraphics[width=0.11\linewidth]{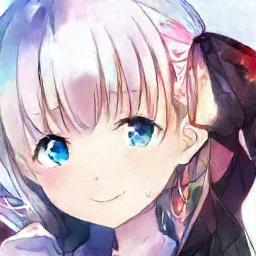}} &
\raisebox{-.5\height}{\includegraphics[width=0.11\linewidth]{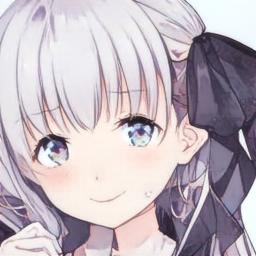}} \\

\raisebox{-.5\height}{\includegraphics[width=0.11\linewidth]{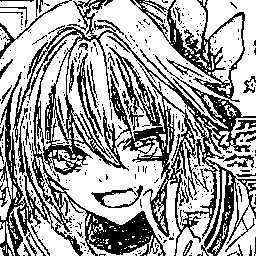}} &
\raisebox{-.5\height}{\includegraphics[width=0.11\linewidth]{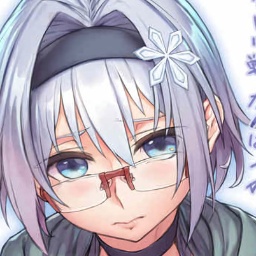}} &
\raisebox{-.5\height}{\includegraphics[width=0.11\linewidth]{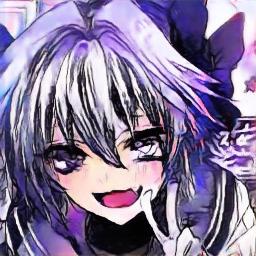}} &
\raisebox{-.5\height}{\includegraphics[width=0.11\linewidth]{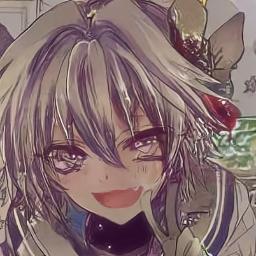}} &
\raisebox{-.5\height}{\includegraphics[width=0.11\linewidth]{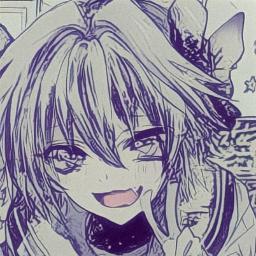}} &
\raisebox{-.5\height}{\includegraphics[width=0.11\linewidth]{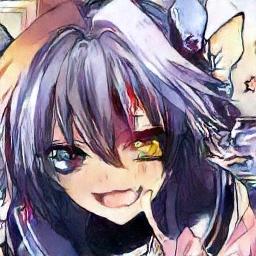}} &
\raisebox{-.5\height}{\includegraphics[width=0.11\linewidth]{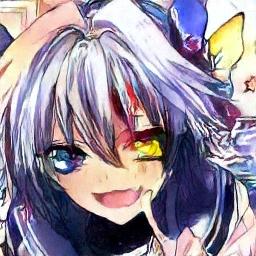}} &
\raisebox{-.5\height}{\includegraphics[width=0.11\linewidth]{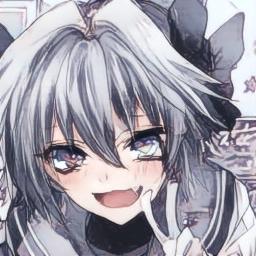}} \\

\raisebox{-.5\height}{\includegraphics[width=0.11\linewidth]{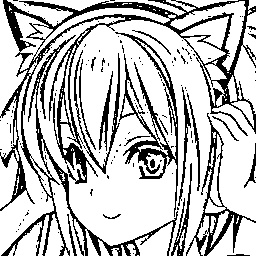}} &
\raisebox{-.5\height}{\includegraphics[width=0.11\linewidth]{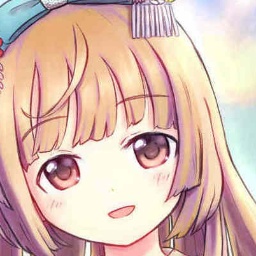}} &
\raisebox{-.5\height}{\includegraphics[width=0.11\linewidth]{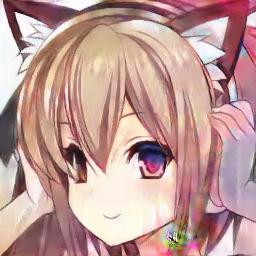}} &
\raisebox{-.5\height}{\includegraphics[width=0.11\linewidth]{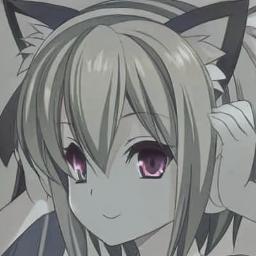}} &
\raisebox{-.5\height}{\includegraphics[width=0.11\linewidth]{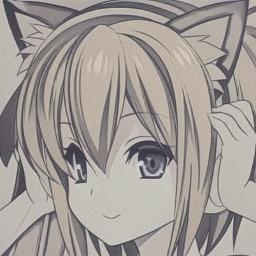}} &
\raisebox{-.5\height}{\includegraphics[width=0.11\linewidth]{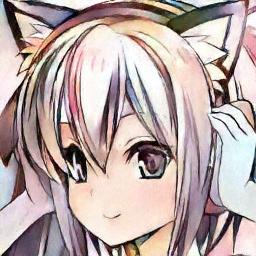}} &
\raisebox{-.5\height}{\includegraphics[width=0.11\linewidth]{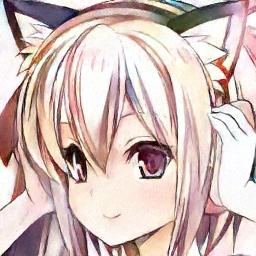}} &
\raisebox{-.5\height}{\includegraphics[width=0.11\linewidth]{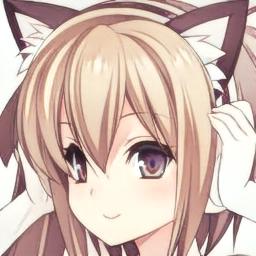}} \\

\raisebox{-.5\height}{\includegraphics[width=0.11\linewidth]{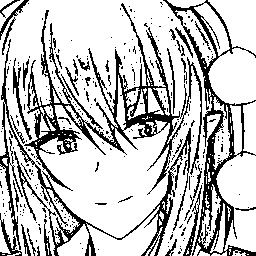}} &
\raisebox{-.5\height}{\includegraphics[width=0.11\linewidth]{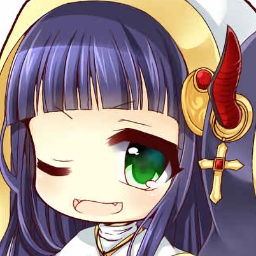}} &
\raisebox{-.5\height}{\includegraphics[width=0.11\linewidth]{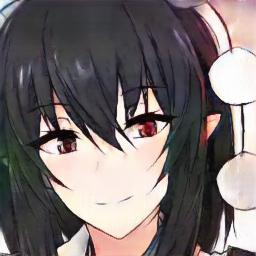}} &
\raisebox{-.5\height}{\includegraphics[width=0.11\linewidth]{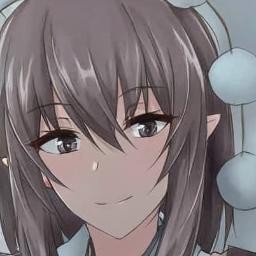}} &
\raisebox{-.5\height}{\includegraphics[width=0.11\linewidth]{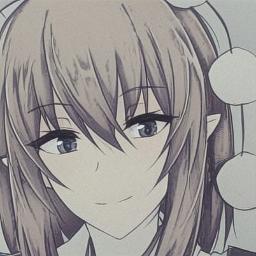}} &
\raisebox{-.5\height}{\includegraphics[width=0.11\linewidth]{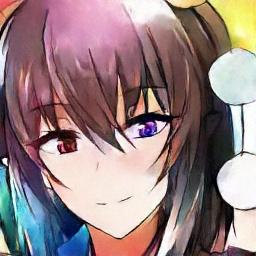}} &
\raisebox{-.5\height}{\includegraphics[width=0.11\linewidth]{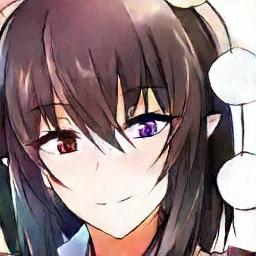}} &
\raisebox{-.5\height}{\includegraphics[width=0.11\linewidth]{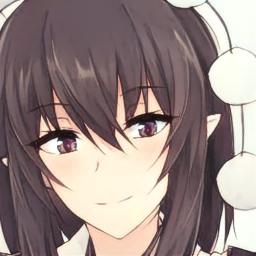}} \\

\raisebox{-.5\height}{\includegraphics[width=0.11\linewidth]{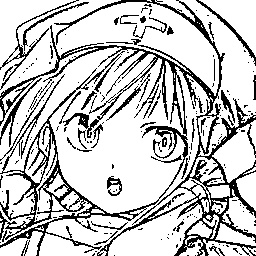}} &
\raisebox{-.5\height}{\includegraphics[width=0.11\linewidth]{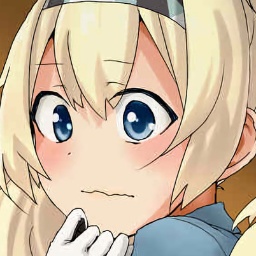}} &
\raisebox{-.5\height}{\includegraphics[width=0.11\linewidth]{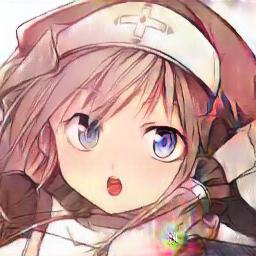}} &
\raisebox{-.5\height}{\includegraphics[width=0.11\linewidth]{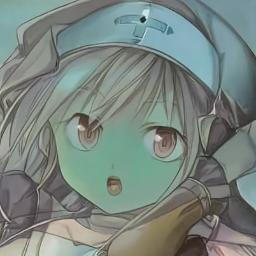}} &
\raisebox{-.5\height}{\includegraphics[width=0.11\linewidth]{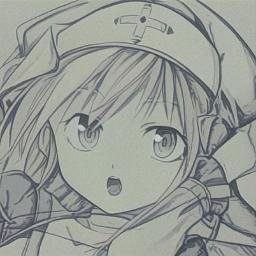}} &
\raisebox{-.5\height}{\includegraphics[width=0.11\linewidth]{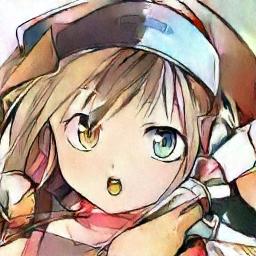}} &
\raisebox{-.5\height}{\includegraphics[width=0.11\linewidth]{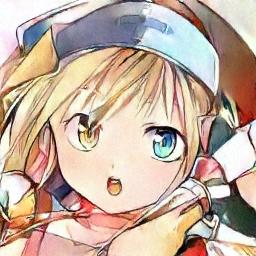}} &
\raisebox{-.5\height}{\includegraphics[width=0.11\linewidth]{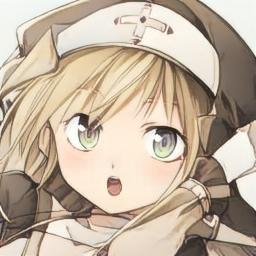}} \\

\raisebox{-.5\height}{\includegraphics[width=0.11\linewidth]{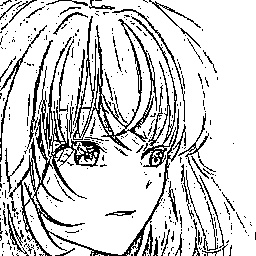}} &
\raisebox{-.5\height}{\includegraphics[width=0.11\linewidth]{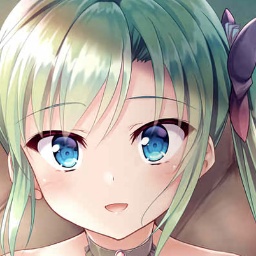}} &
\raisebox{-.5\height}{\includegraphics[width=0.11\linewidth]{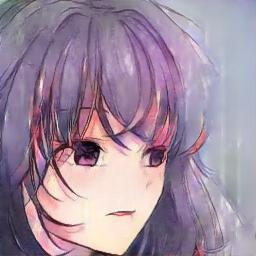}} &
\raisebox{-.5\height}{\includegraphics[width=0.11\linewidth]{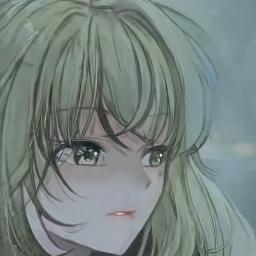}} &
\raisebox{-.5\height}{\includegraphics[width=0.11\linewidth]{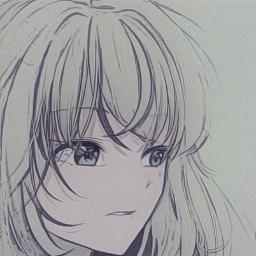}} &
\raisebox{-.5\height}{\includegraphics[width=0.11\linewidth]{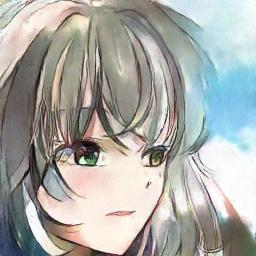}} &
\raisebox{-.5\height}{\includegraphics[width=0.11\linewidth]{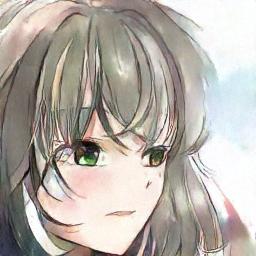}} &
\raisebox{-.5\height}{\includegraphics[width=0.11\linewidth]{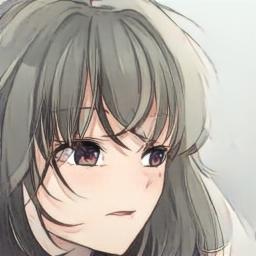}} \\

\raisebox{-.5\height}{\includegraphics[width=0.11\linewidth]{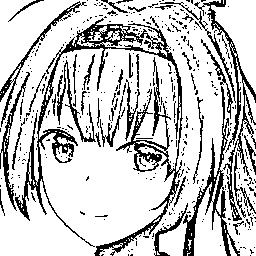}} &
\raisebox{-.5\height}{\includegraphics[width=0.11\linewidth]{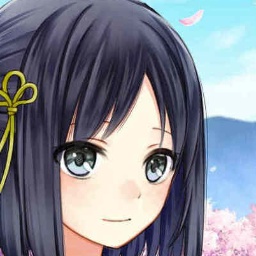}} &
\raisebox{-.5\height}{\includegraphics[width=0.11\linewidth]{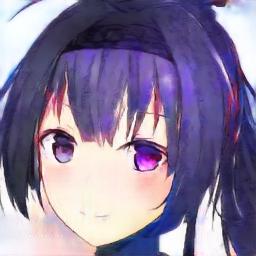}} &
\raisebox{-.5\height}{\includegraphics[width=0.11\linewidth]{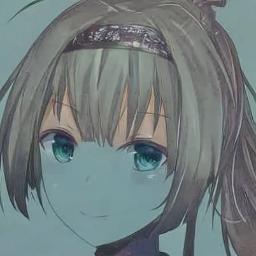}} &
\raisebox{-.5\height}{\includegraphics[width=0.11\linewidth]{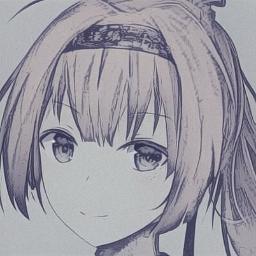}} &
\raisebox{-.5\height}{\includegraphics[width=0.11\linewidth]{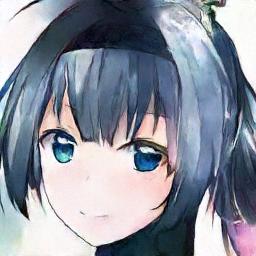}} &
\raisebox{-.5\height}{\includegraphics[width=0.11\linewidth]{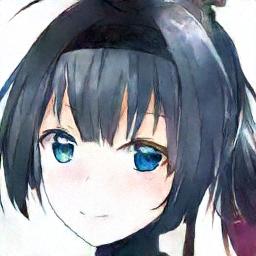}} &
\raisebox{-.5\height}{\includegraphics[width=0.11\linewidth]{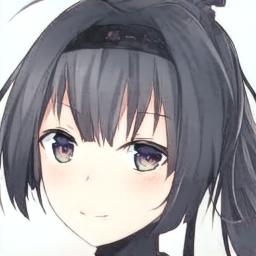}} \\

\raisebox{-.5\height}{\includegraphics[width=0.11\linewidth]{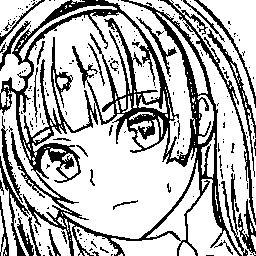}} &
\raisebox{-.5\height}{\includegraphics[width=0.11\linewidth]{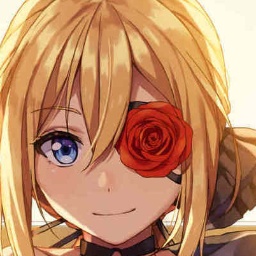}} &
\raisebox{-.5\height}{\includegraphics[width=0.11\linewidth]{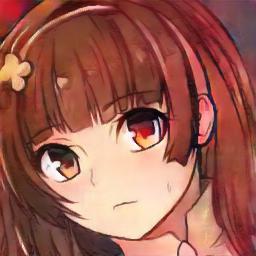}} &
\raisebox{-.5\height}{\includegraphics[width=0.11\linewidth]{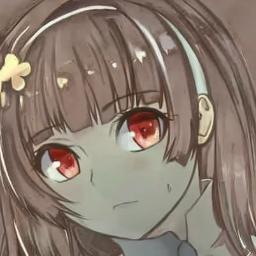}} &
\raisebox{-.5\height}{\includegraphics[width=0.11\linewidth]{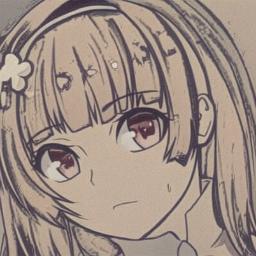}} &
\raisebox{-.5\height}{\includegraphics[width=0.11\linewidth]{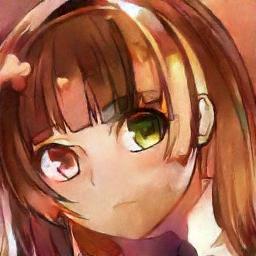}} &
\raisebox{-.5\height}{\includegraphics[width=0.11\linewidth]{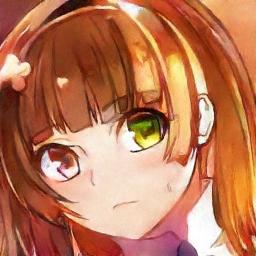}} &
\raisebox{-.5\height}{\includegraphics[width=0.11\linewidth]{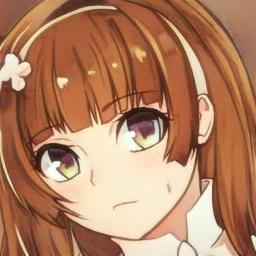}} \\

\raisebox{-.5\height}{\includegraphics[width=0.11\linewidth]{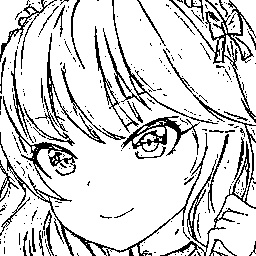}} &
\raisebox{-.5\height}{\includegraphics[width=0.11\linewidth]{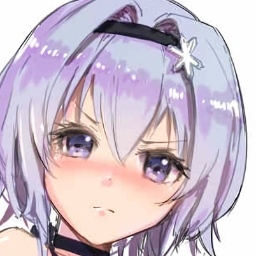}} &
\raisebox{-.5\height}{\includegraphics[width=0.11\linewidth]{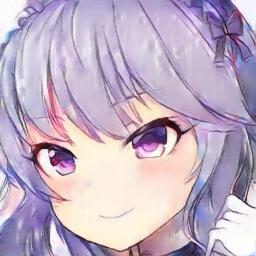}} &
\raisebox{-.5\height}{\includegraphics[width=0.11\linewidth]{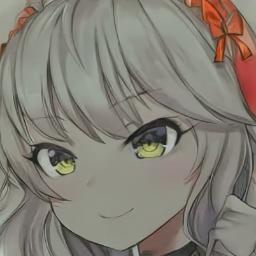}} &
\raisebox{-.5\height}{\includegraphics[width=0.11\linewidth]{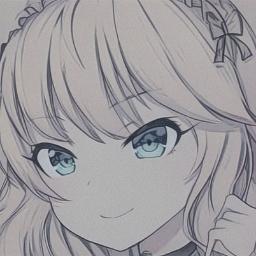}} &
\raisebox{-.5\height}{\includegraphics[width=0.11\linewidth]{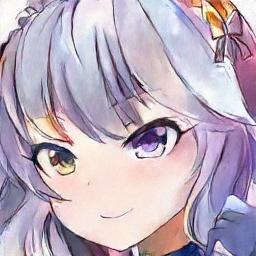}} &
\raisebox{-.5\height}{\includegraphics[width=0.11\linewidth]{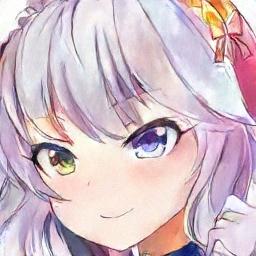}} &
\raisebox{-.5\height}{\includegraphics[width=0.11\linewidth]{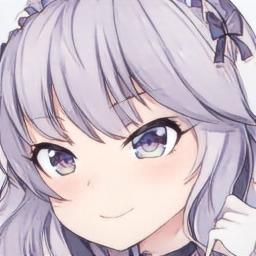}} \\

\raisebox{-.5\height}{\includegraphics[width=0.11\linewidth]{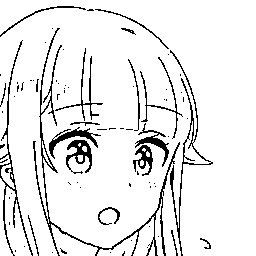}} &
\raisebox{-.5\height}{\includegraphics[width=0.11\linewidth]{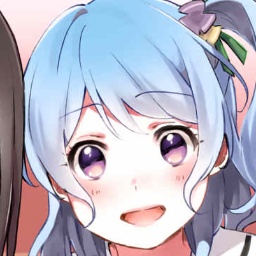}} &
\raisebox{-.5\height}{\includegraphics[width=0.11\linewidth]{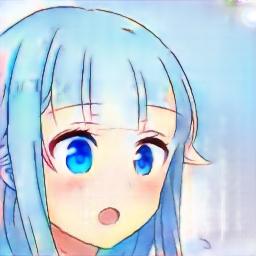}} &
\raisebox{-.5\height}{\includegraphics[width=0.11\linewidth]{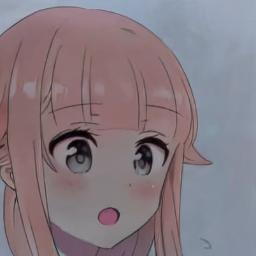}} &
\raisebox{-.5\height}{\includegraphics[width=0.11\linewidth]{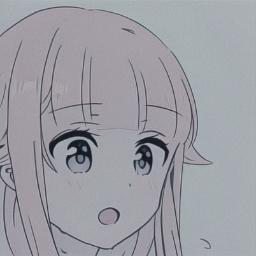}} &
\raisebox{-.5\height}{\includegraphics[width=0.11\linewidth]{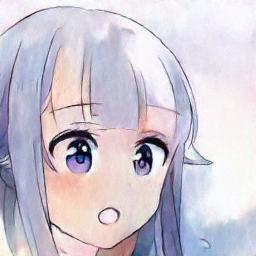}} &
\raisebox{-.5\height}{\includegraphics[width=0.11\linewidth]{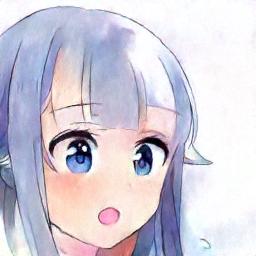}} &
\raisebox{-.5\height}{\includegraphics[width=0.11\linewidth]{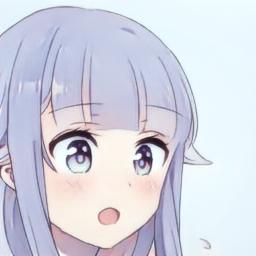}} \\

\raisebox{-.5\height}{\includegraphics[width=0.11\linewidth]{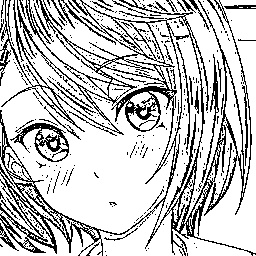}} &
\raisebox{-.5\height}{\includegraphics[width=0.11\linewidth]{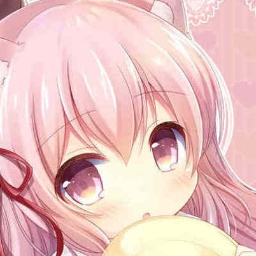}} &
\raisebox{-.5\height}{\includegraphics[width=0.11\linewidth]{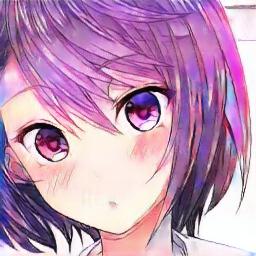}} &
\raisebox{-.5\height}{\includegraphics[width=0.11\linewidth]{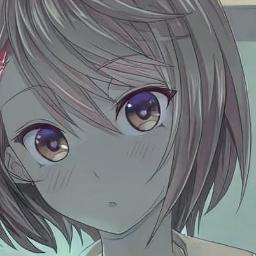}} &
\raisebox{-.5\height}{\includegraphics[width=0.11\linewidth]{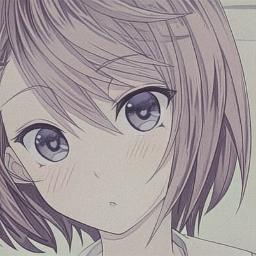}} &
\raisebox{-.5\height}{\includegraphics[width=0.11\linewidth]{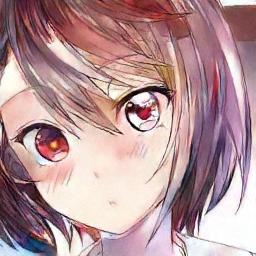}} &
\raisebox{-.5\height}{\includegraphics[width=0.11\linewidth]{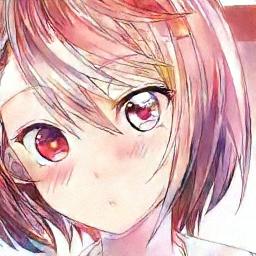}} &
\raisebox{-.5\height}{\includegraphics[width=0.11\linewidth]{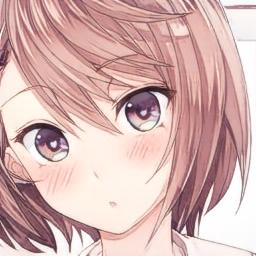}} \\

\end{tabular}

\vspace{0.5em}
\caption{
\textbf{Qualitative comparison under the cross-reference scenario.}
(a) Sketch input. (b) Reference image. (c) SCFT~\cite{Lee2020ReferenceColorization}. (d) AnimeDiffusion~\cite{Cao2024AnimeDiffusion} (pretrained). (e) AnimeDiffusion~\cite{Cao2024AnimeDiffusion} (finetuned). 
(f) AnimeDiffusion (EDM backbone, default $\sigma$-schedule).
(g) Our model (w/o trajectory refinement). (h) Our model (w/ trajectory refinement).
}
\label{fig:more_cross_comparison}
\end{figure}

%% file: sections/appendix/no_rotation_table.tex
\begin{figure}[H]
\centering
\renewcommand{\arraystretch}{0.5}
\setlength{\tabcolsep}{2pt}

\begin{tabular}{ccccccc}
\textbf{(a)} & \textbf{(b)} & \textbf{(c)} & \textbf{(d)} & \textbf{(e)} & \textbf{(f)} & \textbf{(g)} \\

\raisebox{-.5\height}{\includegraphics[width=0.13\linewidth]{images/diff_sketch/6.jpg}} &
\raisebox{-.5\height}{\includegraphics[width=0.13\linewidth]{images/same_reference/6.jpg}} &
\raisebox{-.5\height}{\includegraphics[width=0.13\linewidth]{images/diff_reference/6.jpg}} &
\raisebox{-.5\height}{\includegraphics[width=0.13\linewidth]{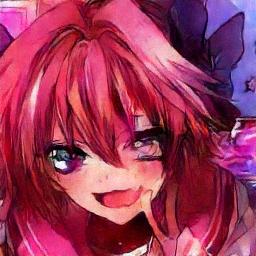}} &
\raisebox{-.5\height}{\includegraphics[width=0.13\linewidth]{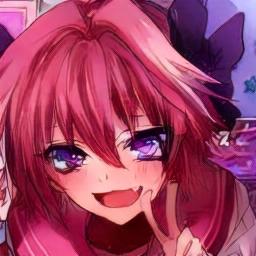}} &
\raisebox{-.5\height}{\includegraphics[width=0.13\linewidth]{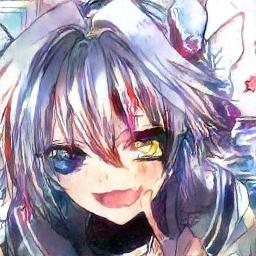}} &
\raisebox{-.5\height}{\includegraphics[width=0.13\linewidth]{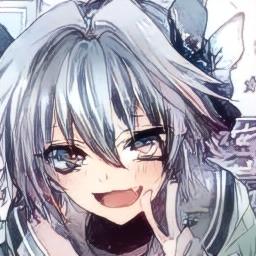}} \\

\raisebox{-.5\height}{\includegraphics[width=0.13\linewidth]{images/diff_sketch/7.jpg}} &
\raisebox{-.5\height}{\includegraphics[width=0.13\linewidth]{images/same_reference/7.jpg}} &
\raisebox{-.5\height}{\includegraphics[width=0.13\linewidth]{images/diff_reference/7.jpg}} &
\raisebox{-.5\height}{\includegraphics[width=0.13\linewidth]{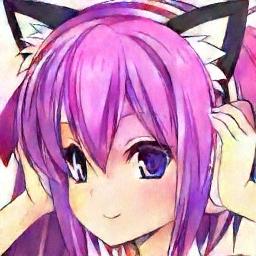}} &
\raisebox{-.5\height}{\includegraphics[width=0.13\linewidth]{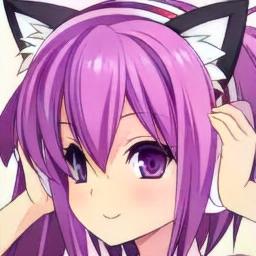}} &
\raisebox{-.5\height}{\includegraphics[width=0.13\linewidth]{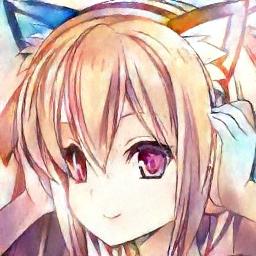}} &
\raisebox{-.5\height}{\includegraphics[width=0.13\linewidth]{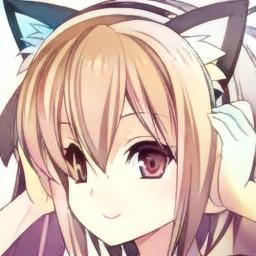}} \\

\raisebox{-.5\height}{\includegraphics[width=0.13\linewidth]{images/diff_sketch/8.jpg}} &
\raisebox{-.5\height}{\includegraphics[width=0.13\linewidth]{images/same_reference/8.jpg}} &
\raisebox{-.5\height}{\includegraphics[width=0.13\linewidth]{images/diff_reference/8.jpg}} &
\raisebox{-.5\height}{\includegraphics[width=0.13\linewidth]{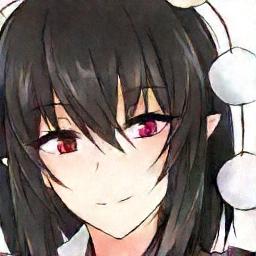}} &
\raisebox{-.5\height}{\includegraphics[width=0.13\linewidth]{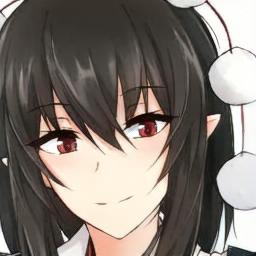}} &
\raisebox{-.5\height}{\includegraphics[width=0.13\linewidth]{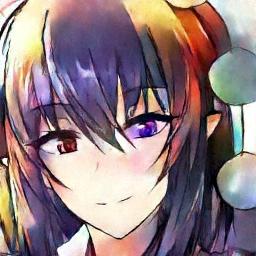}} &
\raisebox{-.5\height}{\includegraphics[width=0.13\linewidth]{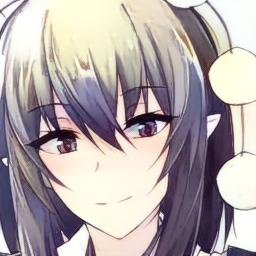}} \\

\raisebox{-.5\height}{\includegraphics[width=0.13\linewidth]{images/diff_sketch/9.jpg}} &
\raisebox{-.5\height}{\includegraphics[width=0.13\linewidth]{images/same_reference/9.jpg}} &
\raisebox{-.5\height}{\includegraphics[width=0.13\linewidth]{images/diff_reference/9.jpg}} &
\raisebox{-.5\height}{\includegraphics[width=0.13\linewidth]{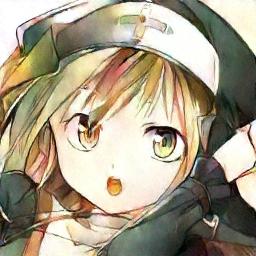}} &
\raisebox{-.5\height}{\includegraphics[width=0.13\linewidth]{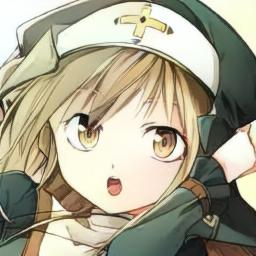}} &
\raisebox{-.5\height}{\includegraphics[width=0.13\linewidth]{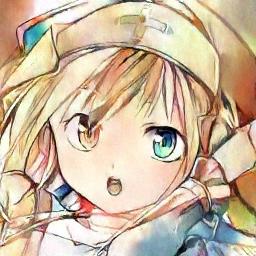}} &
\raisebox{-.5\height}{\includegraphics[width=0.13\linewidth]{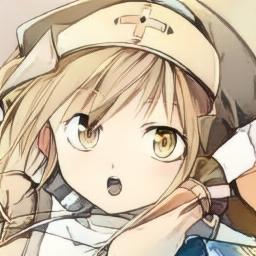}} \\

\raisebox{-.5\height}{\includegraphics[width=0.13\linewidth]{images/diff_sketch/10.jpg}} &
\raisebox{-.5\height}{\includegraphics[width=0.13\linewidth]{images/same_reference/10.jpg}} &
\raisebox{-.5\height}{\includegraphics[width=0.13\linewidth]{images/diff_reference/10.jpg}} &
\raisebox{-.5\height}{\includegraphics[width=0.13\linewidth]{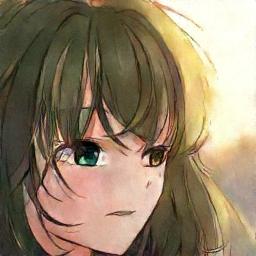}} &
\raisebox{-.5\height}{\includegraphics[width=0.13\linewidth]{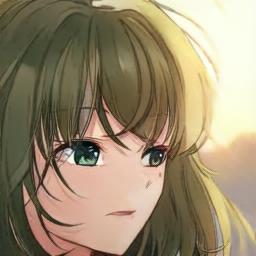}} &
\raisebox{-.5\height}{\includegraphics[width=0.13\linewidth]{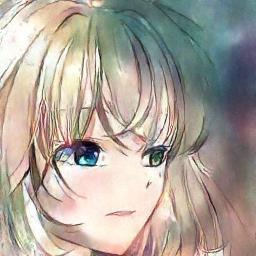}} &
\raisebox{-.5\height}{\includegraphics[width=0.13\linewidth]{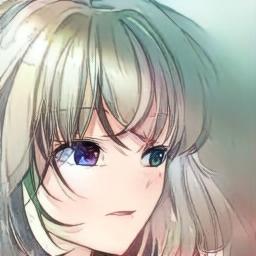}} \\

\raisebox{-.5\height}{\includegraphics[width=0.13\linewidth]{images/diff_sketch/11.jpg}} &
\raisebox{-.5\height}{\includegraphics[width=0.13\linewidth]{images/same_reference/11.jpg}} &
\raisebox{-.5\height}{\includegraphics[width=0.13\linewidth]{images/diff_reference/11.jpg}} &
\raisebox{-.5\height}{\includegraphics[width=0.13\linewidth]{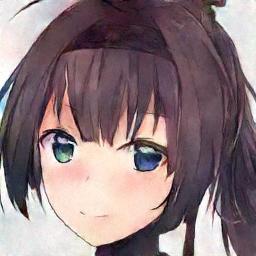}} &
\raisebox{-.5\height}{\includegraphics[width=0.13\linewidth]{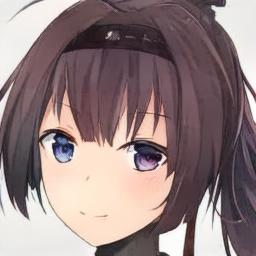}} &
\raisebox{-.5\height}{\includegraphics[width=0.13\linewidth]{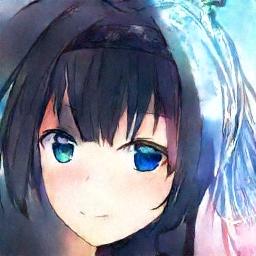}} &
\raisebox{-.5\height}{\includegraphics[width=0.13\linewidth]{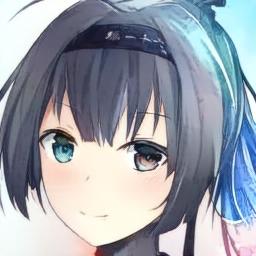}} \\

\raisebox{-.5\height}{\includegraphics[width=0.13\linewidth]{images/diff_sketch/12.jpg}} &
\raisebox{-.5\height}{\includegraphics[width=0.13\linewidth]{images/same_reference/12.jpg}} &
\raisebox{-.5\height}{\includegraphics[width=0.13\linewidth]{images/diff_reference/12.jpg}} &
\raisebox{-.5\height}{\includegraphics[width=0.13\linewidth]{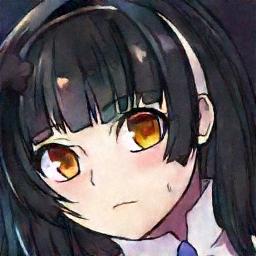}} &
\raisebox{-.5\height}{\includegraphics[width=0.13\linewidth]{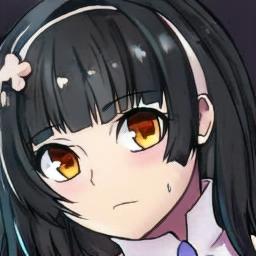}} &
\raisebox{-.5\height}{\includegraphics[width=0.13\linewidth]{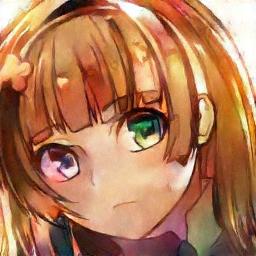}} &
\raisebox{-.5\height}{\includegraphics[width=0.13\linewidth]{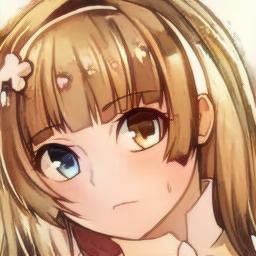}} \\

\raisebox{-.5\height}{\includegraphics[width=0.13\linewidth]{images/diff_sketch/13.jpg}} &
\raisebox{-.5\height}{\includegraphics[width=0.13\linewidth]{images/same_reference/13.jpg}} &
\raisebox{-.5\height}{\includegraphics[width=0.13\linewidth]{images/diff_reference/13.jpg}} &
\raisebox{-.5\height}{\includegraphics[width=0.13\linewidth]{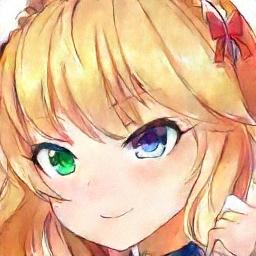}} &
\raisebox{-.5\height}{\includegraphics[width=0.13\linewidth]{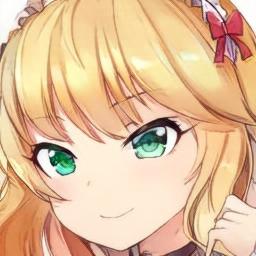}} &
\raisebox{-.5\height}{\includegraphics[width=0.13\linewidth]{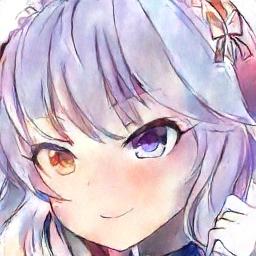}} &
\raisebox{-.5\height}{\includegraphics[width=0.13\linewidth]{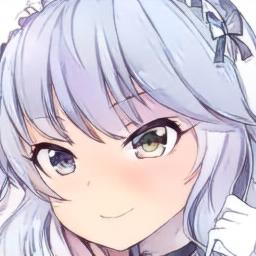}} \\

\raisebox{-.5\height}{\includegraphics[width=0.13\linewidth]{images/diff_sketch/14.jpg}} &
\raisebox{-.5\height}{\includegraphics[width=0.13\linewidth]{images/same_reference/14.jpg}} &
\raisebox{-.5\height}{\includegraphics[width=0.13\linewidth]{images/diff_reference/14.jpg}} &
\raisebox{-.5\height}{\includegraphics[width=0.13\linewidth]{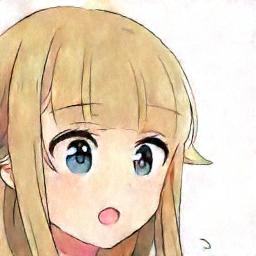}} &
\raisebox{-.5\height}{\includegraphics[width=0.13\linewidth]{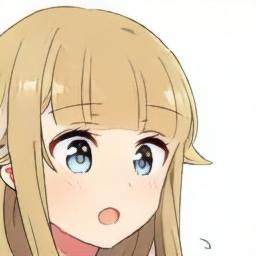}} &
\raisebox{-.5\height}{\includegraphics[width=0.13\linewidth]{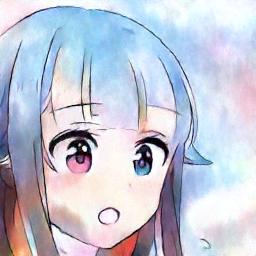}} &
\raisebox{-.5\height}{\includegraphics[width=0.13\linewidth]{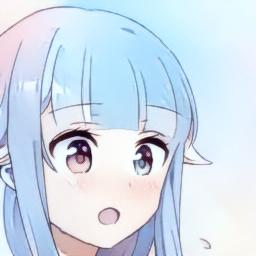}} \\

\raisebox{-.5\height}{\includegraphics[width=0.13\linewidth]{images/diff_sketch/15.jpg}} &
\raisebox{-.5\height}{\includegraphics[width=0.13\linewidth]{images/same_reference/15.jpg}} &
\raisebox{-.5\height}{\includegraphics[width=0.13\linewidth]{images/diff_reference/15.jpg}} &
\raisebox{-.5\height}{\includegraphics[width=0.13\linewidth]{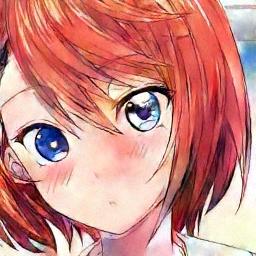}} &
\raisebox{-.5\height}{\includegraphics[width=0.13\linewidth]{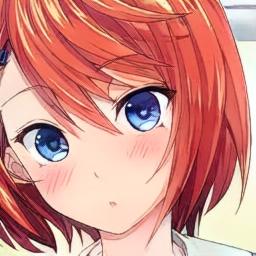}} &
\raisebox{-.5\height}{\includegraphics[width=0.13\linewidth]{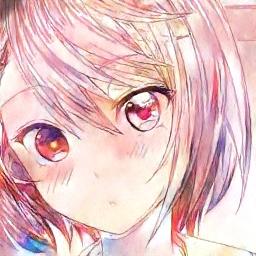}} &
\raisebox{-.5\height}{\includegraphics[width=0.13\linewidth]{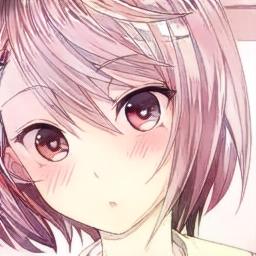}} \\

\end{tabular}

\vspace{0.5em}

\caption{
\textbf{Comparison under same- and cross-reference scenarios without TPS rotation.}
(a) Sketch input. (b) Reference image (same style). (c) Reference image (cross style).
(d–e) Our model under same-reference scenario (w/o and w/ trajectory refinement, no TPS rotation).
(f–g) Our model under cross-reference scenario (w/o and w/ trajectory refinement, no TPS rotation).
Even without explicit rotation-based alignment, our model preserves structural integrity and transfers style consistently across reference domains, outperforming baselines in both scenarios.
}

\label{fig:no_rotation}
\end{figure}
\FloatBarrier

%% file: sections/appendix/implementation_details.tex
\section{Implementation Details}
\label{implementation_details}

To ensure rigorous and reproducible comparisons, we reimplemented each baseline model under a standardized pipeline. All models were trained and evaluated on the same dataset split, using identical image resolution ($256 \times 256$), batch size (32), and consistent data augmentation strategy.

\textbf{Hardware environment :} 2$\times$ NVIDIA H100 SXM5 GPUs with a 128-core AMD EPYC 9354 CPU and 512GB RAM. Experiments were conducted using PyTorch 2.1.0 with AMP-based mixed-precision training.

\textbf{Common hyperparameters :}
\begin{itemize}
    \item Optimizer: AdamW; Learning rate: $1\times10^{-4}$; Weight decay: $1\times10^{-2}$
    \item Scheduler: Cosine decay with 1 epoch warmup
    \item Epochs: 300; Batch size: 32; Gradient clipping: max-norm of 1.0
    \item Distributed training via PyTorch Lightning DDP; 64 data loading workers
\end{itemize}

\subsection{Pretraining Comparisons}

For fair comparison of the \textbf{pretraining phase}, we evaluated models based on their ability to learn from distorted reference inputs and produce structure-preserving reconstructions.

\textbf{SCFT~\cite{Lee2020ReferenceColorization} :}
\begin{itemize}
    \item Dense semantic correspondence-based reference transfer model originally designed for exemplar-guided colorization
    \item Adapted to $256 \times 256$ resolution
    \item Trained from scratch on our dataset with the same optimizer, learning rate schedule, and number of epochs
\end{itemize}

\textbf{AnimeDiffusion~\cite{Cao2024AnimeDiffusion} :}
\begin{itemize}
    \item Diffusion-based colorization model trained with fixed iDDPM-style $\beta$-schedule~\cite{Ho2020DDPM}
    \item Inference conducted using 50 denoising steps with DDIM~\cite{Song2021DDIM}
    \item Official implementation modified for consistent data split and batch size
\end{itemize}

\subsection{Finetuning Comparisons}

\textbf{Finetuning Settings :}
\begin{itemize}
    \item Strategy: MSE, depending on baseline capability
    \item Inference time steps: 50 (Euler or DDIM sampling for diffusion models)
    \item Finetuning conducted with preloaded pretrained weights on the same hardware
\end{itemize}

\textbf{AnimeDiffusion~\cite{Cao2024AnimeDiffusion} :}
\begin{itemize}
    \item MSE-based perceptual finetuning with 50-step DDIM inference~\cite{Song2021DDIM}
    \item Reference and sketch inputs preserved; distorted images created via noise+augmentation
\end{itemize}

\textbf{SSIMBaD (Ours):}
\begin{itemize}
    \item Pretrained with SSIM-aligned $\phi^*(\sigma)$ schedule for uniform perceptual degradation
    \item Finetuned using MSE loss, with explicit control over 50 step inference trajectory
\end{itemize}

\subsection{Evaluation Metrics :}

For both stages, we report PSNR, MS-SSIM~\cite{wang2003multiscale}, and FID~\cite{heusel2017gans}. All models were evaluated using 50-step sampling, and outputs were resized to $256 \times 256$ prior to metric computation.